\documentclass[aps,prb,twocolumn,showpacs,floatfix,superscriptaddress]{revtex4}

\usepackage{amsmath,amssymb}
\usepackage[dvips]{graphics}
\usepackage{epsfig}
\usepackage{times}

\let\up\uparrow
\let\down\downarrow

\newcommand{\varD}{{\mathcal{D}}}
\newcommand{\varE}{{\mathcal{E}}}
\newcommand{\varH}{{\mathcal{H}}}
\newcommand{\varS}{{\mathcal{S}}}

\newcommand{\bfk}{{\mathbf{k}}}
\newcommand{\bfS}{{\mathbf{S}}}
\newcommand{\bfs}{{\mathbf{s}}}
\newcommand{\bfsigma}{{\boldsymbol{\sigma}}}

\newcommand{\Jst}{{J_{\rm I}}}

\newcommand{\Braket}[1]{\mathinner{\langle{\textstyle#1}\rangle}}
\newcommand{\avg}[1]{\left\langle#1\right\rangle}
\newcommand{\ket}[1]{\left|#1\right\rangle}
\newcommand{\rcp}[1]{\frac{1}{#1}}

\newcommand{\eqnref}[1]{Eq.~(\ref{#1})}
\newcommand{\eqnsref}[1]{Eqs.~(\ref{#1})}

\newcommand{\figref}[1]{Fig.~\ref{#1}}
\newcommand{\figsref}[1]{Figs.~\ref{#1}}
\newcommand{\Figref}[1]{Figure~\ref{#1}}
\newcommand{\Figsref}[1]{Figures~\ref{#1}}

\newcommand{\secref}[1]{Sec.~\ref{#1}}

\begin{document}
\title{Josephson effect through a multilevel dot near a singlet-triplet transition}
\author{Minchul Lee}
\affiliation{Department of Applied Physics, College of Applied Science, Kyung Hee University, Yongin 449-701, Korea}
\author{Thibaut Jonckheere}
\affiliation{Centre de Physique Th\'eorique, UMR6207, Case 907, Luminy, 13288 Marseille Cedex 9, France}
\author{Thierry Martin}
\affiliation{Centre de Physique Th\'eorique, UMR6207, Case 907, Luminy, 13288 Marseille Cedex 9, France}
\affiliation{Universit\'e de la M\'editerran\'ee, 13288 Marseille Cedex 9, France}

\begin{abstract}
  We investigate the Josephson effect through a two-level quantum dot with an
  exchange coupling between two dot electrons. We compute the superconducting
  phase relationship and construct the phase diagram in the superconducting
  gap--exchange coupling plane in the regime of the singlet-triplet transition
  driven by the exchange coupling. In our study two configurations for the
  dot-lead coupling are considered: one where effectively only one channel
  couples to the dot, and the other where the two dot orbitals have opposite
  parities. Perturbative analysis in the weak-coupling limit reveals that the
  system experiences transitions from 0 to $\pi$ (negative critical current)
  behavior, depending on the parity of the orbitals and the spin correlation
  between dot electrons.  The strong coupling regime is tackled with the
  numerical renormalization group method, which first characterizes the Kondo
  correlations due to the dot-lead coupling and the exchange coupling in the
  absence of superconductivity. In the presence of superconductivity,
  many-body correlations such as two-stage Kondo effect compete with the
  superconductivity and the comparison between the gap and the relevant Kondo
  temperature scales allows to predict a rich variety of phase diagrams for the
  ground state of the system and for the Josephson current. Numerical
  calculations predicts that our system can exhibit Kondo-driven 0-$\pi$-0 or
  $\pi$-0-$\pi$ double transitions and, more interestingly, that if proper
  conditions are met a Kondo-assisted $\pi$-junction can arise, which is
  contrary to a common belief that the Kondo effect opens a resonant level and
  makes the 0-junction. Our predictions could be probed experimentally for a
  buckminster fullerene sandwiched between two superconductors.

\end{abstract}

\maketitle

\section{Introduction}

The Josephson effect\cite{Josephson,Tinkham96} is one of the most celebrated
manifestation of many body correlations in condensed matter physics: a Cooper
pair current\cite{Bardeen57} between two bulk superconductors separated by an
intermediate region or arbitrary nature can flow even in the absence of an
applied bias. Over the last few decades, the Josephson effect has become a very
active field of
theoretical\cite{Shiba69,Glazman89,Spivak91,Yeyati97,Shimizu98,Rozhkov99,ChoiMS00,Rozhkov01,Makhlin01,Zaikin04,Choi04,Siano04,Lee08,Karrasch08}
and
experimental\cite{Kasumov99,Kasumov01,Scheer01,Ryazanov01,Kontos02,Buitelaar02,Kasumov05,vanDam06,Jarillo-Herrero06,Jorgensen06,Cleuziou06,Eichler07,Sand-Jespersen07,Winkelmann09,Eichler09}
investigation in the context of mesoscopic devices, devices which are small
enough that electron transport occurs in a phase coherent manner.  Because the
tunneling of Cooper pairs through the junction is greatly affected by the
physical properties of the segment between superconducting electrodes, the
study of the Josephson current provides a new way to investigate the electronic
properties of the medium.  In early days, thin layers of insulators and metals
were used to form the Josephson junction.\cite{Golubov04} Advance in
nanofabrication technology can now enable one to make the middle segment small
enough to be considered as a quantum dot (QD), a zero dimensional entity
bridging the two
superconductors.\cite{Buitelaar02,Cleuziou06,Eichler07,Sand-Jespersen07,Eichler09}
Furthermore, even a real (or artificial) molecule can be inserted between two
closely positioned superconducting leads to form a molecular Josephson junction
(MJJ).\cite{Kasumov01,Kasumov05,Winkelmann09} Quantum dots connected to normal
metal leads are known to exhibit the Coulomb blockade phenomenon due to their
large charging energies.\cite{Grabert92} Interestingly, at low enough
temperatures, when an odd number of electrons occupy the QD, one can reach the
Kondo regime.\cite{Goldhaber-Gordon98,Cronenwett98,vanderWiel00} Yet the leads
can be chosen to be superconductors which lead to a competition between Kondo
physics and
superconductivity.\cite{Choi04,Siano04,Lee08,Karrasch08,Winkelmann09,Eichler09}
The purpose of the present work is precisely to study the Josephson effect
through a multilevel quantum dot in this context, with applications to
molecular spintronics.

Indeed, the QD-JJ have received a great theoretical and experimental attention
because they can exhibit an interesting competition between two many-body
correlations: the superconductivity and the Kondo effect. Due to its small
size, the QD has a large Coulomb charging energy. The Kondo effect then emerges
for such a small QD coupled strongly to the leads when the QD has a localized
magnetic moment, that is, nonzero total spin of electrons in it. At
temperatures below the so-called Kondo temperature $T_K$,\cite{Haldane78} the
conduction electrons in the leads screen the localized moment through multiple
cotunneling spin-flip processes, forming a spin-singlet ground state, and
induces a resonance level at the Fermi energy, which increases the linear
conductance up to the unitary-limit value ($=2e^2/h$) that is otherwise
completely suppressed due to the strong Coulomb repulsion. If the leads consist
of $s$-wave superconductors, the conduction electrons form spin-singlet Cooper
pairs incapable of flipping the QD spin. It has been
known,\cite{Shiba69,Glazman89,Spivak91,Rozhkov99,Rozhkov01} and recently
probed,\cite{Ryazanov01,Kontos02,vanDam06} that in the weak dot-lead coupling
limit the large Coulomb repulsion only allows the electrons in a Cooper pair to
tunnel one by one via virtual processes in which the spin ordering of the pair
is reversed, leading to a $\pi$ junction, and that the localized moment remains
unscreened. In the opposite limit where the Kondo temperature exceeds the
superconducting gap $\Delta$, however, the induced Kondo resonance level
restores the 0 junction state of the
supercurrent.\cite{Glazman89,Choi04,Siano04,Karrasch08} As a result, one can
drive a phase transition between spin singlet (0 junction) and doublet ($\pi$
junction) states by changing the relative strengths of $T_K$ and $\Delta$.

Current issues about electronic transport through a QD or a molecule go beyond
the spin-degenerate single-level model and take into account multi-level
structures and/or possible magnetic interactions. For example, the theoretical
prediction that the two-level quantum dots (TLQDs) with spin exchange
interaction coupled to normal-metal leads can experience a quantum phase
transition, specifically the singlet-triplet
transition,\cite{Izumida01,Hofstetter02} was recently confirmed by two
independent experiments.\cite{Quay07,Roch08} The transition was observed to
accompany a drastic change in the transport mechanism, and it was also found
that the spin exchange coupling between electrons could suppress the Kondo
correlation completely or alter its physical nature by changing the screening
mechanism. The influence of such a magnetic interaction on the Josephson
current was also studied for a MJJ where the molecule is modeled by a
single-level QD having spin exchange coupling\cite{Elste05} between spins of QD
electron and a metal ion.\cite{Lee08} It was predicted that the state of the
supercurrent can be switched between 0 and $\pi$ junctions by tuning the
magnetic interaction. On the other hand, theoretical
calculations\cite{Shimizu98,Rozhkov01} and experiments\cite{vanDam06} have
shown that the Josephson junction made of a multi-level quantum dot in the
weak-coupling limit can behave as a $\pi$ junction even when the dot is
nonmagnetic without a localized spin and vice versa. The studies found out the
significant roles of (1) the off-diagonal Cooper pair tunneling
process\cite{Shimizu98} in which two electrons in the pair are transferred via
different orbitals in the QD and (2) the parity of the QD orbital wave
functions\cite{Rozhkov01} that determine the relative sign of the dot-lead
couplings.

In this paper we study the electronic transport through a Josephson junction
having in it a TLQD with the spin exchange interaction between electrons in two
orbital levels. Here we focus on the regime where the doubly-occupied QD
experiences the singlet-triplet transition due to the spin exchange coupling
that is tunable by the gate voltage. The physical properties of the ground
state and the supercurrent-phase relation (SPR) through the junction are
examined as the strengths of the superconductivity and the spin exchange
coupling are varied. In order to study both of the weak- and strong-coupling
limits we exploit the numerical renormalization group (NRG) method which can
take into account the Coulomb interaction in a nonperturbative way. In
additions, the physical understanding of the numerical outcome is supplemented
by the analytical analysis such as fourth-order perturbation theory and scaling
theory.

Our main findings are summarized as follows: (1) The sign of the supercurrent
is determined by the competition between diagonal and off-diagonal tunneling
processes whose strength and sign can be controlled by the parity of the
orbital wave functions and the spin correlation present in the dot. (2) The
origin and physical property of the TLQD-JJ can be explained in terms of the
competition between the superconductivity and the Kondo correlation found from
the normal-lead counterpart of the system. For example, the two-stage Kondo
effect leads to 0-$\pi$-0 or $\pi$-0-$\pi$ double transitions with the exchange
coupling or the superconducting gap. (3) When the superconducting phase
difference between two leads is maximal, the existing Kondo correlation is
greatly affected. Interestingly, we observed that a Kondo-assisted
$\pi$-junction can arise if some conditions are met.

This paper is organized as follows: In Sec.~II we describe the model
Hamiltonian of the TLQD-JJ and specify the regimes that we are interested
in. The weak-coupling limit is studied by using the fourth-order perturbation
analysis in Sec.~III. Section~IV presents the results of the NRG calculations
applied to the weak- and strong-coupling limits and the phase diagrams of the
system with respect to the properties of the SPR. In Sec.~V we summarize our
study.

\section{Model}

\begin{figure}[!b]
  \centering
  \begin{minipage}{4.25cm}
    \centering
    \includegraphics[width=4cm]{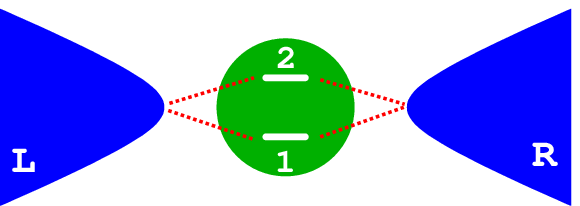}
  \end{minipage}\hfill
  \begin{minipage}{4.25cm}
    \centering
    \includegraphics[width=4cm]{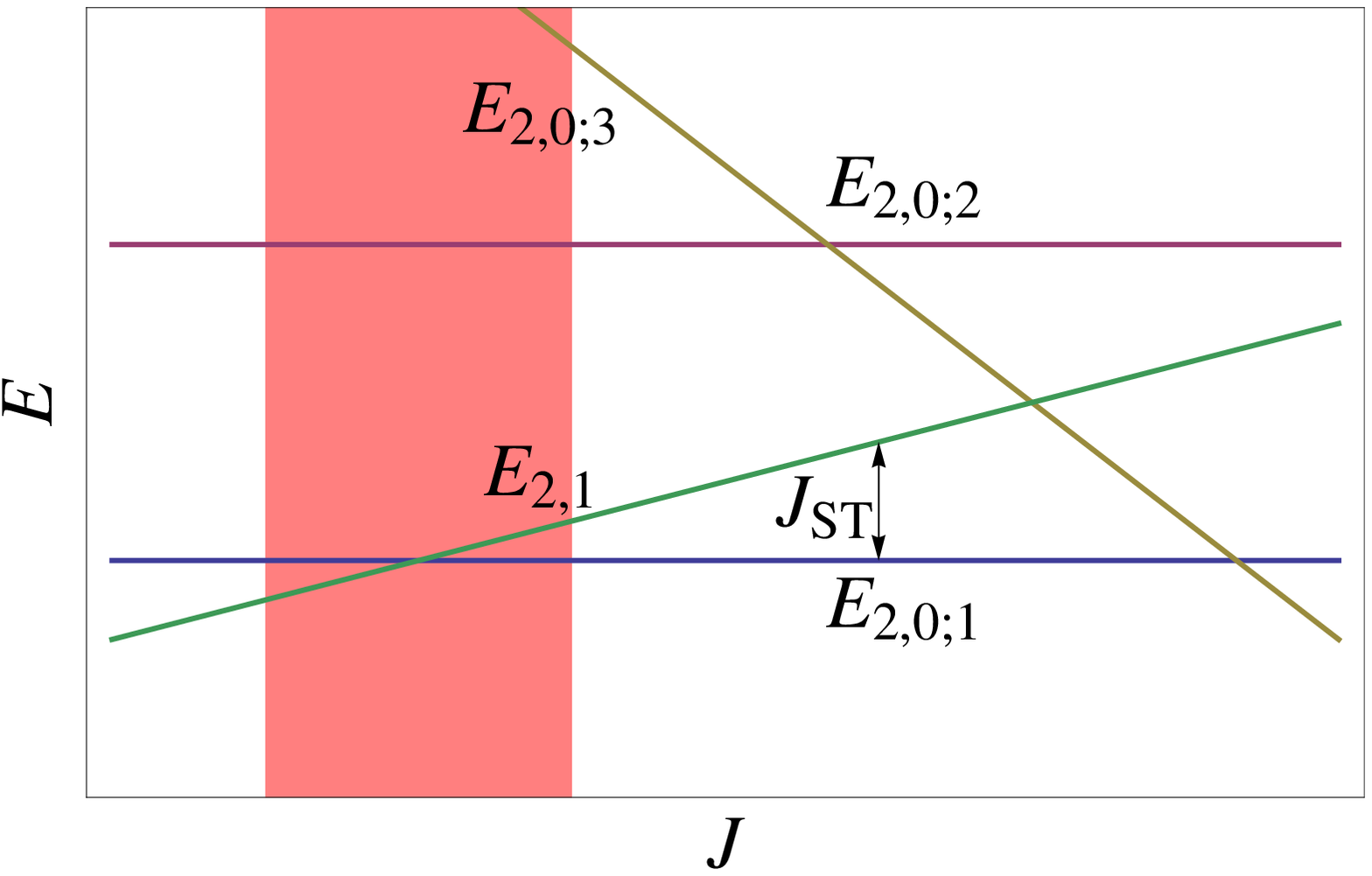}
  \end{minipage}
  \caption{(color online) (LEFT) Sketch of the TLQD connected to two $s$-wave
    superconducting leads. (RIGHT) Energy levels $E_{Q,S;\alpha}$ of
    two-electron $(Q=2)$ states in the TLQD as functions of the exchange
    coupling $J$. The shaded region is of our interest, where the
    singlet-triplet transition occurs in the ground state.}
  \label{fig:tlqd}
\end{figure}

The TLQD connected to two single-channel $s$-wave superconducting leads as
shown in \figref{fig:tlqd} is modeled by the two-impurity Anderson model:
$\varH = \varH_{\rm DD} + \varH_{\rm LD} + \varH_{\rm T}$, where
\begin{align}
  \varH_{\rm DD}
  & =
  \sum_i (\epsilon_i n_i + U n_{i\up} n_{i\down}) + U n_1 n_2
  + J \bfS_1\cdot\bfS_2
  \\
  \varH_{\rm LD}
  & =
  \sum_{\ell\bfk}
  \left[
    \epsilon_\bfk n_{\ell\bfk}
    -
    \left(
      \Delta\, e^{i\phi_\ell} c_{\ell\bfk\up}^\dag c_{\ell-\bfk\down}^\dag
      + (h.c.)
    \right)
  \right]
  \\
  \varH_{\rm T}
  & =
  \sum_{i\ell\bfk\mu}
  \left[t_{i\ell} \, d_{i\mu}^\dag c_{\ell\bfk\mu} + (h.c.)\right].
\end{align}
Here $c_{\ell\bfk\mu}$ ($d_{i\mu}$) destroys an electron with energy
$\epsilon_\bfk$ ($\epsilon_i$) with respect to the fermi level and spin $\mu$
on lead $\ell=L,R$ (in orbital $i=1,2$ on the dot); $n_{\ell\bfk} \equiv
\sum_\mu c_{\ell\bfk\mu}^\dag c_{\ell\bfk\mu}$ and $n_i \equiv \sum_\mu
d_{i\mu}^\dag d_{i\mu}$ are occupation operators for the leads and the dot
orbitals. The Coulomb energy of the strength $U$ is assumed to depend on the
total number of electrons in the dot. The Hund's rule in the dot results in the
ferromagnetic exchange coupling denoted as $J\,(<0)$ between the electron spins
$\bfS_i = \frac12 \sum_{\mu\mu'} d_{i\mu}^\dag \bfsigma_{\mu\mu'} d_{i\mu'}$,
where $\bfsigma$ are Pauli matrices. The left and right leads are assumed to
have identical dispersion energy $\epsilon_\bfk$ and superconducting gap
$\Delta$, while a finite phase difference $\phi=\phi_L-\phi_R$ is applied
between them. The energy-independent dot-lead tunneling amplitudes $t_{i\ell}$
hybridize the electron states between the dot and the leads, which are well
characterized by tunneling rates $\Gamma_{i\ell} = \pi \rho |t_{i\ell}|^2$,
where $\rho$ is the density of states of the leads at the Fermi energy.

Since we are interested in the regime of the singlet-triplet transition of an
isolated dot, we focus on the parameter region in which the dot is doubly
occupied. In addition, we consider the nondegenerate case with a finite
splitting $\delta\epsilon \equiv \epsilon_2 - \epsilon_1 > 0$ between two
orbitals. \Figref{fig:tlqd} displays the energy levels of two-electron states
of the isolated dot as functions of $J$: three singlet states,
$\ket{2,0,0;\alpha}$ with $\alpha=1,2,3$ and three triplet states
$\ket{2,1,M}$, where the states are labeled as $\ket{Q,S,M}$ with the charge
number $Q$, the spin $S$, and the $z$ component of the spin $M$. The singlet
states,
\begin{subequations}
  \begin{align}
    \ket{2,0,0;1} & = d_{1\up}^\dag d_{1\down}^\dag \ket{0}
    \\
    \ket{2,0,0;2} & = d_{2\up}^\dag d_{2\down}^\dag \ket{0}
    \\
    \ket{2,0,0;3}
    & =
    \frac1{\sqrt2}
    (d_{2\up}^\dag d_{1\down}^\dag - d_{2\down}^\dag d_{1\up}^\dag)
    \ket{0}
  \end{align}
\end{subequations}
have the energies, $E_{0;1} = 2\epsilon_1 + U$, $E_{0;2} = 2\epsilon_2 + U$,
$E_{0;3} = \epsilon_1 + \epsilon_2 + U - 3J/4$, respectively, and the triplet
states,
\begin{subequations}
  \begin{align}
    \ket{2,1,1} & = d_{2\up}^\dag d_{1\up}^\dag \ket{0}
    \\
    \ket{2,1,0}
    & =
    \frac1{\sqrt2}
    (d_{2\up}^\dag d_{1\down}^\dag + d_{2\down}^\dag d_{1\up}^\dag) \ket{0}
    \\
    \ket{2,1,-1} & = d_{2\down}^\dag d_{1\down}^\dag \ket{0}
  \end{align}
\end{subequations}
are degenerate with the energy $E_1 = \epsilon_1 + \epsilon_2 + U + J/4$. Due
to the finite splitting $\delta\epsilon > 0 $ and the existence of the
inter-orbital Coulomb interaction, the singlet-triplet transition is driven by
the competition between the states $\ket{2,0,0;1}$ and $\ket{2,1,M}$ [see
\figref{fig:tlqd}]. The bare singlet-triplet splitting is then defined by
\begin{align}
  \Jst^{(0)} \equiv E_1 - E_{0;1} = \delta\epsilon + \frac{J}{4}.
\end{align}
The external gate voltage $V_g$ can tune the singlet-triplet splitting by
affecting the level splitting $\delta\epsilon$,\cite{Kogan03,Holm08} the
exchange coupling strength $J$,\cite{Quay07,Roch08} or both of them. For
simplicity, we assume that the gate-voltage dependency is implemented only
through $J = J(V_g)$ and that $\delta\epsilon$ or $\epsilon_i$ are independent
of $V_g$. Our simplification can still capture the main physics of the system
as long as the regime close to the singlet-triplet transition is concerned.

The configuration of the dot-lead coupling is another important source that can
govern the physics of the system. First, the number of the effective channels
coupled to the dot can be controlled.\cite{Hofstetter02} If the condition,
\begin{align}
  \label{eq:onechannel}
  \frac{t_{1{\rm L}}}{t_{1{\rm R}}}
  =
  \frac{t_{2{\rm L}}}{t_{2{\rm R}}}
\end{align}
is satisfied, the dot-lead coupling matrix has a zero eigenvalue, and one of
the two channels can be completely decoupled from the dot under a proper
unitary transformation, resulting in a one-channel problem. This reduction of
the effective channels then affects the Kondo effect greatly, which will be
discussed later. Secondly, the phase of the coupling coefficients has an
influence on the interference and consequently on the electron transport
through the dot.\cite{Rozhkov01,vanDam06} Even though no magnetic field is
applied in our system, the (real-valued) coupling coefficients can acquire an
additional phase $\pi$ depending on the parity of the orbital wave functions on
the dot.\cite{vanDam06} Two distinctive cases can then be conceived: $t_{1{\rm
    L}} t_{1{\rm R}} t_{2{\rm L}} t_{2{\rm R}} > 0$ when two orbitals have the
same parity and $t_{1{\rm L}} t_{1{\rm R}} t_{2{\rm L}} t_{2{\rm R}} < 0$ when
they have the opposite parities. Taking into account the essential impacts of
the dot-lead coupling and focusing on the consequent qualitative features of
system states and electron transport, we consider two representative cases in
this paper:
\begin{align}
  \label{eq:conf}
  \begin{array}{ll}
    \text{case I}: & t_{1\ell} = t,~ t_{2\ell} = \gamma t
    \\
    \text{case II}: & t_{1\ell} = t,~ t_{2{\rm L}} = - t_{2{\rm R}} = \gamma t
  \end{array}
\end{align}
with $\gamma\le1$. The case I deals with the effective one-channel problem with
\eqnref{eq:onechannel} satisfied, while the case II reflects the two-channel
problem with the negative product of coupling coefficients. The effect of
asymmetric coupling with respect to the orbitals is also examined by setting
$\gamma<1$. Another kind of asymmetric junction such as $t_{i{\rm L}} \ll
t_{i{\rm R}}$ that can happen frequently in realistic experimental setups like
break junctions\cite{Roch08} is not considered in our study because this
asymmetry is observed to make no qualitative impact on the Josephson current.

Finally, since we are interested in the low temperature behavior, we
concentrate for the most part on the Kondo regime. The hybridizations
$\Gamma_{i\ell}$ are chosen to be far smaller than the particle or hole
excitations with respect to the two-electron states in order to suppress the
resonant tunneling. Specifically, throughout our study, we choose $\epsilon_1 =
-1.6 D$, $\epsilon_2 = -1.4D$, $U = D$, and $\Gamma = \pi \rho |t|^2 = 0.05 D$,
where the half band width $D$ is taken as the unit of energy. Here we have also
used the particle-hole symmetry condition $\epsilon_1 + \epsilon_2 + 3U = 0$.

\section{Weak Coupling Limit: $\Delta\gg T_K$\label{sec:wcl}}

\subsection{Fourth-Order Perturbation Theory\label{sec:pt}}

\begin{figure}[!t]
  \centering
  \includegraphics[width=3.5cm]{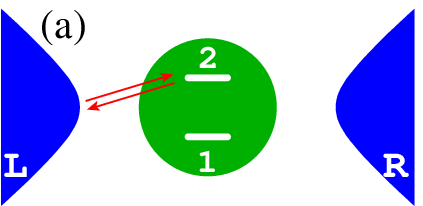}
  \includegraphics[width=3.5cm]{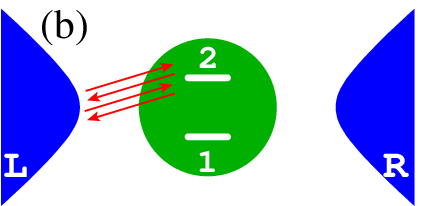}\\
  \includegraphics[width=3.5cm]{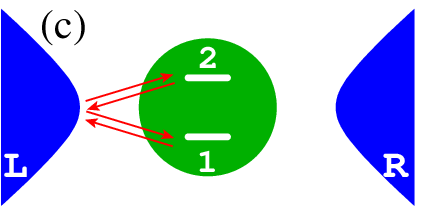}
  \includegraphics[width=3.5cm]{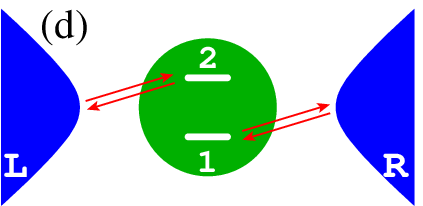}\\
  \includegraphics[width=3.5cm]{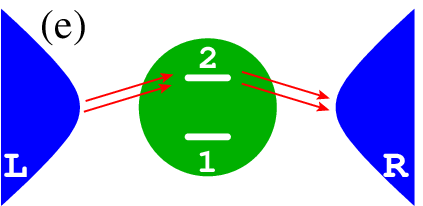}
  \includegraphics[width=3.5cm]{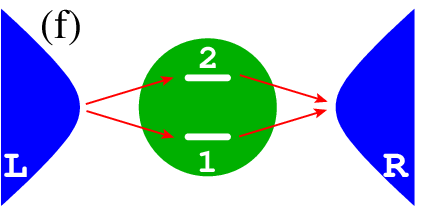}
  \caption{(color online) Listing of typical virtual tunneling processes
    contributing to $\delta E_a$. The arrows indicate the direction of the
    charge transfers for processes contributing to (a) $\beta_{a0}$, (b)
    $\beta_{a1}$, (c) $\beta_{a2}$, (d) $\beta_{a3}$, (e)
    $\beta^{(\prime)}_{a4}$, and (f) $\beta^{(\prime)}_{a5}$.}
  \label{fig:vt}
\end{figure}

First, we consider the weak coupling limit where the superconducting gap
$\Delta$ is much larger than the Kondo temperature $T_K$, which will be defined
in \secref{sec:scl}. In this case the supercurrent can be calculated via
fourth-order perturbation theory in $\varH_{\rm
  T}$.\cite{Rozhkov01,Spivak91,ChoiMS00} We apply degenerate perturbation
theory that takes into account the singlet state $\ket{2,0,0;1}$ and the
triplet states $\ket{2,1,M}$ simultaneously since they are almost degenerate
close to the singlet-triplet transition point of isolated dot. Unlike the
single-level quantum dot studies\cite{Rozhkov01,Spivak91} where it is enough to
collect only terms that depend on the phase difference $\phi$, on the other
hand, one must keep track of all the $\phi$-independent terms in the TLQD study
because they contribute to the renormalization of the singlet-triplet
splitting,\cite{ChoiMS00} and the transition point is shifted from its
unnormalized position, $\Jst^{(0)} = 0$. Due to the singlet nature of the
Cooper pair, there exists no coupling between the singlet and the triplet
states to any order of the perturbation, and the energy of each state is
separately shifted: $E_a = E_a^{(0)} + \delta E_a(\phi)$ for $a=S,T$ with
$E_S^{(0)} = E_{0;1}$ and $E_T^{(0)} = E_1$.  The energy shifts are given by
\begin{align}
  \nonumber
  \delta E_a
  & =
  \beta_{a0} \Gamma \sum_{i\ell} \gamma_{i\ell}^2
  + \frac{\Gamma^2}{\Delta}
  \Big[
  \beta_{a1} \sum_{i\ell} \gamma_{i\ell}^4
  +
  \beta_{a2} \sum_\ell \gamma_{1\ell}^2 \gamma_{2\ell}^2
  \\
  \nonumber
  & \quad\mbox{}
  +
  \beta_{a3} \sum_{\ell\ne\ell'} \gamma_{1\ell}^2 \gamma_{2\ell'}^2
  +
  (\beta_{a4} - \beta'_{a4}\cos\phi)
  \sum_i \gamma_{i{\rm L}}^2 \gamma_{i{\rm R}}^2
  \\
  \label{eq:pert}
  & \quad\mbox{}
  +
  (\beta_{a5} - \beta'_{a5}\cos\phi)
  \gamma_{1{\rm L}} \gamma_{1{\rm R}} \gamma_{2{\rm L}} \gamma_{2{\rm R}}
  \Big],
\end{align}
where we have defined $\gamma_{i\ell} \equiv t_{i\ell}/t$. \Figref{fig:vt}
shows typical virtual hopping processes that contribute to each term in
\eqnref{eq:pert}. The detailed expressions for the coefficients $\beta_{ai}$
can be found in the Appendix.

The effective singlet-triplet splitting then becomes
\begin{align}
  \Jst(\phi)
  = E_T(\phi) - E_S(\phi)
  = \Jst^{(0)} + \delta \Jst(\phi)
\end{align}
with
\begin{align}
  \delta \Jst
  \equiv \delta E_T - \delta E_S
  \equiv \delta \Jst^{(2)} + \delta \Jst^{(4)},
\end{align}
where $\delta \Jst^{(2)}$ and $\delta \Jst^{(4)}$ consist of the terms that are
proportional to $\Gamma$ and $\Gamma^2$, respectively. We find that $\delta
\Jst$ is mostly positive in the parameter regime of our interest, favoring the
singlet formation. The singlet-triplet transition point $J_c$ when $\Jst(J=J_c)
= 0$, which now becomes $\phi$-dependent, is then shifted from its bare value
$J_c^{(0)}=-4\delta\epsilon = -0.8D$ to a more negative value. It should be
noted that the second-order contribution to $\delta \Jst$
\begin{align}
  \delta \Jst^{(2)}
  =
  (\beta_{T0} - \beta_{S0}) \Gamma \sum_{i\ell} \gamma_{i\ell}^2
\end{align}
is finite in contrast to the previous study of parallel double-dot
system\cite{ChoiMS00} where the leading contribution is found to be of the
order of $\Gamma^2$. The main difference comes from the characteristics of the
singlet states in two systems. In the double-dot system studied by Choi
\textit{et al.}, the two quantum dots, each of which is singly occupied, are
identical and have no Coulomb interaction between them so the lowest-lying
singlet state is $\ket{2,0,0;3}$, while it is $\ket{2,0,0;1}$ in our system due
to the existence of the finite splitting and the inter-orbital Coulomb
interaction. The singlet state $\ket{2,0,0;3}$ has the same charge distribution
as the triplet states $\ket{2,1,M}$, so the second-order perturbation does not
give rise to any additional splitting between two states [see
\figref{fig:vt}~(a)]. On the other hand, having $\ket{2,0,0;1}$ as the
lowest-lying singlet states, our system can exhibit a rather huge
renormalization of the singlet-triplet splitting that is of the order of
$\Gamma$. This second-order term $\delta \Jst^{(2)}$ is numerically found to
increase as $\Delta$ is decreased. This tendency is opposite to the expectation
that the renormalization, which is due to the tunneling of Cooper pairs whose
amplitude increases with $\Delta$, should be weakened as $\Delta$ decreases: in
other words, $\lim_{\Delta\to0} \delta \Jst = 0$. This discrepancy should be
resolved by the higher-order terms of the order of $(\Gamma/\Delta)^n$ that are
more involved as $\Delta$ decreases: In fact, the fourth-order term $\delta
\Jst^{(4)}$ is observed to become negative for smaller $\Delta$ so that the
renormalization is diminished. Owing to this opposite $\Delta$-dependencies of
$\delta \Jst^{(2)}$ and the other higher-order terms, $\Jst$ varies
non-monotonically with $\Delta$, which in turns implies that the transition
point $J_c$ also displays a non-monotonic dependency on $\Delta$: see
\figsref{fig:wcpdi} and \ref{fig:wcpdii}.

The supercurrent can be calculated via the derivative of the energy with
respect to the phase difference $\phi$:
\begin{align}
  I_a
  = \frac{2e}{\hbar} \frac{\partial E_a}{\partial\phi}
  = \hat{I}_a \sin\phi
\end{align}
with
\begin{align}
  \label{eq:J}
  \frac{\hat{I}_a}{I_c^{\rm short}}
  =
  2 \left(\frac{\Gamma}{\Delta}\right)^2\!\!
  \left(
    \beta'_{a4}
    \sum_i \gamma_{i{\rm L}}^2 \gamma_{i{\rm R}}^2
    +
    \beta'_{a5}
    \gamma_{1{\rm L}} \gamma_{1{\rm R}} \gamma_{2{\rm L}} \gamma_{2{\rm R}}
  \right)
\end{align}
As a matter of fact, only the virtual processes in \figsref{fig:vt}~(e) and (f)
contribute to the Cooper pair tunneling. The $\beta'_{a4}$-term [(e)] arises
from the diagonal processes where both electrons in a Cooper pair travel
through either the orbital 1 or 2, while the $\beta'_{a5}$-term [(f)] from the
off-diagonal processes with one electron traveling through the orbital 1 and
the other traveling through the orbital 2. Depending on the order of the
sequence of electron tunneling and the spin correlation of dot electrons, the
coefficients $\beta'_{ai}$ can acquire a relative minus sign owing to Fermi
statistics. For the singlet state, one can find that
\begin{align}
  \beta'_{S4} > 0
  \quad\text{and}\quad
  \beta'_{S5} < 0.
\end{align}
The negative sign for $\beta'_{S5}$ is attributed to the processes with one
electron traveling through a filled level (orbital 1) and the other electron
through an empty level (orbital 2).  It should be noted that it is necessary to
take into account the dot electron correlation exactly in order to determine
the supercurrent sign correctly. Not all the processes contributing to
$\beta'_{S5}$ acquire the $\pi$ phase: For example, the processes with the
intermediate state $\ket{2,0,0;3}$ acquire no phase at all [see
\eqnref{eq:bs5}], while their amplitudes are always smaller than those of the
other processes, and finally $\beta'_{S5}$ is negative. For the triplet state,
\begin{align}
  \beta'_{T4} < 0
  \quad\text{and}\quad
  \beta'_{T5} < 0
\end{align}
because of the presence of local magnetic moments in both
orbitals.\cite{Spivak91} Apart from the sign, we have found numerically that
the off-diagonal processes usually have larger amplitude than the diagonal
ones:
\begin{align}
  2|\beta'_{a4}| < |\beta'_{a5}|\qquad \text{for}\ a=S,T.
\end{align}
Hence, when the product $\gamma_{1{\rm L}} \gamma_{1{\rm R}} \gamma_{2{\rm L}}
\gamma_{2{\rm R}}$ is comparable to $\sum_i \gamma_{i{\rm L}}^2 \gamma_{i{\rm
    R}}^2$ in magnitude, the sign of the supercurrent dictates the sign of the
off-diagonal term, or that of the product $- \gamma_{1{\rm L}} \gamma_{1{\rm
    R}} \gamma_{2{\rm L}} \gamma_{2{\rm R}}$ regardless of the spin state: the
current exhibits the 0($\pi$)-junction for the negative (positive)
product. Otherwise, that is, if $|\gamma_{1{\rm L}} \gamma_{1{\rm R}}
\gamma_{2{\rm L}} \gamma_{2{\rm R}}| \ll \sum_i \gamma_{i{\rm L}}^2
\gamma_{i{\rm R}}^2$, the diagonal term prevails in determining the sign of the
supercurrent so that the singlet (triplet) state features the 0($\pi$)-junction
behavior regardless of the sign of the product.

In the following sections, we identify the system state according to its
ground-state spin and the sign of the supercurrent in the $\Delta$-$J$
plane. We use the labels $S$ and $T$ to denote the spin singlet and triplet
state, respectively. Since the phase transition depends on the superconducting
phase difference $\phi$ as well, the phase boundaries are located at three
different values of $\phi$: 0 (red line), $\pi/2$ (green line), and $\pi$ (blue
line).  Between $\phi=0$ and $\pi$ boundaries the system is in the intermediate
state having a stable ground state and a meta-stable state. The intermediate
states are tagged with a subscript that represents the meta-stable state spin.
For example, the ground state in the state $T_S$ is mostly of the spin triplet,
while it is of the spin singlet at and near $\phi = 0$, and the system
experiences a phase transition from spin doublet to singlet as $\phi$ is varied
from 0 to $\pi$.  The state identification is then supplemented by the SPR
calculated from \eqnref{eq:J}, classifying whether it is of either
$0^{(\prime)}$ or $\pi^{(\prime)}$ junctions. Two states with same ground-state
spin can be distinguished if their SPRs are different and the boundary between
them will be colored in yellow line.

\subsection{Case I: $\gamma_{1{\rm L}} \gamma_{1{\rm R}} \gamma_{2{\rm L}} \gamma_{2{\rm R}} > 0$}

\begin{figure}[!t]
  \centering
  \includegraphics[width=7cm]{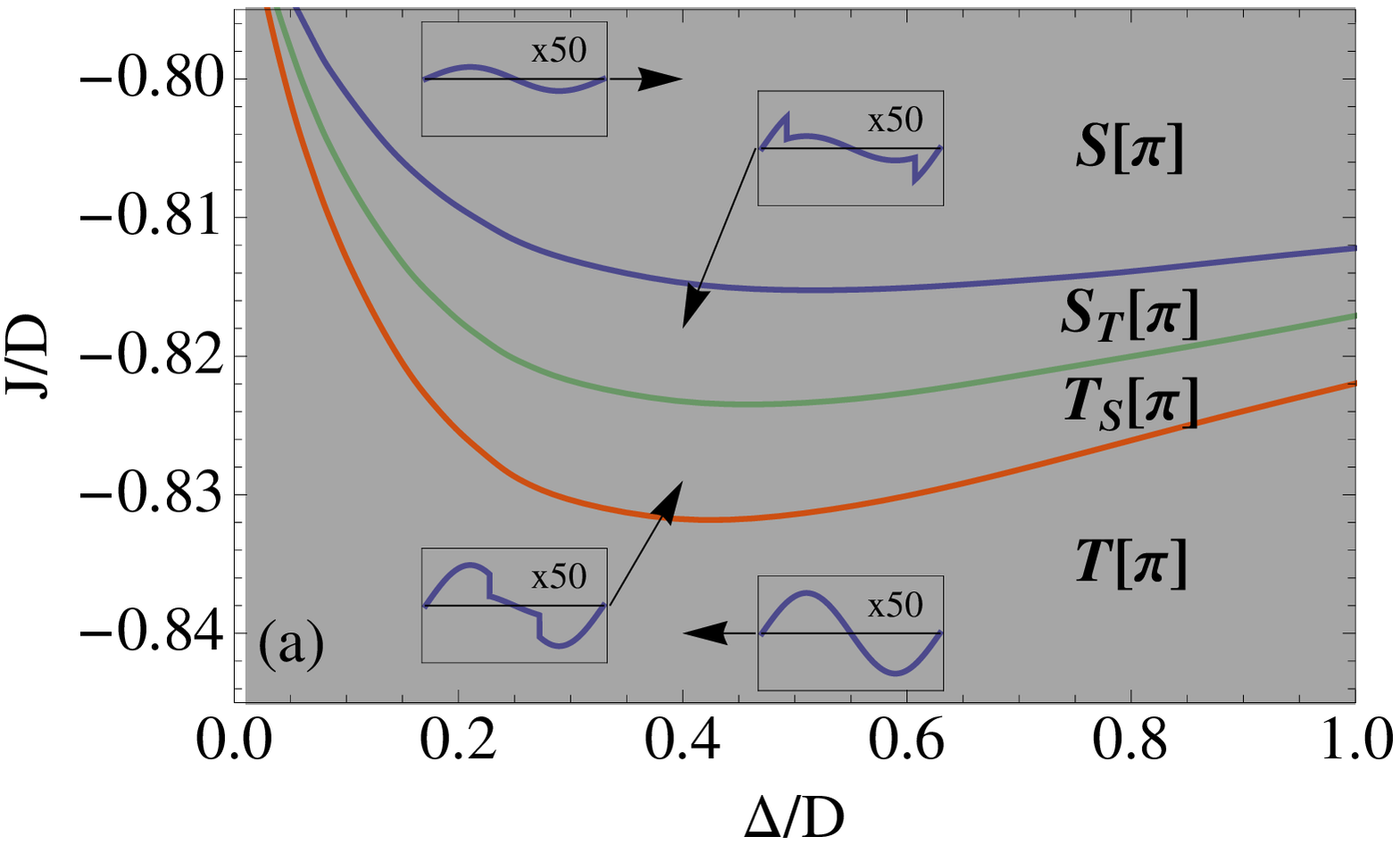}\\
  \includegraphics[width=7cm]{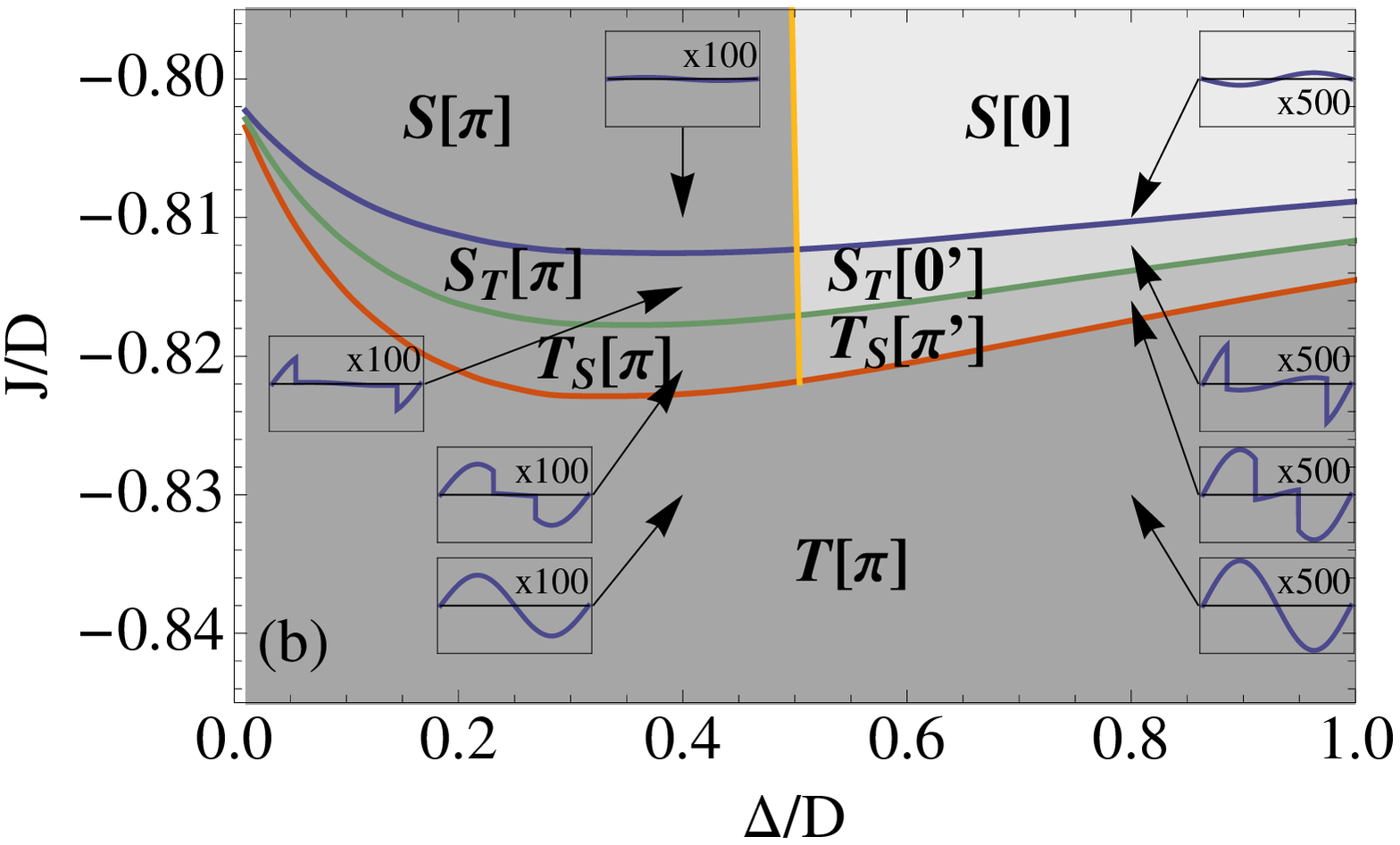}\\
  \includegraphics[width=7cm]{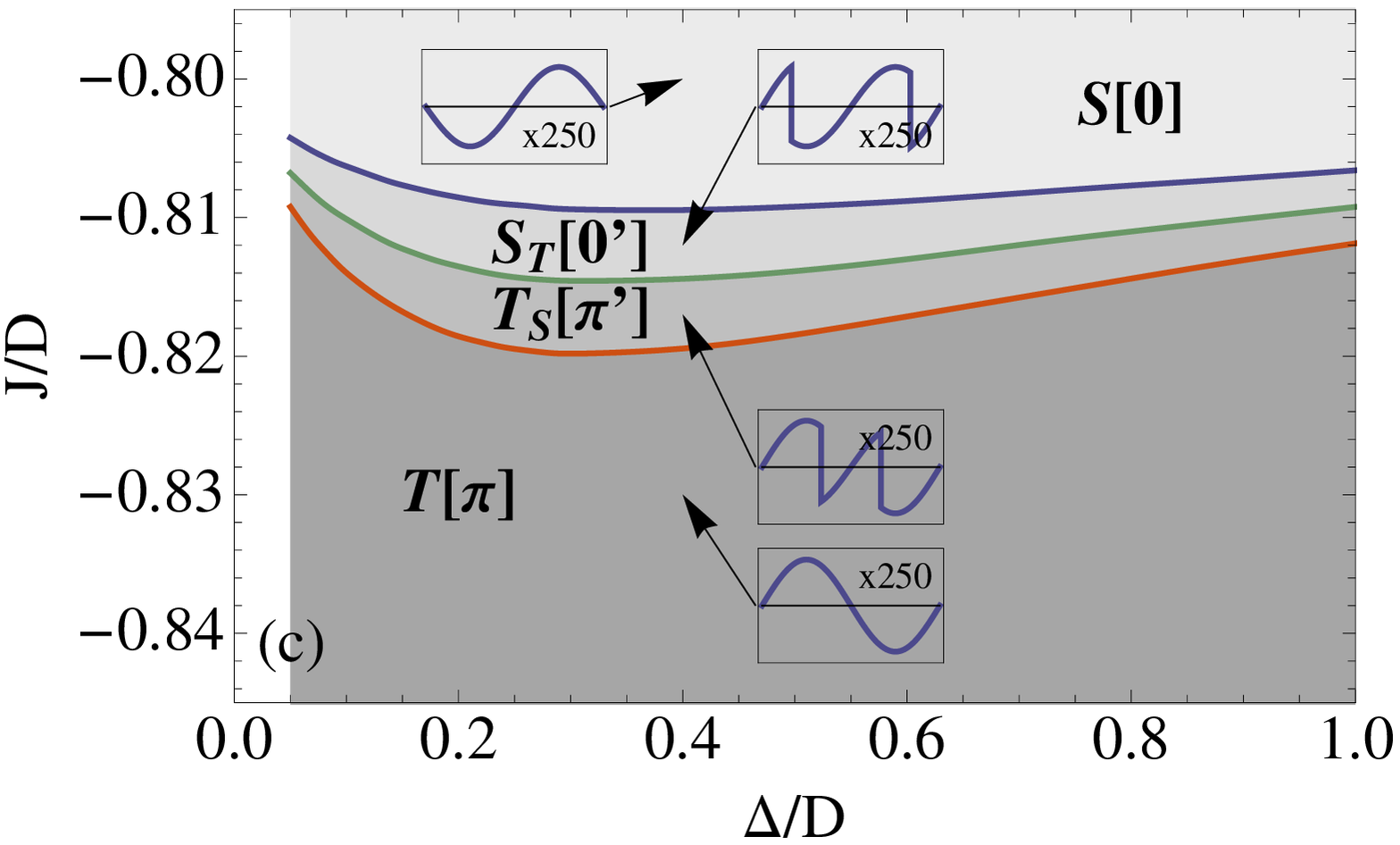}
  \caption{(color online) Phase diagrams in the $\Delta$-$J$ plane for the case
    I with $\gamma = 1$ [(a)], 0.6 [(b)], and 0.1 [(c)]. The phase boundaries
    are located when the ground-state spin is changed at $\phi = 0$ (red line),
    $\pi/2$ (green line), and $\pi$ (blue line). The yellow line separates two
    states with same ground-state spin but different SPRs. Each phase is shaded
    in gray scale according to its SPR: lighter gray for 0 junction and darker
    gray for $\pi$ junction. Refer the detailed classification of the states to
    the text.  The insets show the SPR for $\phi\in[-\pi,\pi]$ at the points
    indicated by the arrows. Here the value of $\Delta$ is swept from $\Gamma$
    to $D$, and for $\Delta<\Gamma$ the above diagrams are not valid.}
  \label{fig:wcpdi}
\end{figure}

\begin{figure}[!t]
  \centering
  \includegraphics[width=6.5cm]{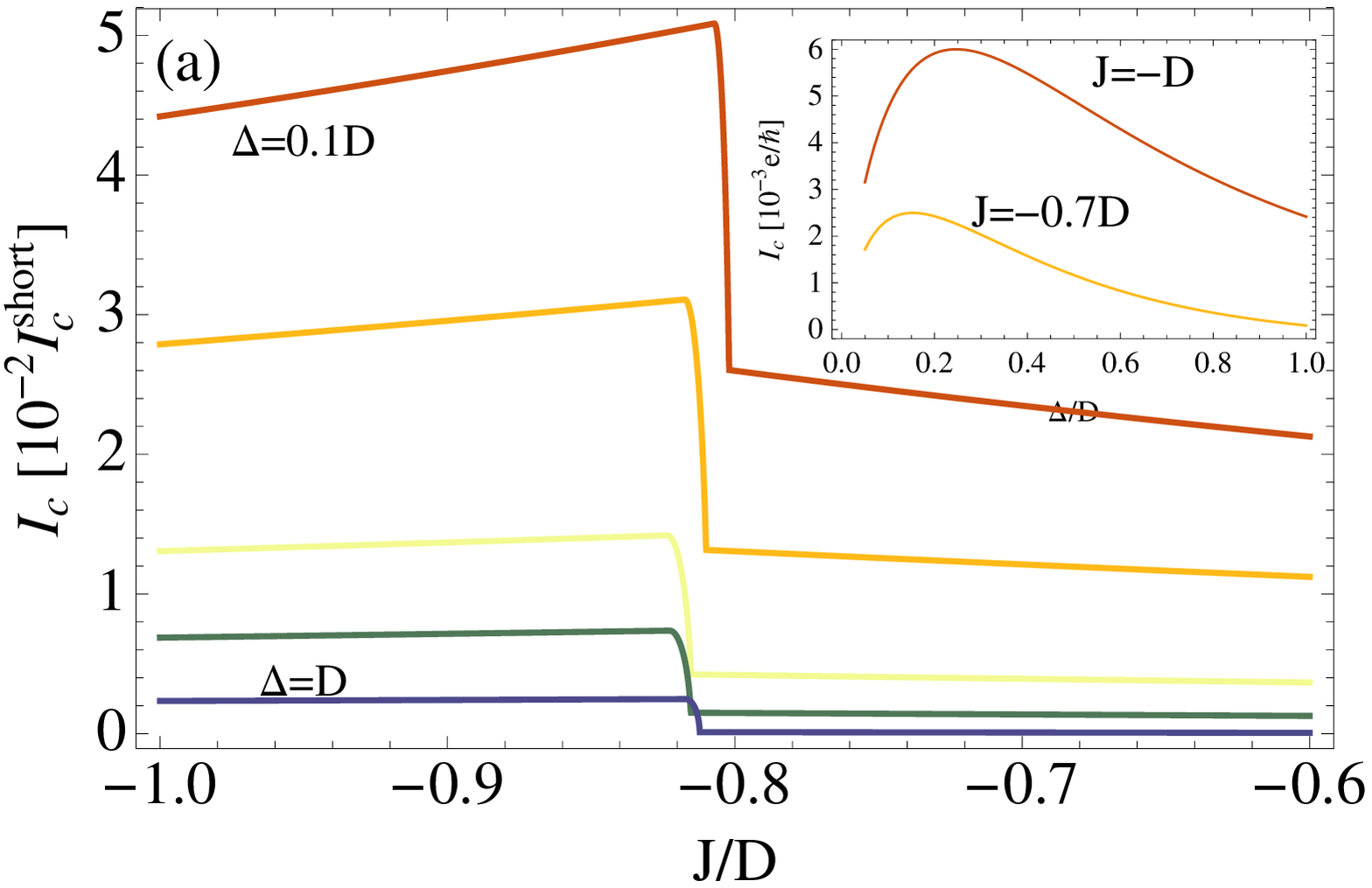}\\
  \includegraphics[width=6.5cm]{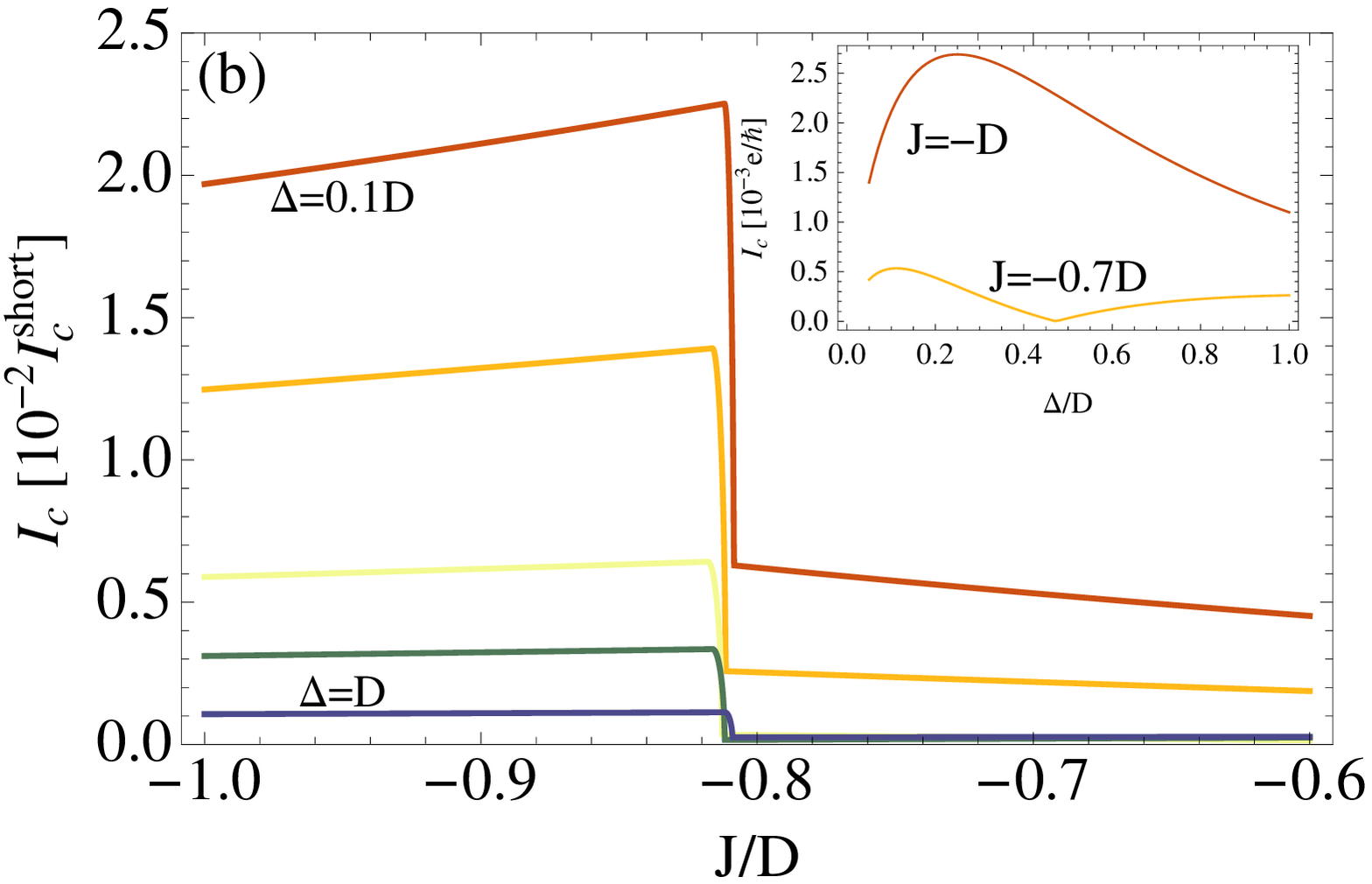}\\
  \includegraphics[width=6.5cm]{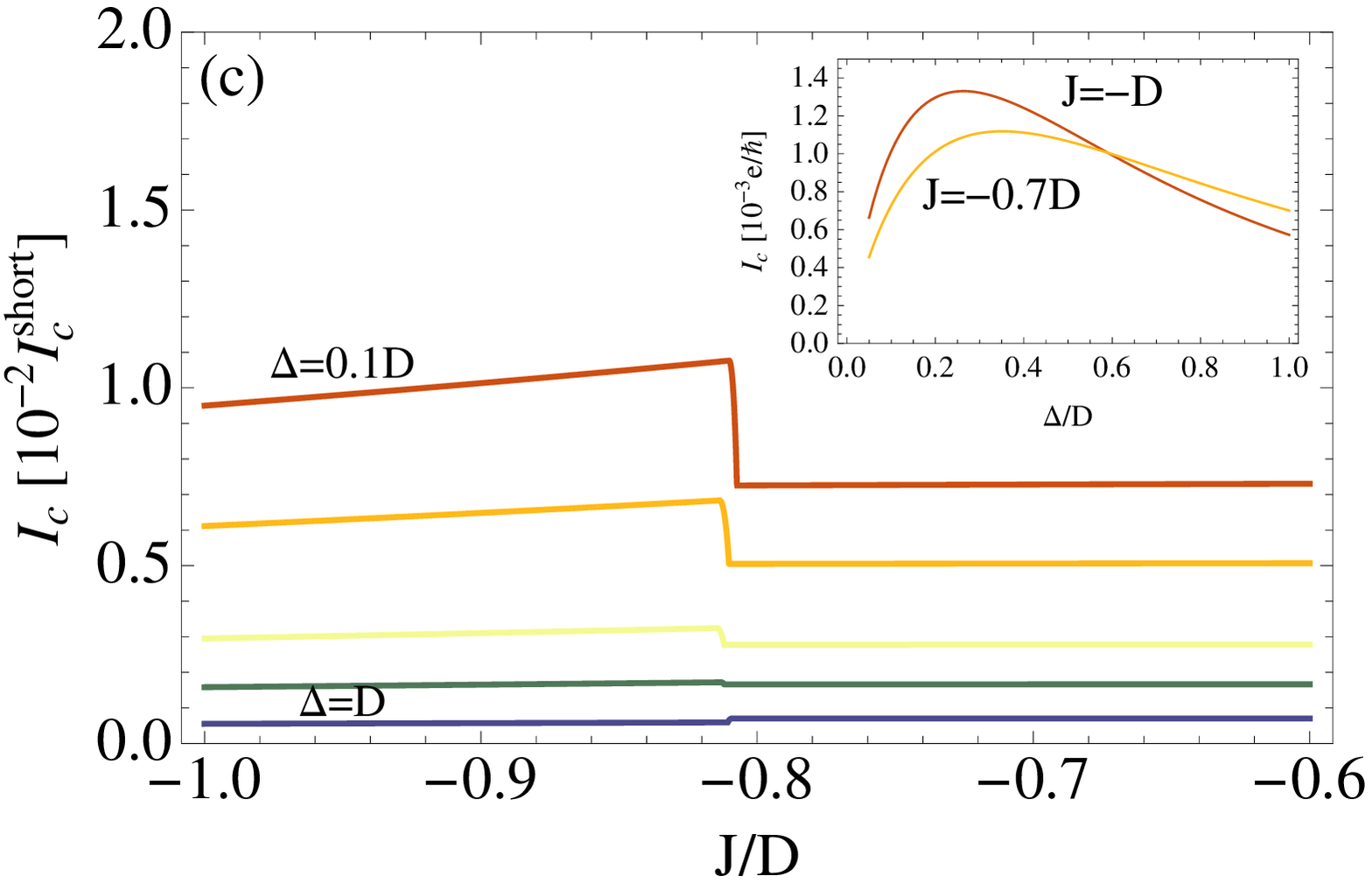}
  \caption{(color online) Critical currents as functions of $J$ in units of
    $I_c^{\rm short}$ for the case I with $\gamma = 1$ [(a)], 0.6 [(b)], and
    0.1 [(c)] for various values of $\Delta$: $0.1D$ (top), $0.2D$, $0.4D$,
    $0.6D$, and $D$ (bottom). The insets display the critical currents as
    functions of $\Delta$ in units of $e/\hbar$ for $J = -0.7D$ (in the spin
    triplet state) and $-D$ (in the spin singlet state).}
  \label{fig:wccci}
\end{figure}

\Figref{fig:wcpdi} shows the phase diagrams in the $\Delta$-$J$ plane in the
case I for various values of $\gamma$. The lower bound of $\Delta$ is set to
$\Gamma$ because the perturbation theory works only when $\Gamma\ll\Delta$.
For $\gamma = 1$, the Josephson coupling
\begin{align}
  \frac{\hat{I}_a}{I_c^{\rm short}}
  =
  2(\Gamma/\Delta)^2 (2\beta'_{a4} + \beta'_{a5})
\end{align}
is always negative because $\beta'_{a5} < 0$ and $2|\beta'_{a4}| <
|\beta'_{a5}|$, and the current exhibits the $\pi$-junction behavior, no matter
what values $J$ and $\Delta$ have [see \figref{fig:wcpdi}~(a)]. For
$\gamma\ll1$, on the other hand, the contribution from the off-diagonal term
becomes negligible since
\begin{align}
  \gamma^2
  =
  \gamma_{1{\rm L}} \gamma_{1{\rm R}} \gamma_{2{\rm L}} \gamma_{2{\rm R}}
  \ll
  \sum_i \gamma_{i{\rm L}}^2 \gamma_{i{\rm R}}^2
  = 1 + \gamma^4,
\end{align}
and the sign of $\hat{I}_a$ is governed solely by the $\beta'_{a4}$ term. The
spin singlet state is then of the 0 junction since $\beta'_{S4} > 0$, and the
singlet-triplet transition accompanies the 0-$\pi$ transition with the
intermediate states as shown in \figref{fig:wcpdi}~(c). \Figref{fig:wcpdi}~(b)
shows that for intermediate values of $\gamma$, both of the 0 and $\pi$
junctions can appear in the spin singlet state: the 0 and $\pi$ junctions take
place in the regions with larger and smaller values of $\Delta$,
respectively. The phase boundary separating two regions moves toward the
smaller $\Delta$ as $\gamma$ is decreased.

Different strength of the Josephson coupling in the spin singlet and triplet
states gives rise to a discontinuous change in the SPR in the intermediate
states [see the insets in \figref{fig:wcpdi}] and a rapid change in the
critical current $I_c \equiv |\hat{I}_a|$ across the singlet-triplet transition
as shown in \figref{fig:wccci}. The numerical calculation of the supercurrent
finds that the supercurrent is stronger in the spin triplet state than in the
spin singlet state: $|\hat{I}_T| > |\hat{I}_S|$. In the spin singlet state the
diagonal and the off-diagonal processes make the opposite contributions
($\beta'_{S4} > 0 > \beta'_{S5}$), resulting in a partial cancellation. This is
not the case in the spin triplet state in which both processes contribute to
the $\pi$ junction ($\beta'_{T4}, \beta'_{T5} < 0$). For small $\gamma$, on the
other hand, such a cancellation does not make a significant role since the
$\beta'_{a5}$ term becomes much smaller than the $\beta'_{a4}$ term, so the
critical currents in both spin states become comparable as can be seen in
\figref{fig:wccci}~(c).

The critical current exhibits a non-monotonic dependence on $\Delta$: see the
insets in \figref{fig:wccci}. In two extreme limits, $\Delta\ll\Gamma$ and
$\Delta\gg\Gamma$, the supercurrent should vanish. The supercurrent, induced by
the proximity effect that is proportional to $\Delta$, should vanish in the
limit $\Delta\to0$. In the opposite limit, the high energy cost $\sqrt{\Delta^2
  + \epsilon_\bfk^2}$ of the quasiparticles created during the virtual
processes suppresses the current. Consequently, the critical current has a
maximum as a function of $\Delta$. For the intermediate values of $\gamma$ when
the 0-$\pi$ transition occur in the spin singlet state, the critical current
can become zero at the transition [see the inset in \figref{fig:wccci}~(b)].

\subsection{Case II: $\gamma_{1{\rm L}} \gamma_{1{\rm R}} \gamma_{2{\rm L}} \gamma_{2{\rm R}} < 0$}

\begin{figure}[!t]
  \centering
  \includegraphics[width=7cm]{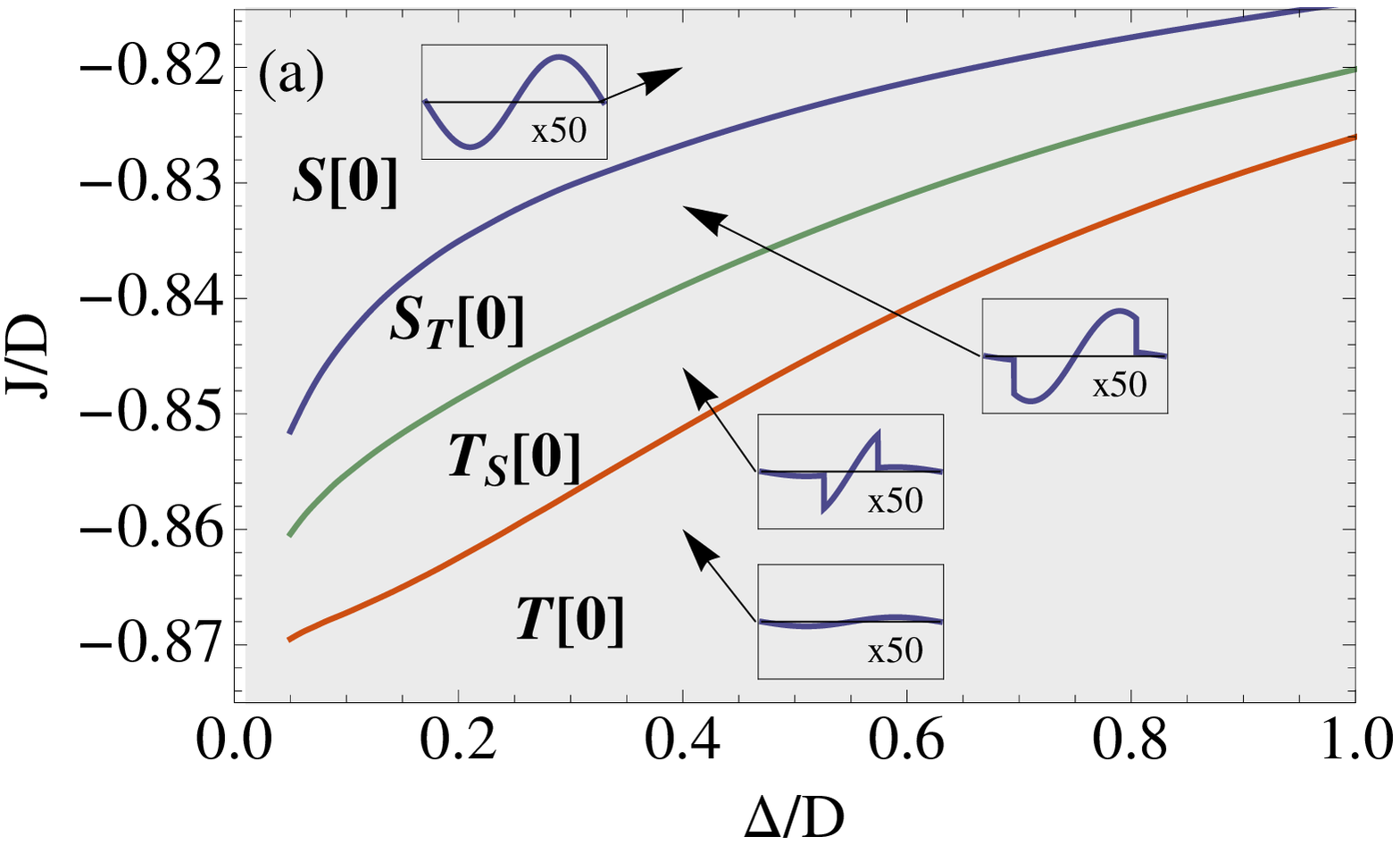}
  \includegraphics[width=7cm]{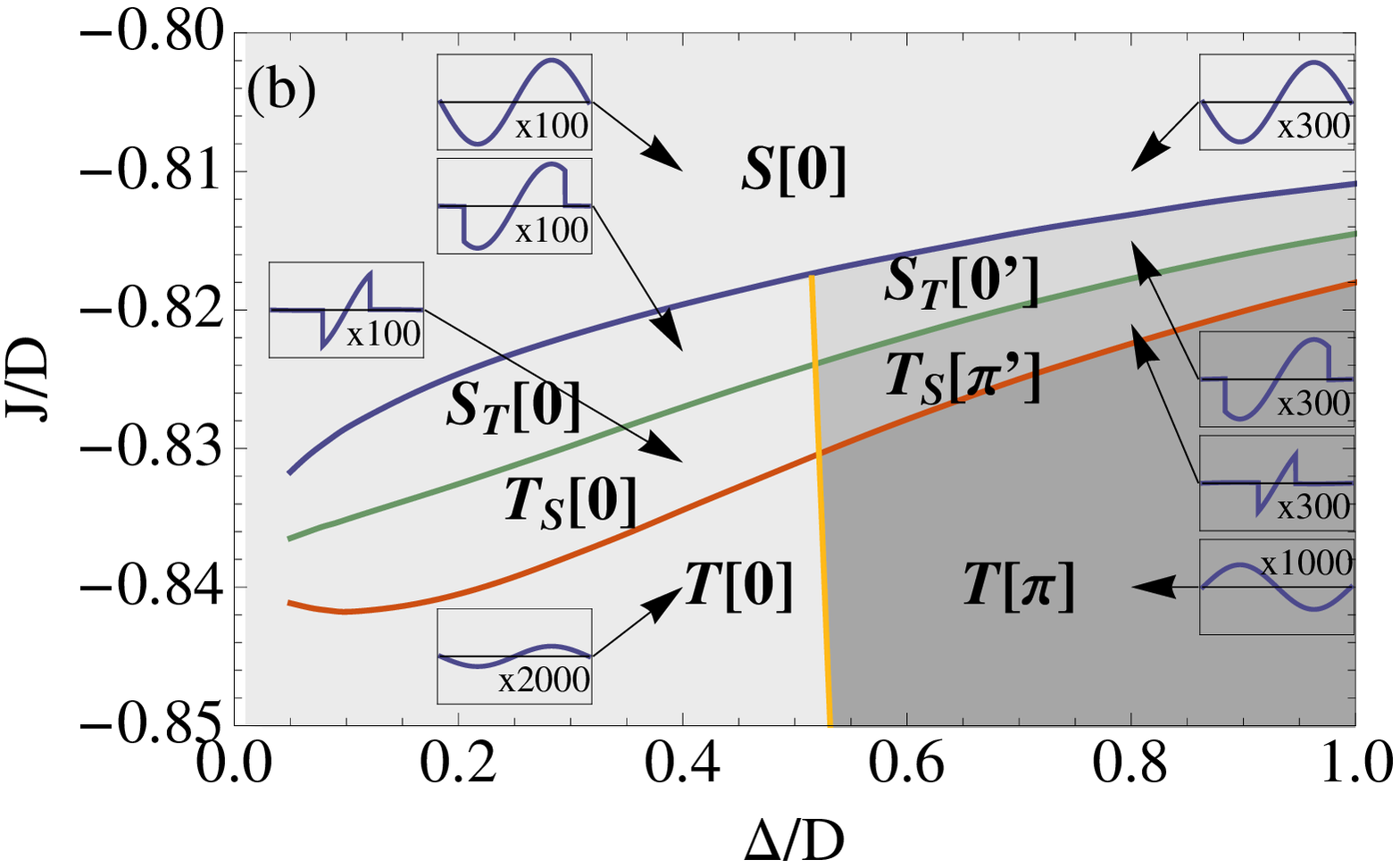}
  \includegraphics[width=7cm]{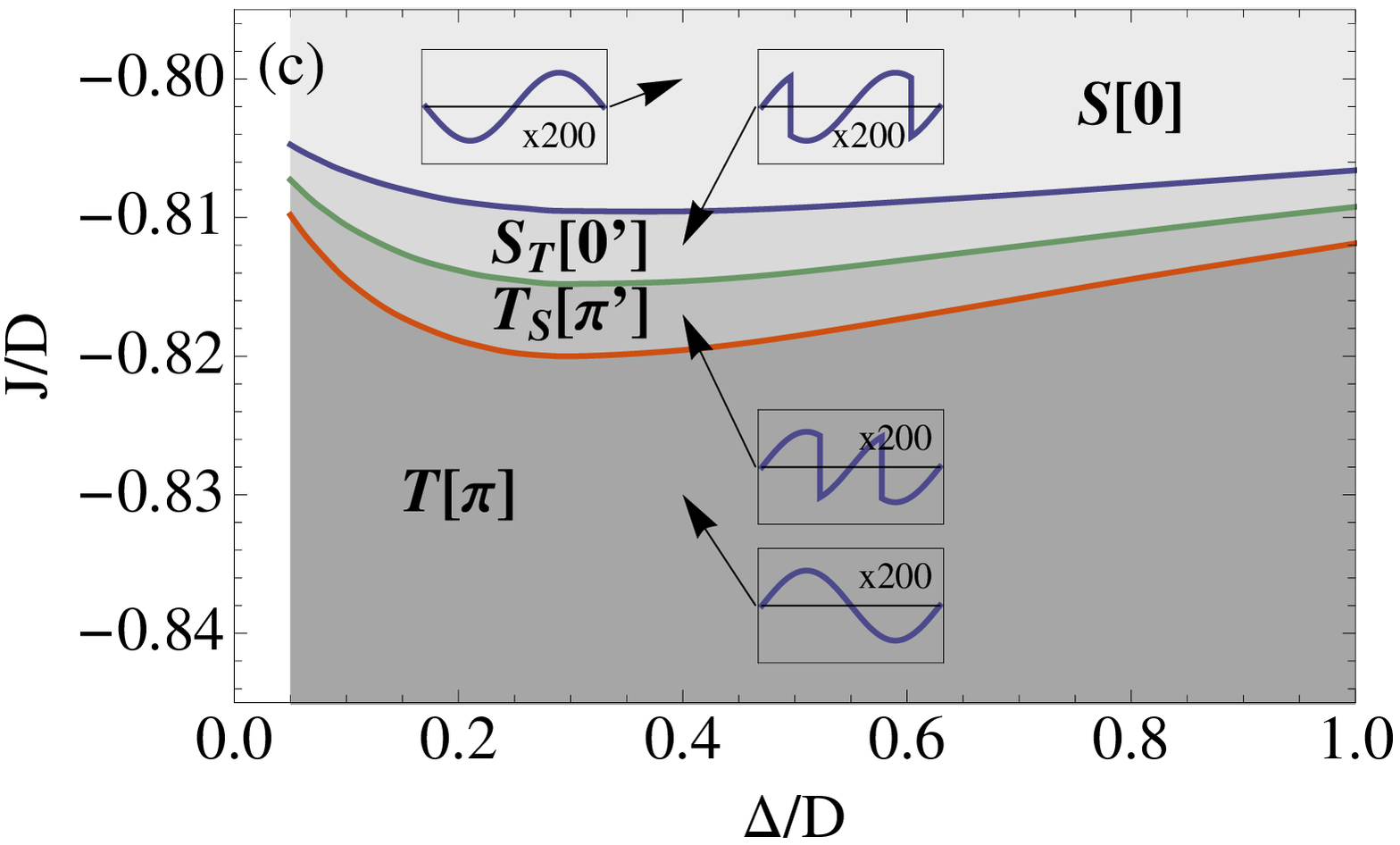}
  \caption{(color online) Phase diagrams in the $\Delta$-$J$ plane for the case
    II with $\gamma = 1$ [(a)], 0.72 [(b)], and 0.1 [(c)]. Refer to
    \figref{fig:wcpdi} for the details.}
  \label{fig:wcpdii}
\end{figure}

\begin{figure}[!t]
  \centering
  \includegraphics[width=6.5cm]{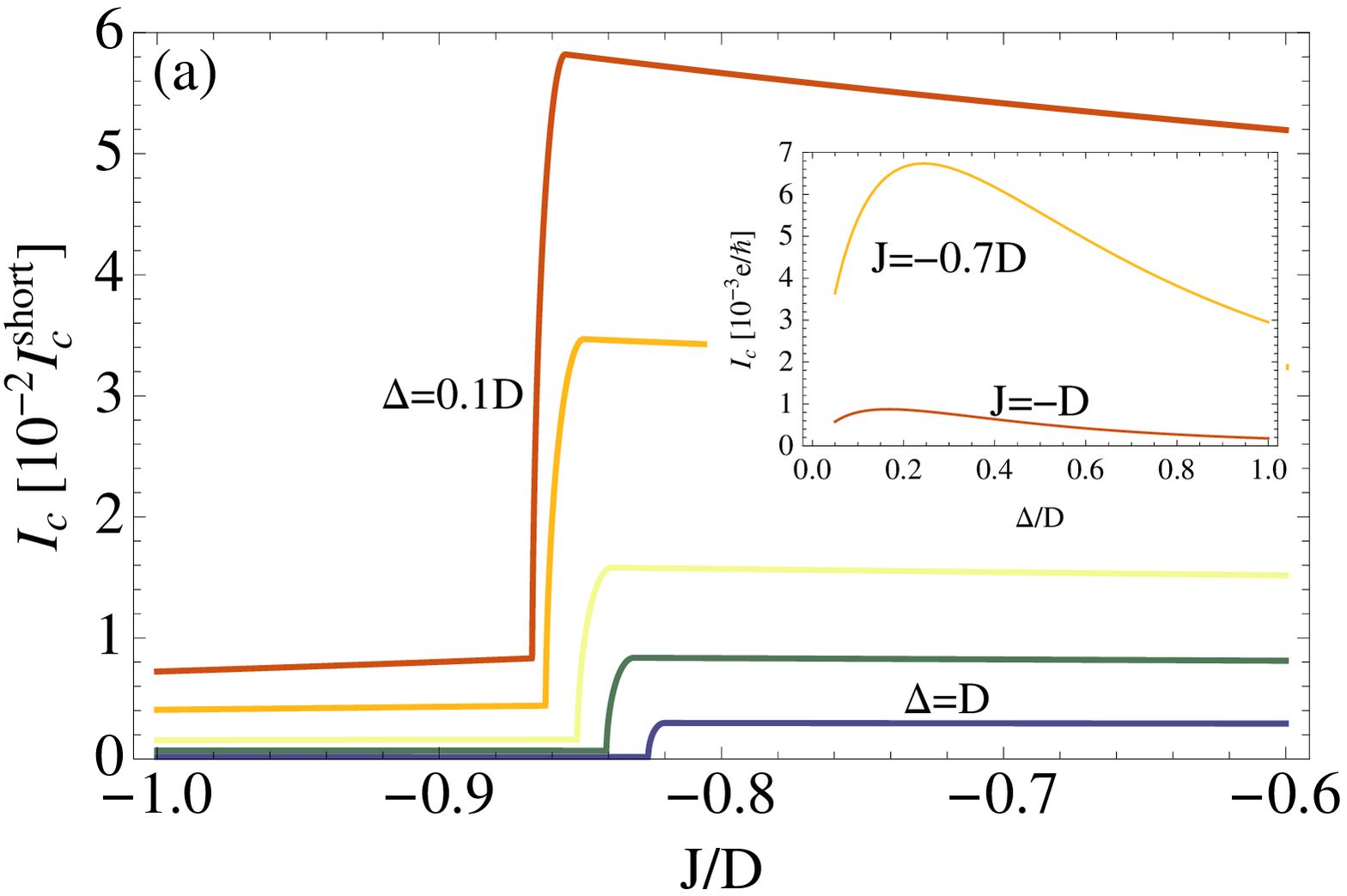}
  \includegraphics[width=6.5cm]{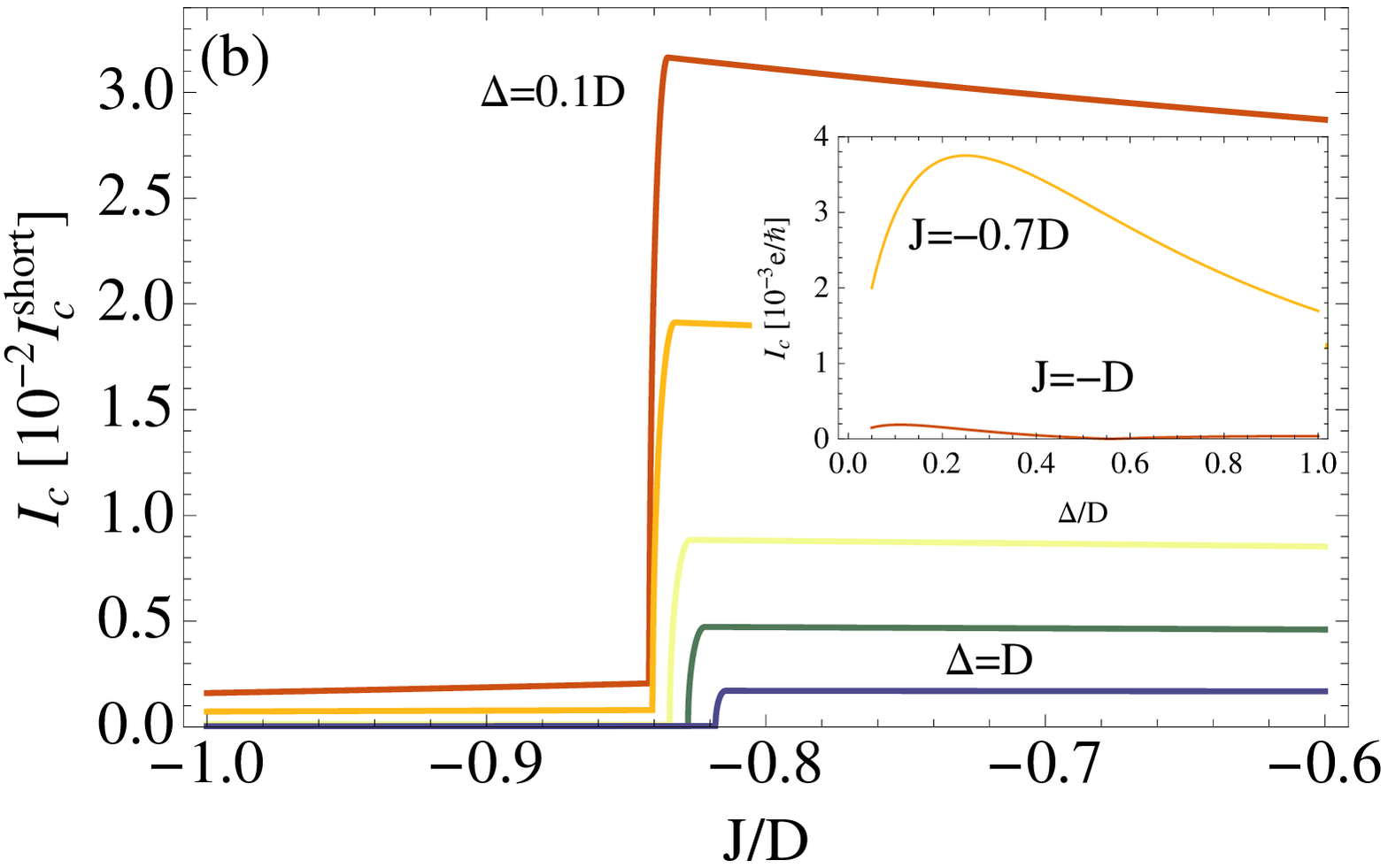}
  \includegraphics[width=6.5cm]{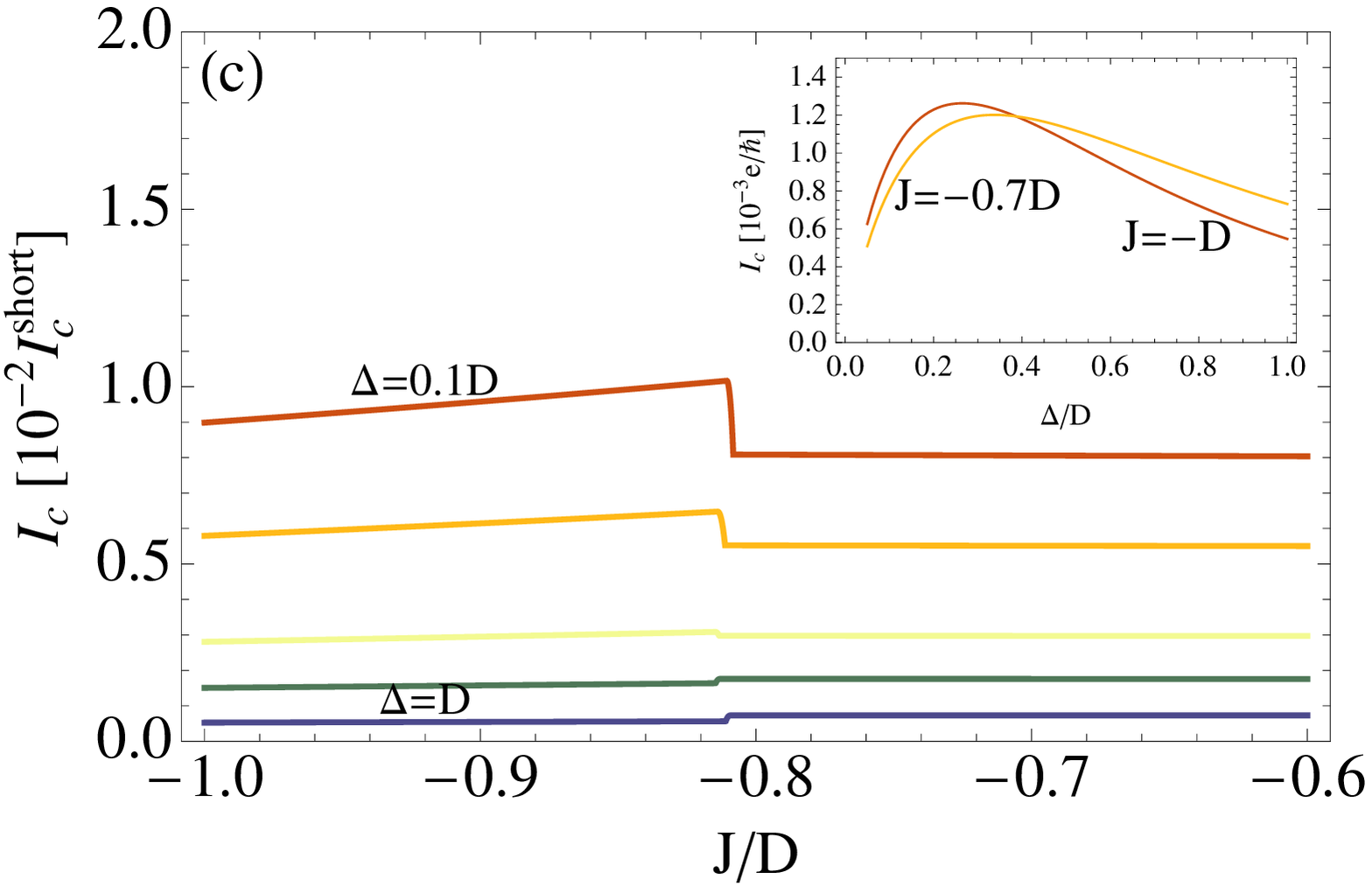}
  \caption{(color online) Critical currents as functions of $J$ in units of
    $I_c^{\rm short}$ for the case II with $\gamma = 1$ [(a)], 0.72 [(b)], and
    0.1 [(c)] for various values of $\Delta$: $0.1D$ (top), $0.2D$, $0.4D$,
    $0.6D$, and $D$ (bottom). Refer the guide for the insets to
    \figref{fig:wcpdii}.}
  \label{fig:wcccii}
\end{figure}

The phase diagram and the critical current in the case II are shown in
\figsref{fig:wcpdii} and \ref{fig:wcccii}, respectively. In this case one of
the dot-lead tunneling amplitude changes its sign, making the product
$\gamma_{1{\rm L}} \gamma_{1{\rm R}} \gamma_{2{\rm L}} \gamma_{2{\rm R}}$
negative and accordingly reversing the sign of the off-diagonal
contributions. It then switches the junction characteristics from $\pi$ to 0
junction for the case $\gamma = 1$ when the off-diagonal term prevail over the
diagonal term: compare \figref{fig:wcpdi}~(a) and
\figref{fig:wcpdii}~(a). However, for $\gamma\ll1$ when the off-diagonal
contributions are negligible, the negative product does not affect the
supercurrent and the phase diagram so much: \figref{fig:wcpdi}~(c) and
\figref{fig:wcpdii}~(c) are almost identical. As a result, for the intermediate
values of $\gamma$, the additional 0-$\pi$ transition now takes place in the
spin triplet state in contrast to the case I: compare \figref{fig:wcpdi}~(b)
and \figref{fig:wcpdii}~(b). Another difference from the case I is that the
critical current is now much larger in the spin singlet state than in the spin
triplet state as long as $\gamma$ is not so small: see \figref{fig:wcccii}. The
same argument used in the case I applies as well: With the negative product,
the $\beta'_{T4}$ term which is negative cuts down the positive contribution
from the $\beta'_{T5}$ term, while in the spin singlet state both of two terms
contributes to the 0 junction.

In addition to the properties of the supercurrent, the shape of the phase
boundaries are also different from those in the case
I. \Figsref{fig:wcpdii}~(a) and (b) show that for the moderate values of
$\gamma$ the phase boundaries are much shifted toward the spin triplet side,
implying that the spin singlet state is being further favored. Furthermore, the
transition point $J_c$ displays a monotonic dependence on $\Delta$ and does not
approach its bare value in the limit $\Delta\to0$, which is contradictory to
our expectation from the previous weak-coupling argument. The inclination to
the spin singlet state is accounted for by looking at the $\beta_{a5}$ term in
$\delta E_a$ that is proportional to $\gamma_{1{\rm L}} \gamma_{1{\rm R}}
\gamma_{2{\rm L}} \gamma_{2{\rm R}}$ [see the last term in
\eqnref{eq:pert}]. Numerical calculations observe $\beta_{S5} > 0$ and
$|\beta_{S5}| > |\beta_{\rm T5}|$, which means that with the negative product
$\gamma_{1{\rm L}} \gamma_{1{\rm R}} \gamma_{2{\rm L}} \gamma_{2{\rm R}} < 0$
the spin singlet state is much lowered than the spin triplet state [see the
Appendix for expressions of $\beta_{S5}$ and $\beta_{T5}$]. This term is also
observed to make the fourth-order splitting term $\delta \Jst^{(4)}$ positive,
which is the cause of the monotonic behavior of $J_c$. One may suspect that
this favoring of the spin singlet state in the limit $\Delta\to0$ is the
artifact of the fourth-order perturbation close to its limit of validity,
$\Gamma/\Delta \sim 1$. However, the non-perturbative NRG study in the
following section finds that the system should be of the spin singlet state in
the vanishing $\Delta$ limit and that it is attributed to the complete
screening of dot spins by the two-channel conduction electrons, which will be
discussed in details in the next section. Considering that the Kondo effect
which is responsible for the screening cannot be correctly captured by the
perturbation theory, it is quite interesting that it still reflects correct
asymptotic behaviors in the limit $\Delta\to0$: the approaching of $J_c$ to its
bare value in the case I and the precursor of the disappearance of the spin
triplet state in the case II.

\section{Strong Coupling Limit: $\Delta\ll T_K$\label{sec:scl}}

In this section we extend our study to the strong-coupling limit by using the
NRG method that is known to be suitable for the non-perturbative study of the
low-temperature properties of the impurity system. Even though the standard NRG
procedure\cite{Wilson75} can be directly applied to the original Hamiltonian,
we have introduced a unitary transformation
\begin{align}
  \begin{bmatrix}
    c_{a\bfk\mu} \\ c_{b\bfk\mu}
  \end{bmatrix}
  =
  \varS
  \begin{bmatrix}
    c_{{\rm L}\bfk\mu} \\ c_{{\rm R}\bfk\mu}
  \end{bmatrix},\
  \begin{bmatrix}
    d_{a\mu} \\ d_{b\mu}
  \end{bmatrix}
  =
  \varD
  \begin{bmatrix}
    d_{1\mu} \\ d_{2\mu}
  \end{bmatrix}
\end{align}
which makes all the matrix elements of the Hamiltonian real in order to boost
up the speed of the numerical computation. Here the unitary matrices are
chosen to be
\begin{align}
  \label{eq:sd}
  \varS
  =
  \rcp{\sqrt2}
  \begin{bmatrix}
    i e^{-i\phi/4} & -i e^{+i\phi/4}
    \\
    e^{-i\phi/4} & e^{+i\phi/4}
  \end{bmatrix},\
  \varD
  =
  \begin{bmatrix}
    1 & 0
    \\
    0 & \chi
  \end{bmatrix},
\end{align}
where $\chi=1$ and $i$ in the cases I and II, respectively. Under the unitary
transformation, each part of the Hamiltonian is transformed into
\begin{align}
  \varH'_{\rm DD}
  & =
  \sum_{s=a,b} (\epsilon_s n_s + U n_{s\up} n_{s\down}) + U n_a n_b
  + J \bfS_a\cdot\bfS_b
  \\
  \varH'_{\rm LD}
  & =
  \sum_{s\bfk}
  \left[
    \epsilon_\bfk n_{s\bfk}
    -
    (-1)^s \Delta
    \left(
      c_{s\bfk\up}^\dag c_{s-\bfk\down}^\dag
      {+} (h.c.)
    \right)
  \right]
  \\
  \varH'_{\rm T}
  & =
  \sum_{ss'\bfk\mu}
  \left[t_{ss'} \, d_{s\mu}^\dag c_{s'\bfk\mu} + (h.c.)\right],
\end{align}
respectively, where $(-1)^a = -1$ and $(-1)^b = 1$. Here the transformed
dot-lead coupling matrix is given by
\begin{align}
  \label{eq:dlc}
  \begin{bmatrix}
    t_{aa} & t_{ab}
    \\
    t_{ba} & t_{bb}
  \end{bmatrix}
  =
  \begin{cases}
    \displaystyle
    \sqrt2 t
    \begin{bmatrix}
      \sin\frac{\phi}{4} & \cos\frac{\phi}{4}
      \\
      \gamma \sin\frac{\phi}{4} & \gamma \cos\frac{\phi}{4}
    \end{bmatrix},
    & \text{case I}
    \\
    \displaystyle
    \sqrt2 t
    \begin{bmatrix}
      \sin\frac{\phi}{4} & \cos\frac{\phi}{4}
      \\
      \gamma \cos\frac{\phi}{4} & - \gamma \sin\frac{\phi}{4}
    \end{bmatrix},
    & \text{case II}
  \end{cases}
\end{align}
The Wilson's NRG technique\cite{Wilson75,Krishnamurthy80} consists of the
logarithmic discretization of the conduction bands, the mapping onto a
semi-infinite chain, and the iterative diagonalization of the properly
truncated Hamiltonian.  Following the standard NRG procedures extended to
superconducting leads,\cite{Yoshioka00} we evaluate various physical quantities
from the recursion relation
\begin{align}
  \nonumber
  \widetilde{\varH}_{N+1}
  & =
  \sqrt{\Lambda} \widetilde{\varH}_N
  +
  \xi_N \sum_{s\mu} (f_{sN\mu}^\dag f_{sN{+}1\mu} + (h.c.))
  \\
  & \quad\mbox{}
  -
  \sum_s (-1)^s \widetilde{\Delta}
  (f_{sN{+}1\up}^\dag f_{sN{+}1\down}^\dag + (h.c.))
\end{align}
for $N\ge0$ with the initial Hamiltonian given by
\begin{align}
  \nonumber
  \widetilde{\varH}_0
  & =
  \frac{1}{\sqrt{\Lambda}}
  \Bigg[
  \widetilde{\varH}_{\rm D}
  +
  \sum_{ss'} \sqrt{\widetilde{\Gamma}_{ss'}}
  \sum_\mu (d_{s\mu}^\dag f_{s'0\mu} + (h.c.))
  \\
  & \qquad\qquad\mbox{}
  - \sum_{s\mu} (-1)^s \widetilde{\Delta}
  (f_{s0\up}^\dag f_{s0\down}^\dag + (h.c.))
  \Bigg].
\end{align}
Here the fermion operators $f_{sN\mu}$ have been introduced as a result of the
logarithmic discretization of the conduction bands and the accompanying
tridiagonalization, $\Lambda$ is the logarithmic discretization parameter (we
choose $\Lambda = 4$), and
\begin{align}
  \xi_N
  & =
  \frac{1 - \Lambda^{-(N+1)}}%
  {\sqrt{(1 - \Lambda^{-(2N+1)})(1 - \Lambda^{-(2N+3)})}},
  \\
  \widetilde{\varH}_{\rm D}
  & =
  \frac{\varH'_{\rm D}}{\varE D}, \quad
  \widetilde{\Delta}
  =
  \frac{\Delta}{\varE D},\quad
  \sqrt{\widetilde{\Gamma}_{ss'}}
  =
  \frac{1}{\varE} \sqrt{\frac{2\Gamma}{\pi D}} \frac{t_{ss'}}{t}
\end{align}
with $\varE = (1+\Lambda^{-1})/2$.  The original Hamiltonian is recovered by
\begin{align}
  \frac{\varH'}{D}
  =
  \lim_{N\to\infty} \varE \Lambda^{-(N-1)/2} \widetilde{\varH}_N.
\end{align}
It has been known\cite{Krishnamurthy80,Campo05} that the logarithmic
discretization underestimates the coupling between the conduction-band
electrons and the dot electrons. In order to avoid this problem, we multiply
$\widetilde\Gamma$ by a correction factor $A_\Lambda$ given
by\cite{Krishnamurthy80,Campo05}
\begin{align}
  A_\Lambda = \frac{\ln\Lambda}{2} \frac{\Lambda + 1}{\Lambda - 1}.
\end{align}

Within the NRG procedure, the spin of the ground state, the occupation
$\avg{n_i}$, and the spin correlation $\avg{\bfS_1\cdot\bfS_2}$ can be directly
calculated from the expectation values of the corresponding operators. The
supercurrent can be also obtained by calculating the expectation value
\begin{align}
  I = \frac{e}{2} \left\langle \dot{N}_L - \dot{N}_R \right\rangle,
\end{align}
where $N_\ell = \sum_\bfk n_{\ell\bfk}$. In terms of the fermion operators
$f_{s0\mu}$, the current expectation value is expressed as
\begin{align}
  \label{eq:jc}
  \frac{I}{I_c^{\rm short}}
  =
  \frac{D}{\Delta} \sqrt{\frac{2\Gamma}{\pi D}}
  \sum_{ss'\mu} \avg{i_{ss'} d_{s\mu}^\dag f_{s'0\mu} + (h.c.)}
\end{align}
with the current matrix defined by
\begin{align}
  \begin{bmatrix}
    i_{aa} & i_{ab}
    \\
    i_{ba} & i_{bb}
  \end{bmatrix}
  =
  \rcp{2t}
  \begin{bmatrix}
    t_{ab} & - t_{aa}
    \\
    t_{bb} & - t_{ba}
  \end{bmatrix}.
\end{align}
The Andreev levels are located from the subgap many-body excitations which are
identified as the poles of the dot Green's functions.

\subsection{Normal Leads: $\Delta = 0$}

In the presence of Coulomb interaction and spin exchange coupling, strong
dot-lead coupling can induce nontrivial many-body correlations that may compete
with and even suppress superconductivity. A promising candidate of such
many-body correlations in the QD system is the Kondo effect. In order to
identify nontrivial correlations in our system and to elaborate the analysis of
the strongly-coupled Josephson junction, it is quite useful to investigate the
normal-lead case with $\Delta = 0$. The NRG procedure described above is then
applied by setting $\Delta = 0$ and $\phi = 0$: The latter condition, though
not being essential, is imposed in order to simplify the dot-lead coupling
matrix [\eqnref{eq:dlc}]. The normal-lead version of our system has been well
studied in the literature, so we briefly summarize the known theoretical
analyses and present relevant numerical results in our parameter regime for
comparison with the superconducting case.

\subsubsection{Case I: Single Channel\label{sec:nlI}}

\begin{figure}[!b]
  \centering
  \includegraphics[width=4.25cm]{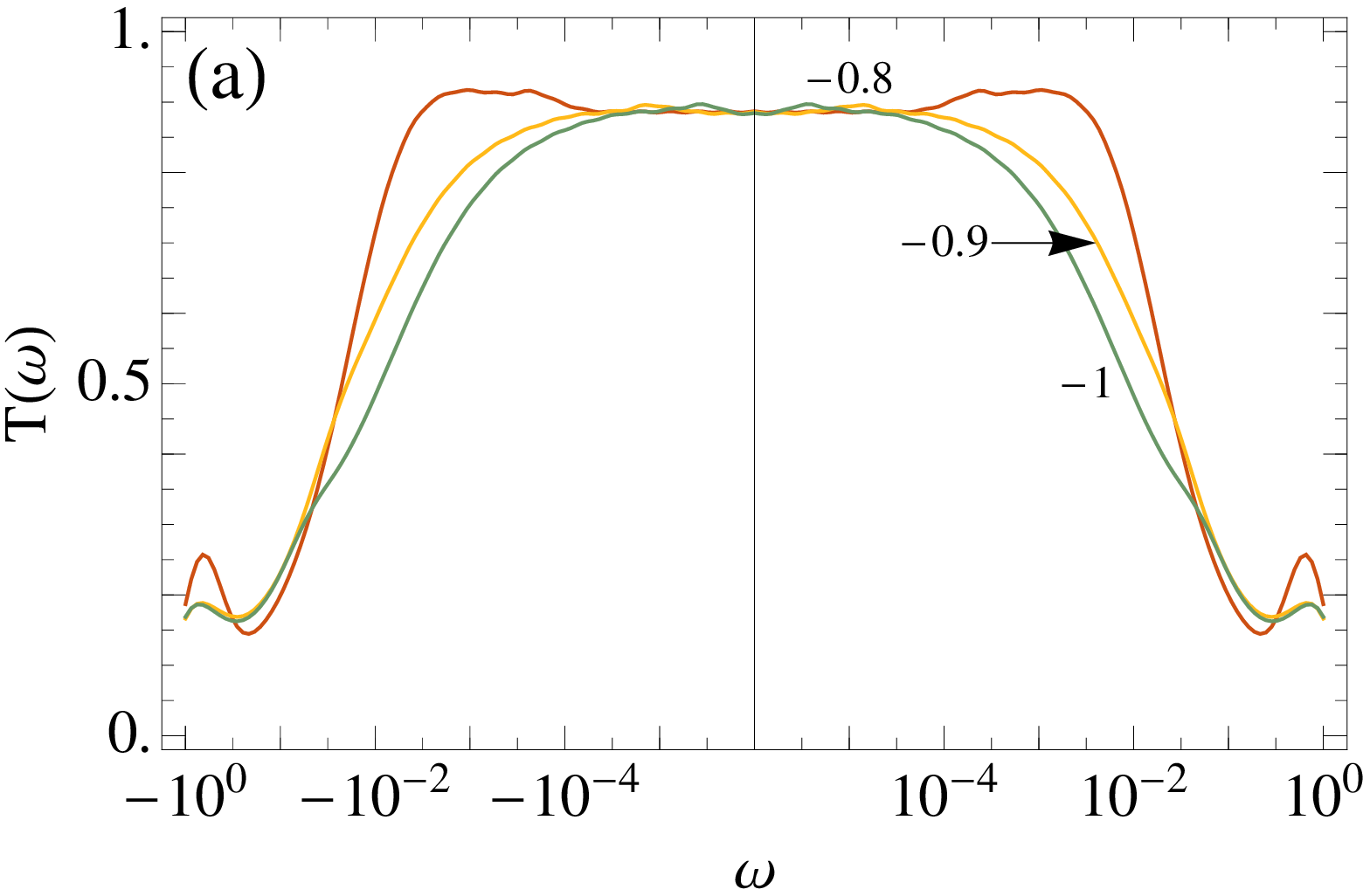}%
  \includegraphics[width=4.25cm]{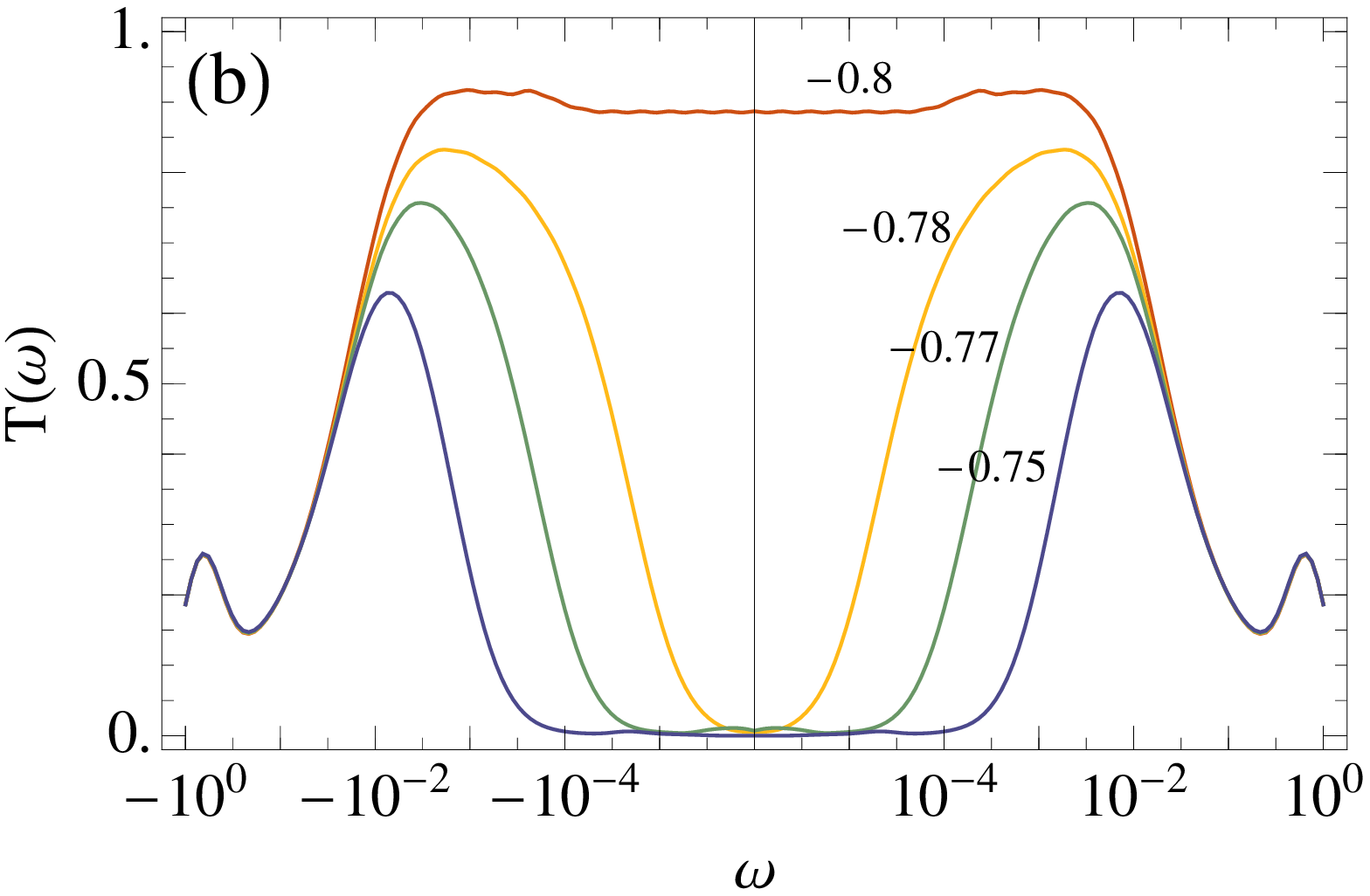}
  \caption{(color online) Energy-resolved transmission coefficient $T(\omega)$
    for a two-level QD coupled to normal leads (case I) with ferromagnetic
    $(\Jst < J_c)$ [(a)] and antiferromagnetic $(\Jst > J_c)$ [(b)] exchange
    coupling for various values of $J/D$ (as annotated).}
  \label{fig:nit}
\end{figure}

In the case I only the lead-$b$, that is, the symmetrized conduction-band
channel is coupled to the dot, with the other channel completely detached: see
\eqnref{eq:dlc}. The two-level QD system attached to a single conduction
channel has been well studied in the context of the quantum phase transition in
a vicinity of singlet-triplet degenerate point.\cite{Hofstetter02} In this case
the system can be mapped onto an exchange-coupled $S=1/2$ Kondo model through a
Schrieffer-Wolff transformation:\cite{Schrieffer66}
\begin{align}
  \label{eq:km}
  \varH_{\rm eff}
  =
  \varH_{\rm LD}
  + J_a\, \widetilde\bfS_a\cdot\bfs_{bb}
  + J_b\, \widetilde\bfS_b\cdot\bfs_{bb}
  + \Jst\, \widetilde\bfS_a\cdot\widetilde\bfS_b\,.
\end{align}
Here the Kondo spins $\widetilde\bfS_a$ and $\widetilde\bfS_b$ are fictitious
QD spins defined on the basis of the spin singlet state $\ket{2,0,0;1}$ and the
spin triplet states $\ket{2,1,M}$. Both of the Kondo spins are coupled to the
localized spin of the conduction channel $\bfs_{bb}$ associated to the
symmetrized combination of left and right leads [see \eqnref{eq:sd}]: Here we
define the localized spins of conduction-band electron spins as
\begin{align}
  \bfs_{ss'}
  \equiv
  \frac12 \sum_{\bfk\bfk'} \sum_{\mu\mu'}
  c_{s\bfk\mu}^\dag \bfsigma_{\mu\mu'} c_{s'\bfk'\mu'}.
\end{align}
The effective spin couplings are, up to linear order in $\Gamma$,
\onecolumngrid
\begin{subequations}
  \begin{align}
    J_{a/b}
    & =
    \frac{2\Gamma}{\pi\rho}
    \left(
      \frac{\gamma^2 \mp \sqrt2 \gamma}{-\epsilon_2 - U - J/4}
      +
      \frac{1 \mp \sqrt2 \gamma}{\epsilon_1 + 2U - J/4}
      +
      \rcp{-\epsilon_1 - U - J/4}
      +
      \frac{\gamma^2}{\epsilon_2 + 2U - J/4}
    \right)
    \\
    \label{eq:Jst}
    \Jst
    & =
    \delta\epsilon + \frac{J}{4}
    + \frac{4D\Gamma}{\pi}
    \left(
      \frac{2}{-\epsilon_1-U} - \frac{\gamma^2}{-\epsilon_2-U-J/4}
      - \rcp{-\epsilon_1-U-J/4}
    \right).
  \end{align}
\end{subequations}
\twocolumngrid
\noindent
Note that one has $J_a \ne J_b$ as long as $\gamma\ne0$.

The ground state of the Kondo Hamiltonian, \eqnref{eq:km} can be of the spin
singlet or doublet depending on the strength of the effective exchange coupling
$\Jst$ and is known to undergo a phase transition at the critical coupling
$\Jst = J_c$, which is of the Kosterlitz-Thouless-type.\cite{Vojta02} The
ferromagnetic side $(\Jst < J_c)$ corresponds to an underscreened $S=1$ Kondo
model where the conduction electrons screen one of the Kondo spins and the
remaining $S=1/2$ spin then couples ferromagnetically to the conduction band
and becomes asymptotically free at low energies.\cite{Cragg79} The
corresponding Kondo temperature $T_K(J)$ decreases with increasing $|\Delta J|
\equiv |\Jst - J_c|$ [see \figref{fig:nit}~(a)]. On the other hand, on the
antiferromagnetic side $(\Jst > J_c)$, a two-stage Kondo effect takes place for
small $\Delta J$.\cite{Hofstetter02,SCQD,vanderWiel02} First, the Kondo effect
leads to a screening of one of the Kondo spins, $\bfS_a$ with the larger
coupling (for example we assume $J_a > J_b$) which therefore defines the larger
Kondo temperature $T_K$. For temperatures lower than $T_K$, the second spin
$\bfS_b$ is decoupled from the conduction band. At a much lower energy scale
(denoted as $T_K^I$), the effective antiferromagnetic exchange coupling $\Jst$
between $\bfS_a$ and $\bfS_b$ then induces the second screening due to the
local Fermi liquid that is formed on the first spin. $T_K^I$ is then the Kondo
temperature of the second spin screened by electrons of a bandwidth $\sim T_K$
and density of states $\sim 1/(\pi T_K)$:\cite{Hofstetter02,SCQD}
\begin{align}
  \label{eq:tki}
  T_K^I \sim T_K \exp\left[-\pi \frac{T_K}{\Delta J}\right].
\end{align}
The second Kondo effect leads to a Fano resonance and makes a dip in the
energy-resolved transmission coefficient\cite{Hofstetter02}
\begin{align}
  T(\omega)
  =
  - \sum_{ii'\mu} \frac{\Gamma}{2} \gamma_{i'i} {\rm Im} G_{ii'\mu}(\omega),
\end{align}
where we have introduced the retarded QD Green's functions $G_{ii'\mu}(t) =
-i\Theta(t) \Braket{\{d_{i\mu}(t),d_{i'\mu}^\dag\}}$ and a coupling matrix
$\gamma_{ii'}$ with $\gamma_{11} = 1$, $\gamma_{12} = \gamma_{21} = \gamma$,
and $\gamma_{22} = \gamma^2$. As shown in \figref{fig:nit}~(b), the dip becomes
widened with increasing $\Delta J$ and eventually overrides the Kondo peak
until $T_K^I \approx T_K$ at which the Kondo effect completely vanishes.

\subsubsection{Case II: Two Channels\label{sec:nlII}}

Unless the zero-eigenvalue condition, \eqnref{eq:onechannel} is satisfied, the
dot is always coupled to both of the two conduction-band channels. The
low-energy physics of the system is then governed by the two-channel
two-impurity Kondo model with an exchange coupling. Similarly to the case I, the
effective spin model can be derived via the Schrieffer-Wolff transformation:
\begin{align}
  \nonumber
  \varH_{\rm eff}
  & =
  \varH_{\rm LD}
  +
  \widetilde\bfS_a\cdot
  \left(J_a\,\bfs_{aa} + J_b\,\bfs_{bb} + J_{ab} (\bfs_{ab} + \bfs_{ba})\right)
  \\
  \nonumber
  & \quad\mbox{}
  +
  \widetilde\bfS_b\cdot
  \left(J_a\,\bfs_{aa} + J_b\,\bfs_{bb} - J_{ab} (\bfs_{ab} + \bfs_{ba})\right)
  \\
  \label{eq:HeffII}
  & \quad\mbox{}
  + \Jst\,\widetilde\bfS_a\cdot\widetilde\bfS_b
  + 2iJ_{ab} (\widetilde\bfS_a\times\widetilde\bfS_b)
  \cdot(\bfs_{ab} - \bfs_{ba}).
\end{align}
The exchange coupling coefficients are given by
\begin{subequations}
  \begin{align}
    J_a
    & =
    \frac{2\Gamma}{\pi\rho}
    \left(
      \rcp{\epsilon_1 + 2U - J/4} {+} \rcp{-\epsilon_1 - U - J/4}
    \right)
    \\
    J_b
    & =
    \frac{2\gamma^2\Gamma}{\pi\rho}
    \left(
      \rcp{-\epsilon_2 - U - J/4} {+} \rcp{\epsilon_2 + 2U - J/4}
    \right)
    \\
    J_{ab}
    & =
    \frac{\sqrt2\gamma\Gamma}{\pi\rho}
    \left(
      \rcp{-\epsilon_2 - U - J/4} {+} \rcp{\epsilon_1 + 2U - J/4}
    \right),
  \end{align}
\end{subequations}
while one obtains the same expression for $\Jst$ as \eqnref{eq:Jst}.  Here each
of two Kondo spins is coupled to composite localized spins of conduction-band
channels. The effective Hamiltonian, \eqnref{eq:HeffII} is not convenient for
further analysis since it contains cross terms ($\bfs_{ab}$ and $\bfs_{ba}$)
that do not conserve the channel degrees of freedom. We introduce a unitary
transformations that diagonalizes the conduction-band spin operator in the
channel basis that is coupled to $\widetilde\bfS_q$ for $q=a,b$:
\begin{align}
  \begin{bmatrix}
    \tilde{c}_{a\bfk\mu} \\ \tilde{c}_{b\bfk\mu}
  \end{bmatrix}
  =
  \begin{bmatrix}
    \cos\vartheta & \sin\vartheta
    \\
    -\sin\vartheta & \cos\vartheta
  \end{bmatrix}
  \begin{bmatrix}
    c_{a\bfk\mu} \\ c_{b\bfk\mu}
  \end{bmatrix}
\end{align}
with $\vartheta\equiv \pm\frac12 \tan^{-1}[2J_{ab}/(J_a{-}J_b)]$ for $q=a$
(upper sign) and $b$ (lower sign), respectively. In terms of rotated
conduction-band spins $\widetilde\bfs_{ss'} \equiv \frac12 \sum_{\bfk\bfk'}
\sum_{\mu\mu'} \tilde{c}_{s\bfk\mu}^\dag \bfsigma_{\mu\mu'}
\tilde{c}_{s'\bfk'\mu'}$, the spin exchange terms in the effective Hamiltonian
read
\begin{align}
  \label{eq:Heffrot}
  & \widetilde\bfS_q\cdot
  \left(J_1\,\widetilde\bfs_{aa} + J_2\,\widetilde\bfs_{bb}\right)
  \\
  \nonumber
  & \mbox{}
  +
  \widetilde\bfS_{\bar{q}}\cdot
  \left(
    J_3\,\widetilde\bfs_{aa} + J_4\,\widetilde\bfs_{bb}
    + J_5\, (\widetilde\bfs_{ab} + \widetilde\bfs_{ba})
  \right)
  +
  \Jst\,\widetilde\bfS_a\cdot\widetilde\bfS_b,
\end{align}
where $\bar{q} = a (b)$ for $q=b (a)$ denotes the index of the Kondo spin for
which the coupled conduction-band spin operator is not diagonalized, and the
coefficients are given by
\begin{subequations}
  \begin{align}
    J_1 & = \bar{J} + \delta J \sec2\vartheta,
    &
    J_3 & = \bar{J} + \delta J \cos4\vartheta \sec2\vartheta,
    \\
    J_2 & = \bar{J} - \delta J \sec2\vartheta,
    &
    J_4 & = \bar{J} - \delta J \cos4\vartheta \sec2\vartheta,
    \\
    J_5 & = - 2 \delta J \sin2\vartheta
  \end{align}
\end{subequations}
with $\bar{J} \equiv (J_a+J_b)/2$ and $\delta J = (J_a-J_b)/2$. The index $q$
is chosen between $a$ and $b$ such that either $J_1$ or $J_2$ is the largest
among the coefficients.
Now the scaling analysis is ready with \eqnref{eq:Heffrot}. Suppose that $J_1$
is the largest one. Upon decreasing temperature, the Kondo spin
$\widetilde{S}_q$ is first screened by the conduction-band spin
$\widetilde\bfs_{aa}$, which defines a Kondo temperature $T_{K,1}$. Below this
Kondo temperature, the spins $\widetilde{S}_q$ and $\widetilde{s}_{aa}$ are
energetically frozen so that the remaining degrees of freedom is approximately
governed by the exchange coupling,
\begin{figure}[!t]
  \centering
  \includegraphics[width=4.25cm]{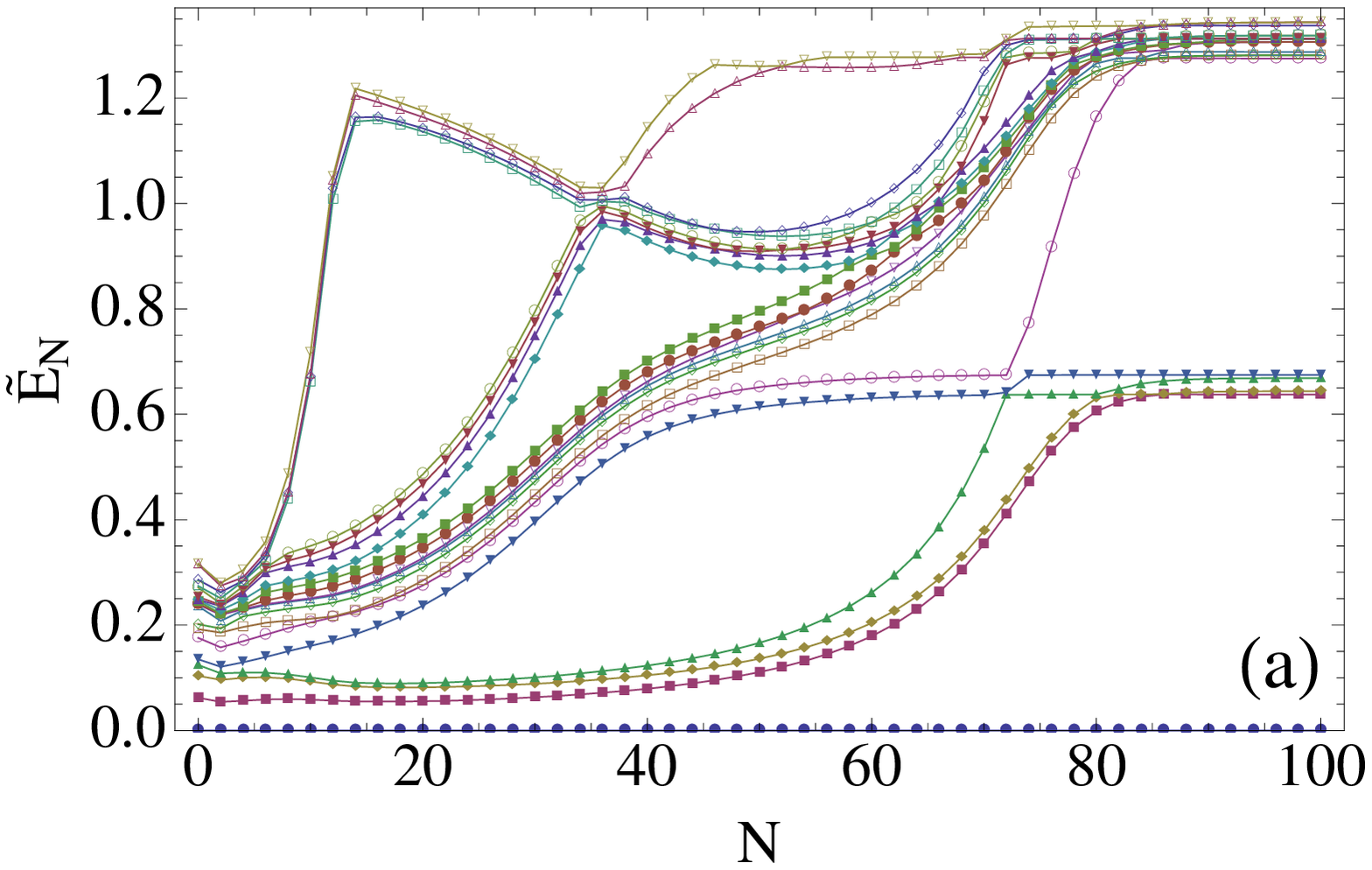}%
  \includegraphics[width=4.25cm]{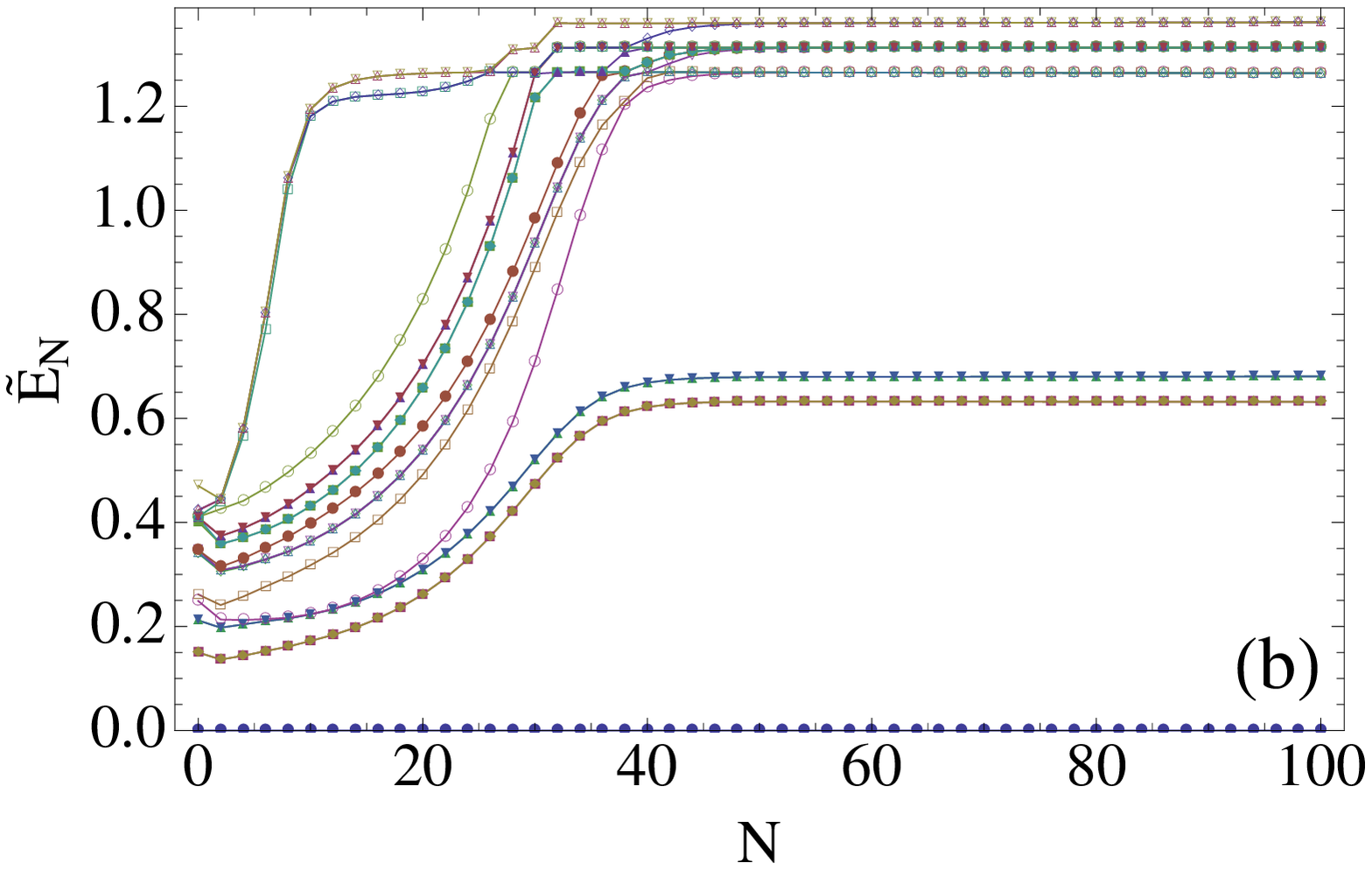}
  \caption{(color online) Scaled NRG eigenenergy flows with the iteration $N$
    for a two-level QD coupled to normal leads (case II) with (a) $\gamma =
    0.5$ and $J/D = -0.9$ and (b) $\gamma = 1$ and $J/D = -1.2$.}
  \label{fig:niiflow}
\end{figure}
\begin{align}
  J_4 \widetilde{S}_{\bar{q}}\cdot \widetilde{s}_{bb}.
\end{align}
The antiferromagnetic coupling will eventually screen out the remaining Kondo
spin $\widetilde{S}_{\bar{q}}$ at a lower Kondo temperature $T_{K,2}$ since
$J_4 < J_1$. Hence the system undergoes two-stage Kondo
effects\cite{Pustilnik01} as the temperature goes down: Two Kondo spins are
screened out one by one since their couplings to relevant conduction-band
degrees of freedom are different in magnitude. The ground state is of the spin
singlet at low temperatures $(T < T_{K,2})$ due to complete screening, while
the partial screening in the intermediate temperature $T_{K,2} < T < T_{K,1}$
leaves the system in the spin doublet. We have confirmed this scaling analysis
numerically by examining the RG flow of the scaled low-lying eigenenergies in
the NRG procedure. \Figref{fig:niiflow}~(a) clearly shows that the flow is in
the high-temperature regime attracted by an unstable fixed point and then goes
to the stable fixed point in the lower temperature.

The spin exchange coupling $\Jst\,\widetilde\bfS_a\cdot\widetilde\bfS_b$ can
interrupt the Kondo correlation. In fact, we have observed that the two-stage
Kondo effect ceases to happen if the exchange coupling $\Jst$ is so
antiferromagnetic that $\Jst > k_B T_{K,1}$. In this regime, the Kondo spins
are frozen to form a spin singlet by themselves before the conduction-band
electrons screen them out. Hence, at zero temperature the system undergoes a
transition between a Kondo state and an antiferromagnetic state as the exchange
coupling $\Jst$ is varied. In contrast to the case I, however, the transition
does not involve any change in the spin state: The ground state in both states
is of the spin singlet. Note that the two-stage Kondo effect arises in the
ferromagnetic side in the two-channel case while the one in the single-channel
case happens in the antiferromagnetic side.

It may be interesting to consider a special case when the two Kondo
temperatures are equal to each other: $T_{K,1} = T_{K,2}$. This can happen when
$J_a = J_b$ so that $J_1 = J_4 = J_a + J_{ab}$, $J_2 = J_3 = J_a - J_{ab}$, and
$J_5 = 0$, giving rise to the exchange Hamiltonian:
\begin{align}
  &
  J_1
  \left(
    \widetilde\bfS_q\cdot\widetilde\bfs_{aa}
    +
    \widetilde\bfS_{\bar{q}}\cdot\widetilde\bfs_{bb}
  \right)
  +
  J_2
  \left(
    \widetilde\bfS_q\cdot\widetilde\bfs_{bb}
    +
    \widetilde\bfS_{\bar{q}}\cdot\widetilde\bfs_{aa}
  \right)
  \\
  \nonumber
  & \mbox{}
  +
  \Jst\,\widetilde\bfS_a\cdot\widetilde\bfS_b.
\end{align}
Since $J_1 > J_2$, the Kondo spins $\widetilde\bfS_q$ and
$\widetilde\bfS_{\bar{q}}$ are simultaneously screened by the localized spins
$\widetilde\bfs_{aa}$ and $\widetilde\bfs_{bb}$, respectively, defining a same
Kondo temperature $T_K$. The RG flow in the NRG procedure confirms that there
exist no unstable fixed point and that only one Kondo temperature governs the
flow: see \figref{fig:niiflow}~(b). In our system, the condition $J_a = J_b$ is
satisfied with $\gamma = 1$ under the particle-hole symmetry condition.

\begin{figure}[!t]
  \centering
  \includegraphics[width=4.25cm]{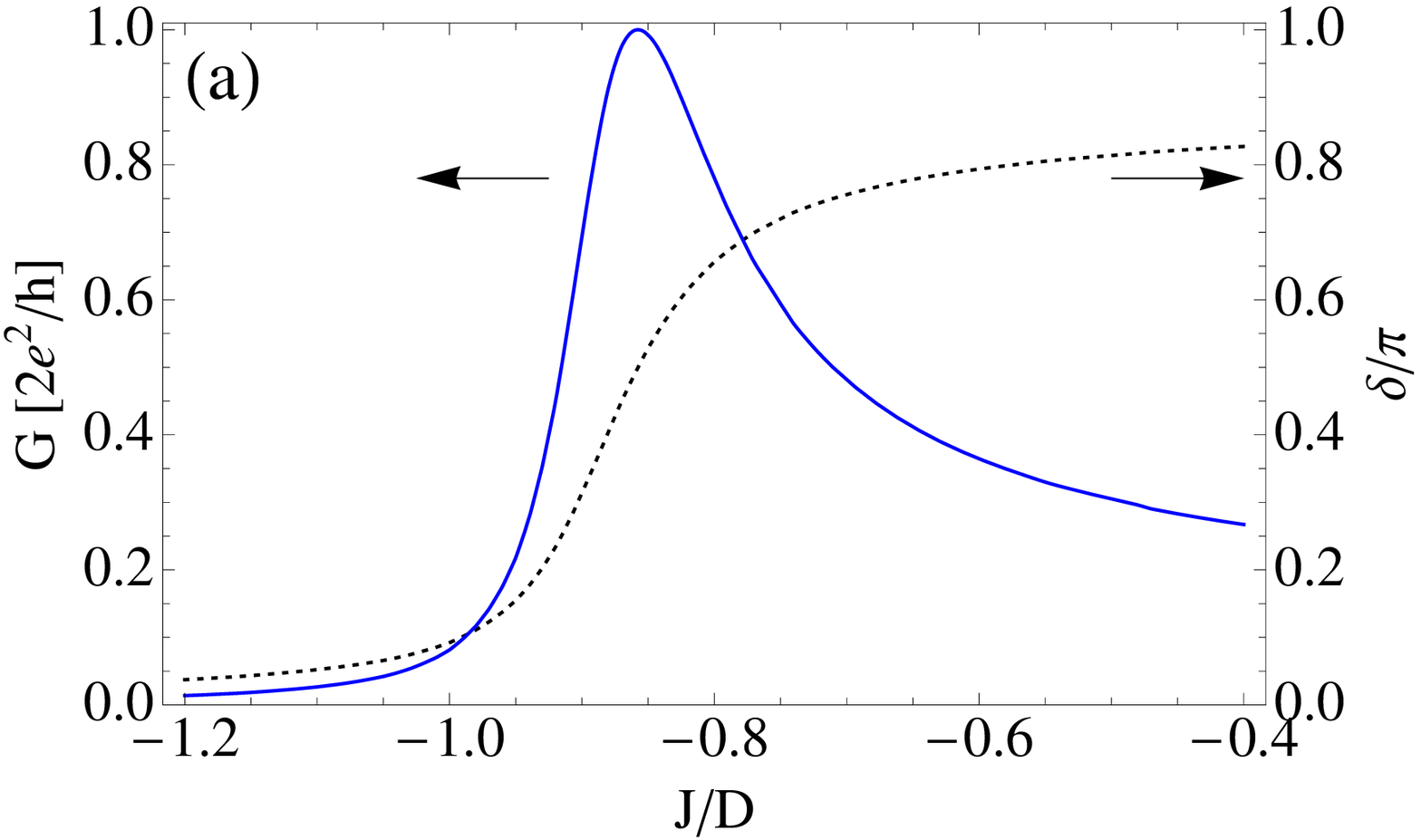}%
  \includegraphics[width=4.25cm]{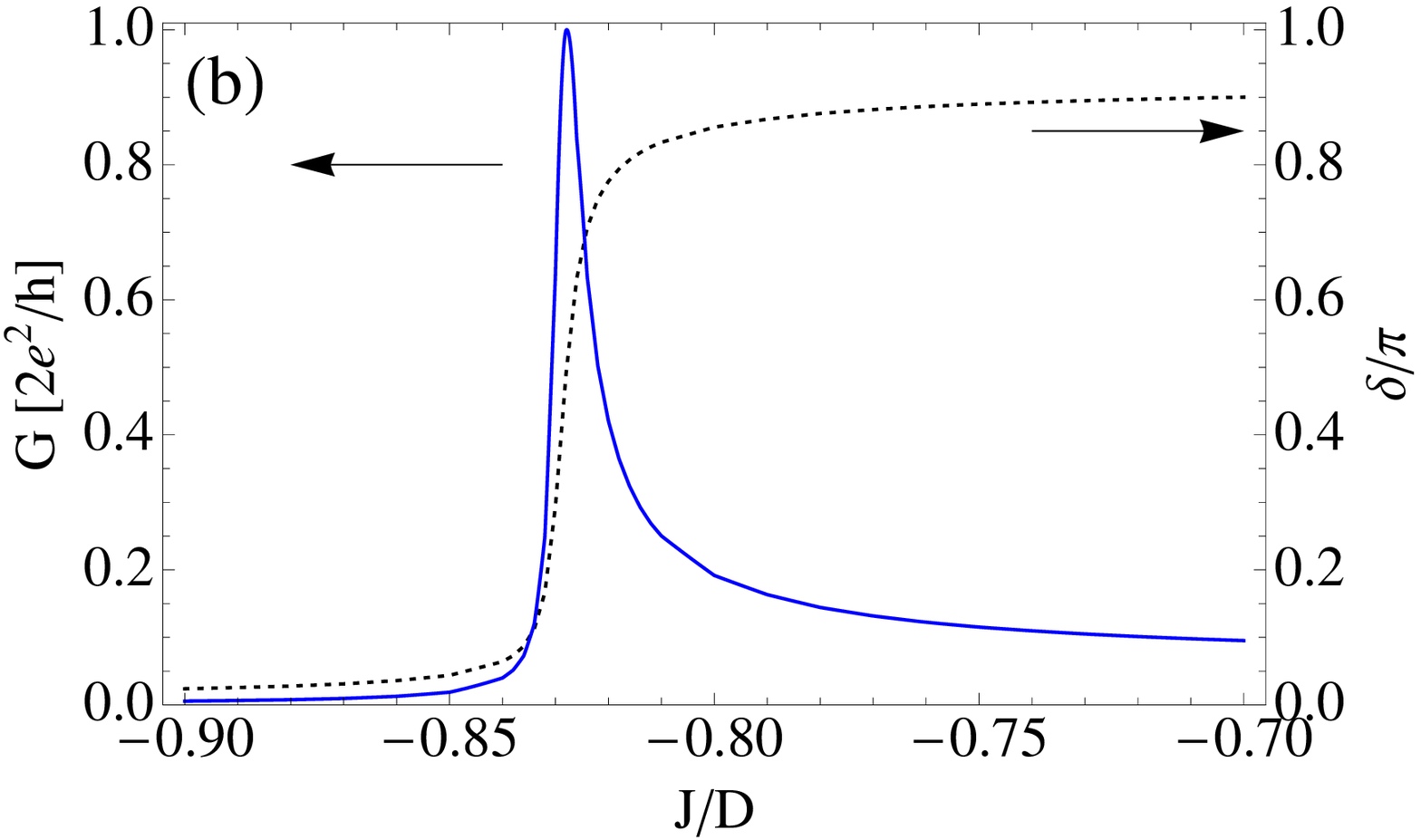}
  \caption{(color online) Phase shift and linear conductance as functions of
    $J$ for a two-level QD coupled to normal leads (case II) with $\gamma = 1$
    [(a)] and 0.5 [(b)].}
  \label{fig:niic}
\end{figure}

The transport in the vicinity of the singlet-triplet transition of the isolated
dot and on both of the antiferromagnetic and ferromagnetic sides can be
analyzed by measuring the linear conductance from the NRG
calculations. According to the Landauer-B\"uttiker formula in terms of the
scattering matrix,\cite{Pustilnik01} the zero-temperature linear conductance
can be expressed in terms of the phase shift $\delta_s$ for each channel:
\begin{align}
  \label{eq:G}
  G = \frac{2e^2}{h} \sin^2\delta,
\end{align}
with phase difference $\delta = \delta_a - \delta_b$. We have extracted the
phase shifts from the energy spectrum in the NRG procedure by using the fact
that the fixed point is described by a non-interacting Fermi
liquid.\cite{Hofstetter04} On the ferromagnetic side, the Kondo screening forms
a resonance level on each channel, which corresponds to a phase shift $\pi/2$
in both channels and $\delta=0$. On the antiferromagnetic side, on the other
hand, both QD electrons occupy the orbital 1 so that $\delta_a = \delta_1 =
\pi$ and $\delta_b = 0$, resulting in $\delta=\pi$. It implies that the
conductance, \eqnref{eq:G} must approach zero on both sides of the
singlet-triplet transition of the isolated dot while it has a maximum near the
transition when $\delta = \pi/2$. \Figref{fig:niic} shows that the phase
difference increases rapidly from zero to $\pi$ near the transition point
$(J\approx -0.8D)$ and that the conductance reaches the unitary limit when
$\delta=\pi/2$. The maximal conductance point is shifted with respect to the
bare singlet-triplet transition point $J = -0.8D$ since the dot-lead
correlation favors the spin singlet state energetically [see \eqnref{eq:Jst}].
As can be seen from \figref{fig:niic}, the zero-temperature linear conductance
does not reflect the presence of two different Kondo scales: the qualitative
feature of the conductance is same for $\gamma=1$ and $\gamma<1$. Inclusion of
Zeeman splitting,\cite{Hofstetter04} finite temperatures, or superconductivity
can, however, alter the low-temperature transport property dramatically if the
relevant energy scale is between two Kondo temperatures and one of the Kondo
correlation with lower Kondo temperature is suppressed. In the next section, we
study how it happens in Josephson junctions.

\subsection{Superconducting Leads: $\Delta \ne 0$}

Now we investigate the TLQD Josephson junction with $\Delta\ne0$, considering
the single- and two-channel cases separately as in the study of normal-lead
case. The system state is identified as in the study of the weak-coupling
regime: refer to \secref{sec:pt} for the definitions of labels of the system
state and phase boundaries. Here we introduce a new label $D$ to denote the
spin-doublet ground state which is missing in the weak-coupling regime.

A series of studies on the single-level QD Josephson
junction\cite{Shiba69,Glazman89,Spivak91,Rozhkov99,Ryazanov01,Kontos02,Choi04,Siano04,Karrasch08}
have already revealed that the competition between the Kondo effect and the
superconductivity can lead to a Kondo-driven phase transition between the
Kondo-dominant and superconductivity-dominant states and that the transition
can be driven by tuning the relative strength between the Kondo temperature
$T_K$ and the superconducting gap $\Delta$. Strong conductivity $(\Delta\gg
T_K)$ in the leads enforces the conduction electrons to form Copper pairs by
themselves and does not interfere the spin correlation between the QD
electrons. In the opposite limit ($\Delta\ll T_K)$, however, the
conduction-band electrons in the leads screen out the QD spins through
spin-flip processes.  Hence one can expect that as $\Delta$ is decreased the
system undergoes a phase transition from states that prevail in the
weak-coupling limit to other states governed by the Kondo effect. Below we find
out that the superconducting gap introduces an infrared energy cutoff to the
system and acts like a coherent probe for the Kondo excitation
spectrum. Therefore, we expect that distinct Kondo effects in the two cases --
cases I and II -- should lead to different phase diagrams in the presence of
the superconductivity even at zero temperature.

\subsubsection{Case I: Single Channel}

\begin{figure}[!t]
  \centering
  \includegraphics[width=7cm]{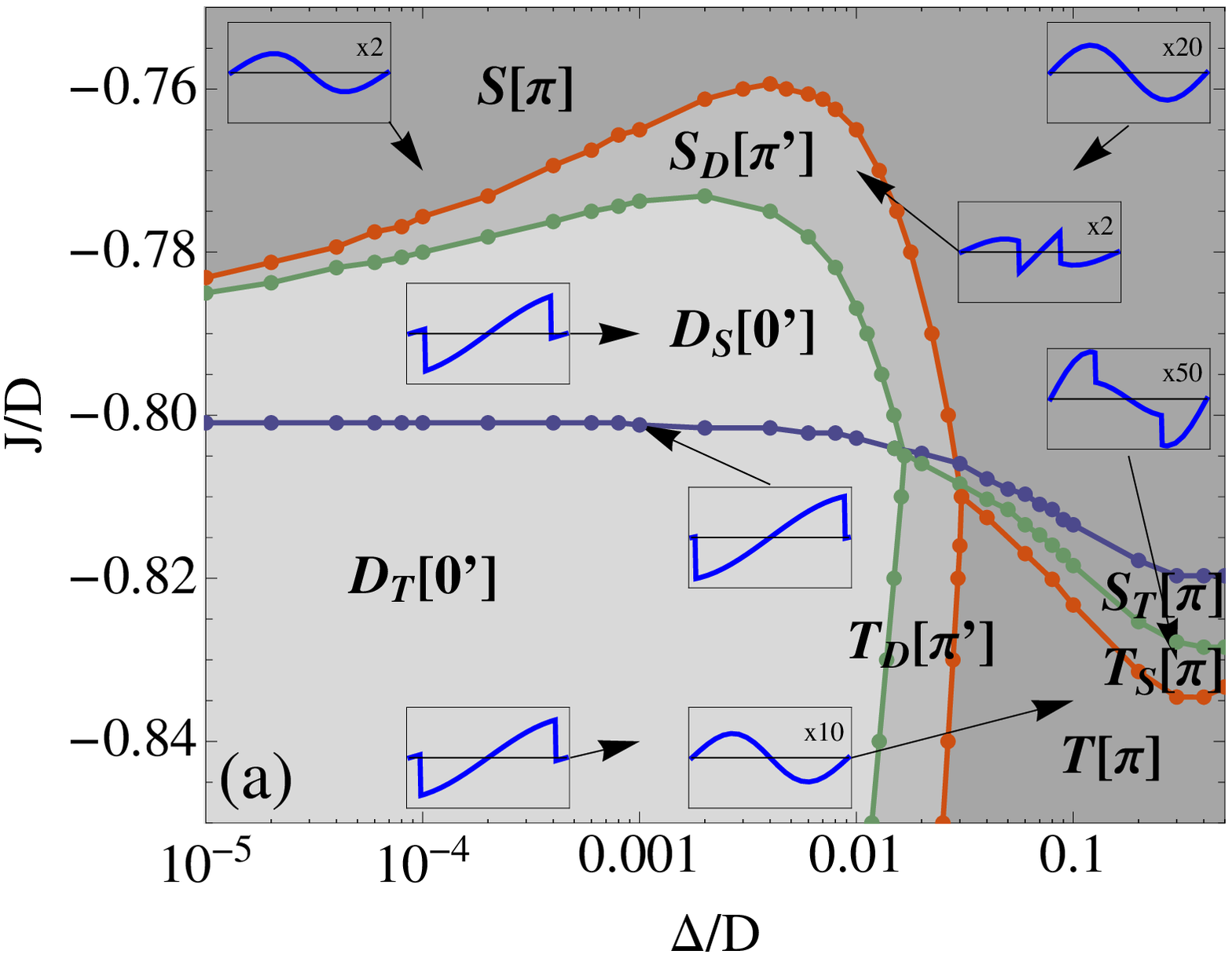}
  \includegraphics[width=7cm]{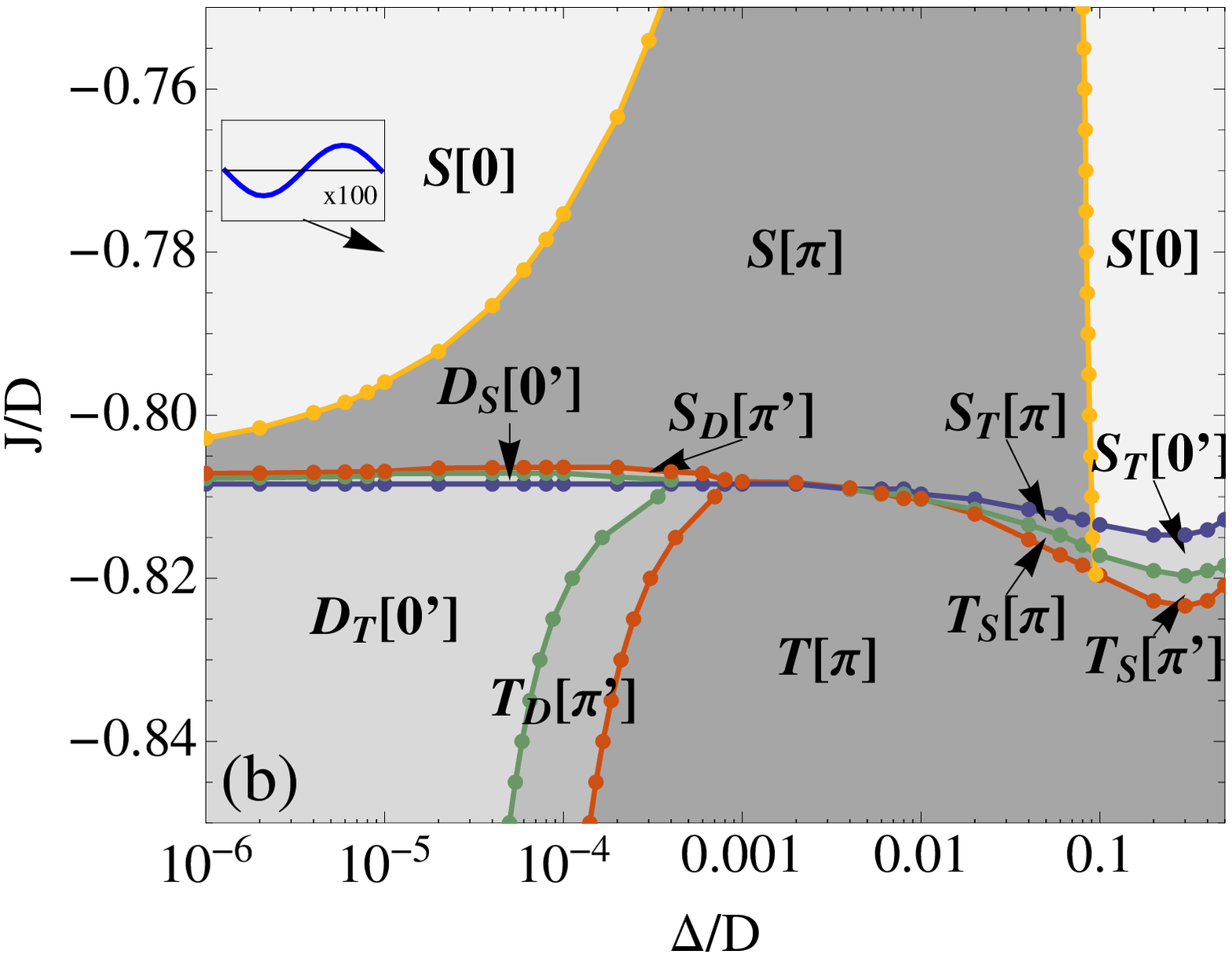}
  \includegraphics[width=7cm]{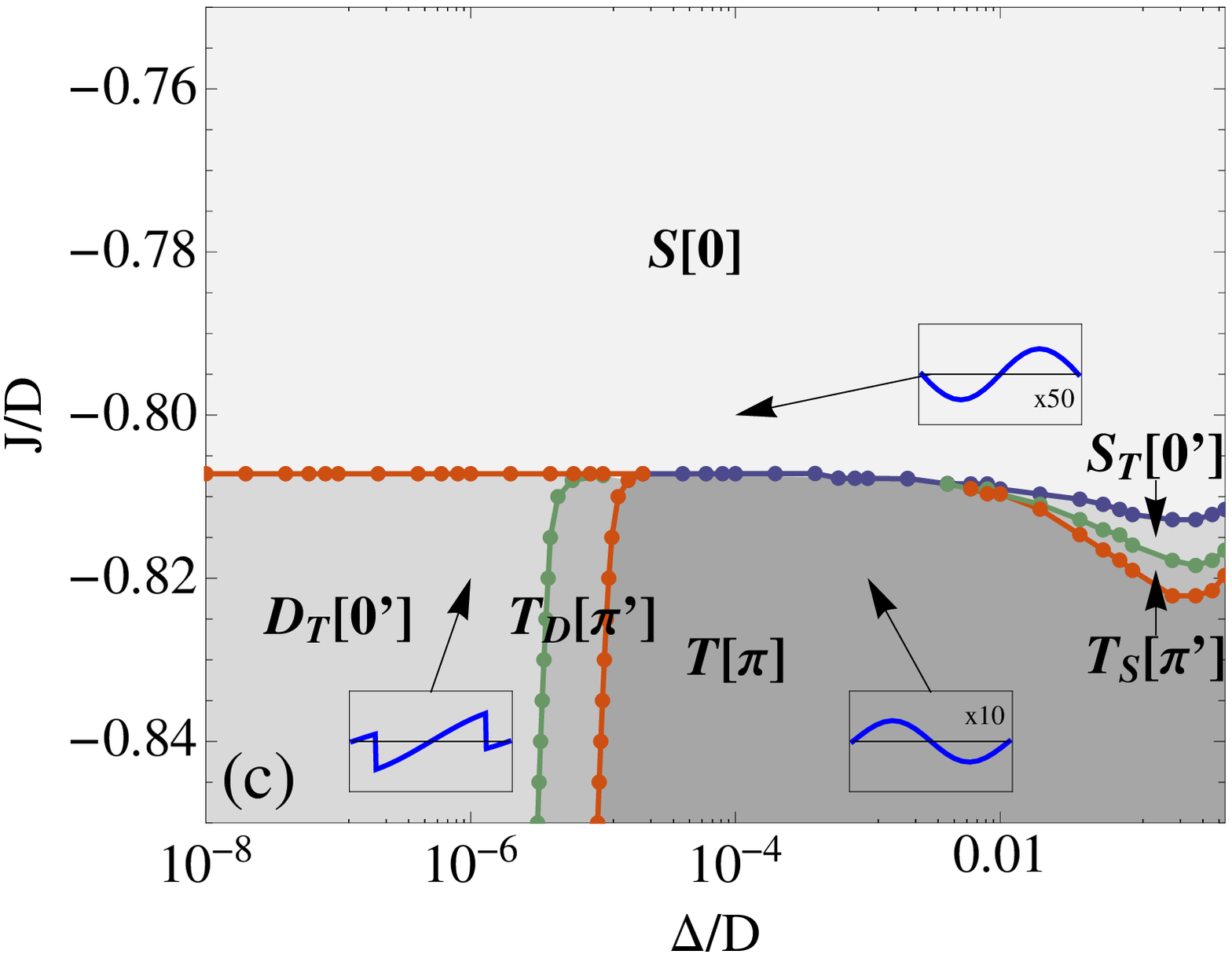}
  \caption{(color online) Phase diagrams in the $\Delta$-$J$ plane for the case
    I with $\gamma = 1$ [(a)], 0.4 [(b)], and 0.1 [(c)]. The phase boundaries
    are located when the ground-state spin is changed at $\phi = 0$ (red line),
    $\pi/2$ (green line), and $\pi$ (blue line). The yellow line separates two
    states with same ground-state spin but different SPRs. Refer the detailed
    classification of the states to the text. The insets show the SPRs for
    $\phi\in[-\pi,\pi]$ at the points indicated by the arrows. Here the solid
    lines are guide for eyes.}
  \label{fig:pdi}
\end{figure}

\textit{Weak Coupling Regime}--- \Figref{fig:pdi} shows the phase diagrams in
the $\Delta$-$J$ plane in the case I.  As expected, the NRG calculations
confirm the results of the perturbation theory in the weak coupling limit
($\Delta/T_K\gg1$). The system undergoes the singlet-triplet transition through
intermediate states, $S_T$ and $T_S$, as $J$ is tuned. The phase boundaries
between them are in perfect agreement with ones found from the perturbation
theory: compare \figsref{fig:wcpdi} and \ref{fig:pdi}. Not only the
superconductivity-induced renormalization of the singlet-triplet splitting is
well reproduced, but also its asymptotic behavior ($\delta \Jst\to0$) is
correctly predicted in the limit $\Delta\to0$ where the perturbation theory
breaks down. Note that the normal-lead contribution to $\Jst$ [see
\eqnref{eq:Jst}] becomes effective for $\Delta\ll\Gamma$, leaving $\delta\Jst$
finite.  The SPRs calculated from \eqnref{eq:jc} also clearly follow those of
the perturbative results: compare the insets of \figsref{fig:wcpdi} and
\ref{fig:pdi}. At $\gamma=1$, the SPR is of the $\pi$-junction regardless of
the spin of the ground state, while that of the spin singlet state becomes of
the 0-junction as $\gamma$ is decreased.

\textit{Strong Coupling Regime}--- For smaller $\Delta$, on the other hand, the
transition to the spin doublet state takes place, which is clearly ascribed to
the Kondo effect. The conduction-band electrons in the effective single channel
screen out one of the two QD spins, leaving the other unscreened. We have
observed that the transition takes place at values of $\Delta \approx T_K$ [see
red lines in \figref{fig:pdi}], where $T_K$ is the Kondo temperature estimated
from the width of the transmission coefficient in the normal-lead case [refer
to \figref{fig:nit} for $\gamma = 1$]. The phase transition is highly dependent
on the values of $J$ and exhibits asymmetric structure with respect to the sign
of $\Jst$. On the ferromagnetic side $(\Jst<J_c)$ the system experiences a
transition from the spin triplet to the spin doublet state with decreasing
$\Delta$, while on the antiferromagnetic side $(\Jst>J_c)$ the
singlet-doublet-singlet double transition is observed. In the strongly
antiferromagnetic side ($\Jst\gg J_c$) there exists no transition at all. This
$J$-dependence of the transition originates from the fact that the spin
exchange coupling affects the Kondo effect as discussed in
\secref{sec:nlI}. First, on the ferromagnetic side $(\Jst < J_c)$, the Kondo
temperature decreases with increasing $|\Delta J|$. It explains the shift of
the $T$-$D$ phase boundary in \figref{fig:pdi} toward smaller $\Delta$ with
increasing $|J|$.\cite{Lee08} On the antiferromagnetic side $(\Jst > J_c)$, a
two-stage Kondo effect with two Kondo temperature $T_K$ and $T_K^I$ takes place
for small $\Delta J$.  As long as $\Delta > T_K^I$, the second Kondo effect
does not appear since the superconducting gap blocks any quasi-particle
excitation within the gap $\Delta$. Therefore, for $T_K^I < \Delta < T_K$, one
Kondo spin remains unscreened, forming the spin doublet state.  For
$\Delta\lesssim T_K^I$, however, Cooper pairs notice the suppression of the
Kondo resonance level, and their tunneling is governed by cotunneling under
strong Coulomb interaction, restoring the weak-coupling supercurrent in the
presence of the spin singlet correlation. Hence the observed shape of the
$S$-$D$ phase boundary and the reentrant behavior are well explained by the
fact that the first Kondo temperature $T_K$ decreases with increasing $\Delta
J$ as in the ferromagnetic side and that the second one $T_K^I$ decreases with
decreasing $\Delta J$ and vanishes as $\Delta J\to0$.

\textit{Finite Phase Difference}--- The transition boundaries depend on the
phase difference $\phi$, which is responsible for the occurrence of the
intermediate states. One should note that in the presence of finite $\Delta$
and $\phi$ both of two conduction-band channels are coupled to the QD: All the
elements of the dot-lead coupling matrix, \eqnref{eq:dlc} become finite. In
addition, the finite $\Delta$ makes it impossible to decouple one channel
completely via any unitary transformation. However, the effect of the second
channel is energetically cut off by the finite gap $\Delta$ itself in a sense
that the Kondo temperature due to the coupling to the second channel is always
smaller than $\Delta$. The single-channel argument is thus sufficient to
account for the phase transitions even at finite $\phi$. The dot-lead coupling
matrix, \eqnref{eq:dlc} then indicates that the couplings $t_{ab}$ and $t_{bb}$
responsible for the Kondo effect are reduced from $\sqrt2 t$ to $\sqrt2 t
\cos\frac{\phi}{4}$, which accordingly lowers the Kondo temperature. The
decrease of the Kondo temperature is clearly demonstrated in the phase
diagrams: The phase boundaries at $\phi=\pi/2$ [see green lines in
\figref{fig:pdi}] are located at smaller $\Delta$ than those at $\phi=0$ and
the regime of double transition is shrunken due to the increase of the second
Kondo temperature $T_K^I$ [see \eqnref{eq:tki}]. However, not only the
diminished dot-lead coupling is responsible for the reduction of the Kondo
temperature. We have found that a scaling analysis with finite $\Delta$
produces exotic terms (like proximity terms) which are missing in the
normal-lead case. Such terms with finite $\phi$ can suppress the Kondo
correlation further by twisting the phase correlation between two leads. We
have observed that at the maximally twisted condition, that is, $\phi=\pi$, no
Kondo state appears at all so that the Kondo state exists only in the
intermediate state.  Such a vulnerability of the Kondo effect at a maximally
twisted phase condition to any finite magnetic perturbation was also observed
in the magnetic molecular JJ.\cite{Lee08} The Kondo state may survive the
maximally twisted condition only if there is no additional magnetic interaction
$(\Jst = 0)$ as in the single-level QD-JJ. In our system, however, no
appearance of the Kondo state at $\phi=\pi$ is observed even along the
$D_T$-$D_S$ boundary where the effective splitting is supposed to vanish $(J
\approx J_c)$. We attribute it to the fact that the superconductivity shifts
the energy levels of the QD spin states and induces the $\phi$-dependent
singlet-triplet splitting, which seems to favor energetically the spin singlet
or triplet states over the Kondo state.

\textit{SPR and Andreev Levels}--- Once the Kondo correlation prevails over the
superconductivity, a resonant level is formed at the Fermi level, and the
Cooper pairs tunnel through the Kondo resonant state, resulting in a ballistic
0-junction. Together with the $\phi$-dependent phase transition, the resonant
tunneling makes the curve of the SPR break into three distinct segments as soon
as the Kondo effect becomes effective, as seen in the insets of
\figref{fig:pdi}. The central segment resembles that of a ballistic short
junctions, while the two surrounding segments are parts of the tunneling SPR
for the spin singlet or triplet states.  Since the Kondo state does not occur
at $\phi=\pm\pi$, the SPR keeps the three-segment structure and does not become
of the perfect ballistic junction that was observed in the single-level
QD-JJ.\cite{Choi04}

\begin{figure}[!t]
  \centering
  \includegraphics[width=7cm]{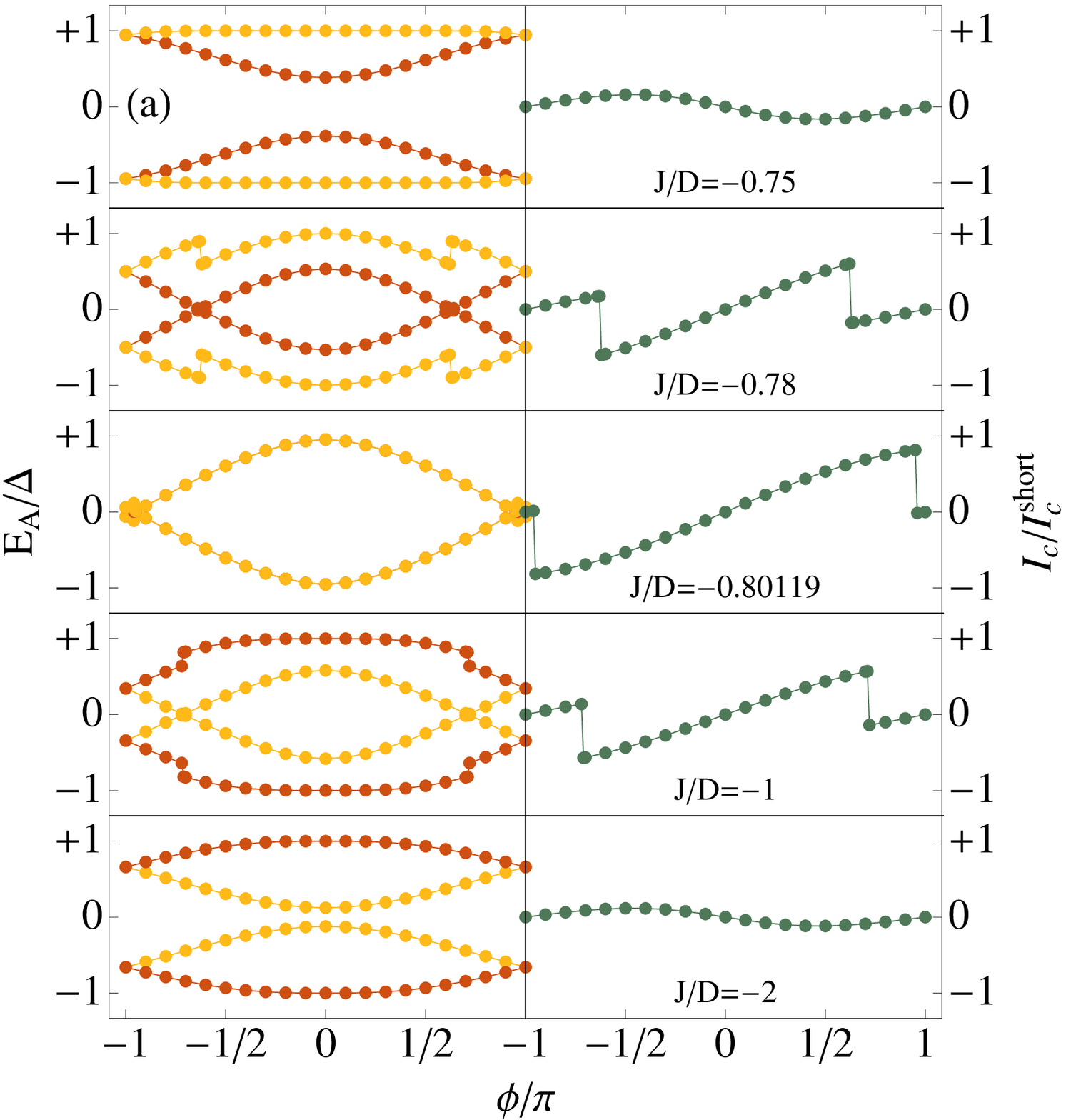}\\
  \includegraphics[width=6.5cm]{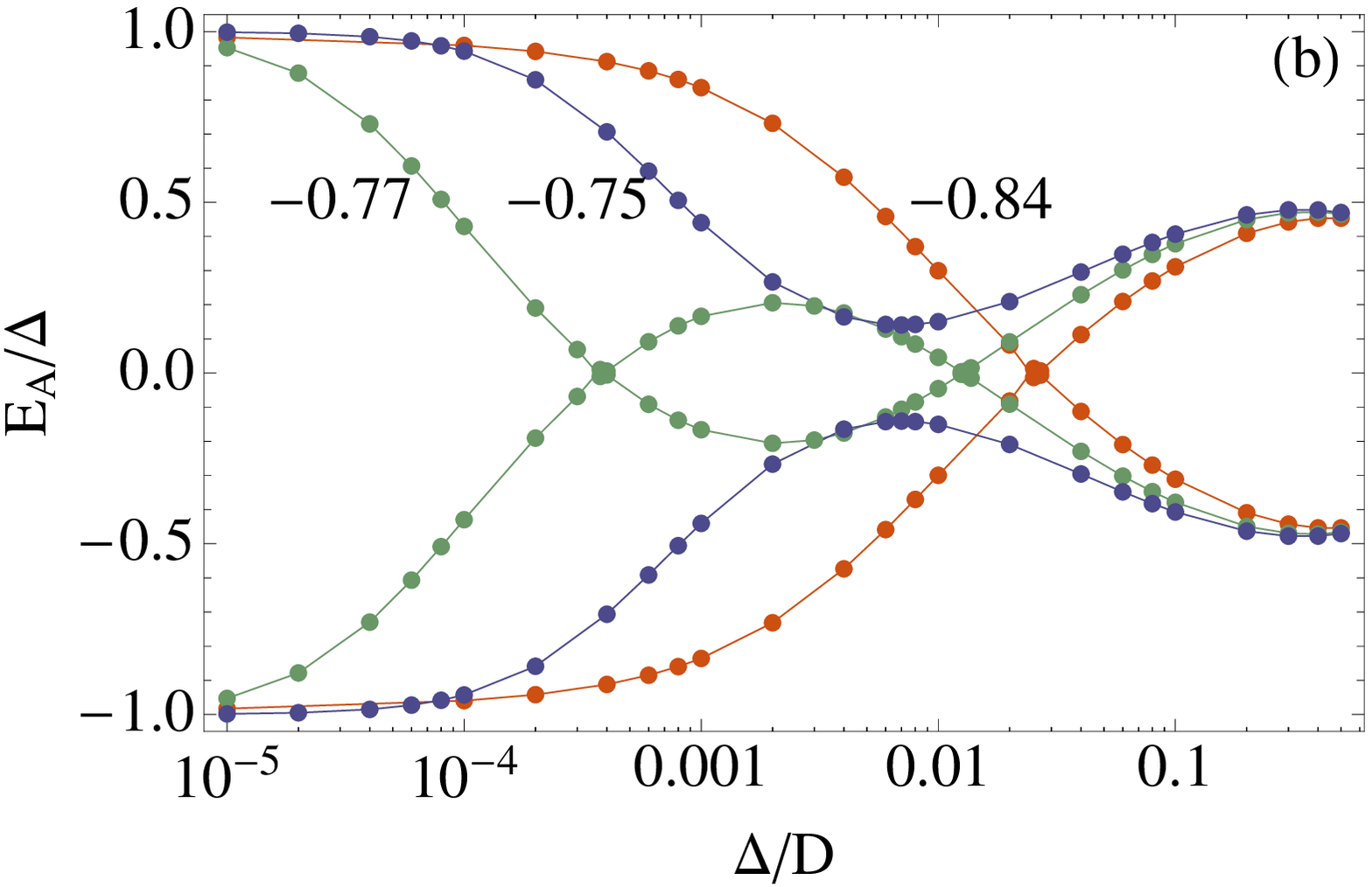}
  \caption{(color online) (a) (LEFT) Andreev levels in units of $\Delta$ and
    (RIGHT) supercurrent $I$ in units of $I_c^{\rm short}$ as functions of
    $\phi$ and $\Delta/D = 10^{-3}$ for various values of $J$ as annotated. Red
    and yellow dots/lines denote the spin singlet and triplet Andreev levels,
    respectively. (b) Andreev levels (close to the Fermi level) in units of
    $\Delta$ as functions $\Delta/D$ at $\phi=0$ for several values of $J$ as
    labeled. Here we used $\gamma = 1$.}
  \label{fig:ali}
\end{figure}

\Figref{fig:ali}~(a) displays typical variations of the Andreev levels and
supercurrent with $J$ at a fixed value of $\Delta/D = 10^{-3}$. On the
$D_T$-$D_S$ boundary with $\Jst \approx J_c$ [see the middle plots], the
spin-singlet(red) and triplet(yellow) Andreev levels are degenerate in the
central segment, while the degeneracy is lifted in side segments around
$\phi=\pm\pi$. Any finite effective singlet-triplet splitting, $\Jst \ne J_c$,
clearly induces a spin splitting in the subgap excitations, lifting the
degeneracy, and consequently shifts the crossing between the ground state and
the lowest excitation toward $\phi=0$, shrinking the central segment. Across
the crossing, the ground-state spin is changed from 1/2 to 0 (1) for $\Jst>J_c$
$(\Jst<J_c)$. Similarly, the Andreev levels exhibit discontinuities in the
spectra; for $\Jst>J_c$ $(\Jst<J_c)$, two outmost Andreev levels with spin 1
(0) in the central segment cannot remain as one-electron excitations with
respect to the spin-0(1) ground state at the transition and are replaced by new
Andreev levels with spin 1/2. In parallel with the abrupt change in the Andreev
levels, the SPR shows a discontinuous sign change (note that $I\propto
-\partial E_a/\partial\phi$, as the continuum-excitation contribution is
negligible), culminating in a transition from the 0 to the $\pi$ state:
accordingly, the states $D_S/D_T$ and $S_D/T_D$ are of the $0'$ and $\pi'$
states, respectively. As $\Jst$ grows in magnitude, the central segment shrinks
and eventually vanishes. The SPR then becomes sinusoidal, which is that of a
tunnel junction. Once the tunneling junction is fully established, stronger
singlet-triplet splitting does not lead to any qualitative change in the SPR.
The observed 0-$\pi$ transition is quite asymmetric with respect to the sign of
$\Jst$. First, the phase transition in the antiferromagnetic region ($\Jst >
J_c$) takes place at $\Jst \lesssim T_K$, while the 0 state survives much
larger ferromagnetic coupling ($\Jst < J_c$). Second, the antiferromagnetic
spin splitting gives rise to a double 0-$\pi$ transition that restores the spin
singlet correlation at small $\Delta$. \Figref{fig:ali}~(b) clearly shows that
for the antiferromagnetic $\Jst$ less than $T_K$ (at $J = -0.77$) the Andreev
levels make double crossings as $\Delta$ is varied and the spin singlet ground
state is restored at small $\Delta$. This is not the case in the ferromagnetic
region (at $J = -0.84$) where only one crossing appears nor in the strongly
antiferromagnetic region (at $J = -0.75$) where there exists no crossing at
all. Note that the crossing takes place only when the spin-doublet ground state
is replaced by the spin-singlet or triplet ones: No crossing appears in
switching between the spin singlet and triplet ground states since each of them
cannot be a single-particle excitation to the other.

\textit{Asymmetric Coupling}--- The phase diagram is also sensitive to the
asymmetry factor $\gamma$ as well.  The decrease of $\gamma$ gives rise to the
shrink of the intermediate states $S_D$ and $D_S$ in the antiferromagnetic side
and the shift of the $T$-$T_D$ and $T_D$-$D_T$ boundaries toward smaller values
of $\Delta$ [compare the phase diagrams in \figref{fig:pdi}]. To understand
this behavior, one should take a look at the typical dependence of the Kondo
temperature $T_K$ on the lead-dot coupling: $T_K \propto
\exp\left[-A/\Gamma_{\rm tot}\right]$, where $\Gamma_{\rm tot} =
2(1+\gamma^2)\Gamma$ is the total hybridization and $A$ is a
$\Gamma$-independent constant. As $\gamma$ is decreased from 1 to 0, the total
hybridization decreases from $4\Gamma$ to $2\Gamma$, which leads to the
exponential decrease of the Kondo temperature. Hence, with the smaller $T_K$,
the transitions from the spin double state to the spin singlet or triplet
states occur at smaller values of $\Jst$ or $\Delta$. The exponential
dependence of $T_K$ and $T_K^I$ on $\gamma$ makes the intermediate states $S_D$
and $D_S$ almost vanish and hard to detect even at $\gamma\approx0.4$.

Another effect of the asymmetry is the appearance of a second 0 state in the
singlet side in small-$\Delta$ region [see \figref{fig:pdi}~(b)].  This 0 state
is not like the one predicted from the perturbation theory which works only in
the large-$\Delta$ limit. In the small-$\Delta$ limit, all the high-order
processes should be taken into account in order to determine the sign of the
supercurrent. Even though the analytical analysis taking all the orders of the
processes is difficult to apply, a rough account can be proposed: From the
lowest-order terms, \eqnsref{eq:bs4} and (\ref{eq:bs5}), one can know that the
sign of the supercurrent due to each tunneling process is determined based on
which intermediate two-electron state appears in the middle of the tunneling
process. For example, processes with an intermediate spin singlet state
$\ket{2,0,0;3}$ make a positive contribution. In higher-order processes,
different kinds of intermediate states will appear one by one, and a negative
contribution can arise only for processes with an odd number of the
intermediate states that invert the sign of the current. Since processes with
an even number of sign-inverting states should outnumber odd-number processes,
one can claim that higher-order processes are likely to make the positive
contribution and consequently to favor the 0 state. We observe that this 0
state takes place for $\gamma < 1$ and that at $\gamma = 1$ the spin singlet
state is purely the $\pi$ state. We guess that the second 0 state also benefits
from the weakening of the negative $\beta'_{S5}$ term with decreasing $\gamma$.
\Figref{fig:pdi}~(b) shows that the second 0 state expands toward
large-$\Delta$ region as $J$ is increased. It can be explained by the argument
that with increasing $J$ the positively contributing processes with the
intermediate state $\ket{2,0,0;3}$ have larger amplitude because the energy
cost $E_{0;3}-E_{0;1} = \delta\epsilon - 3J/4$ diminishes; see
\eqnref{eq:bs5}. At smaller $\gamma$, the two 0 states expand further and
eventually merge to form a single 0 state, as shown in \figref{fig:pdi}~(c).

\begin{figure}[!t]
  \centering
  \includegraphics[width=7cm]{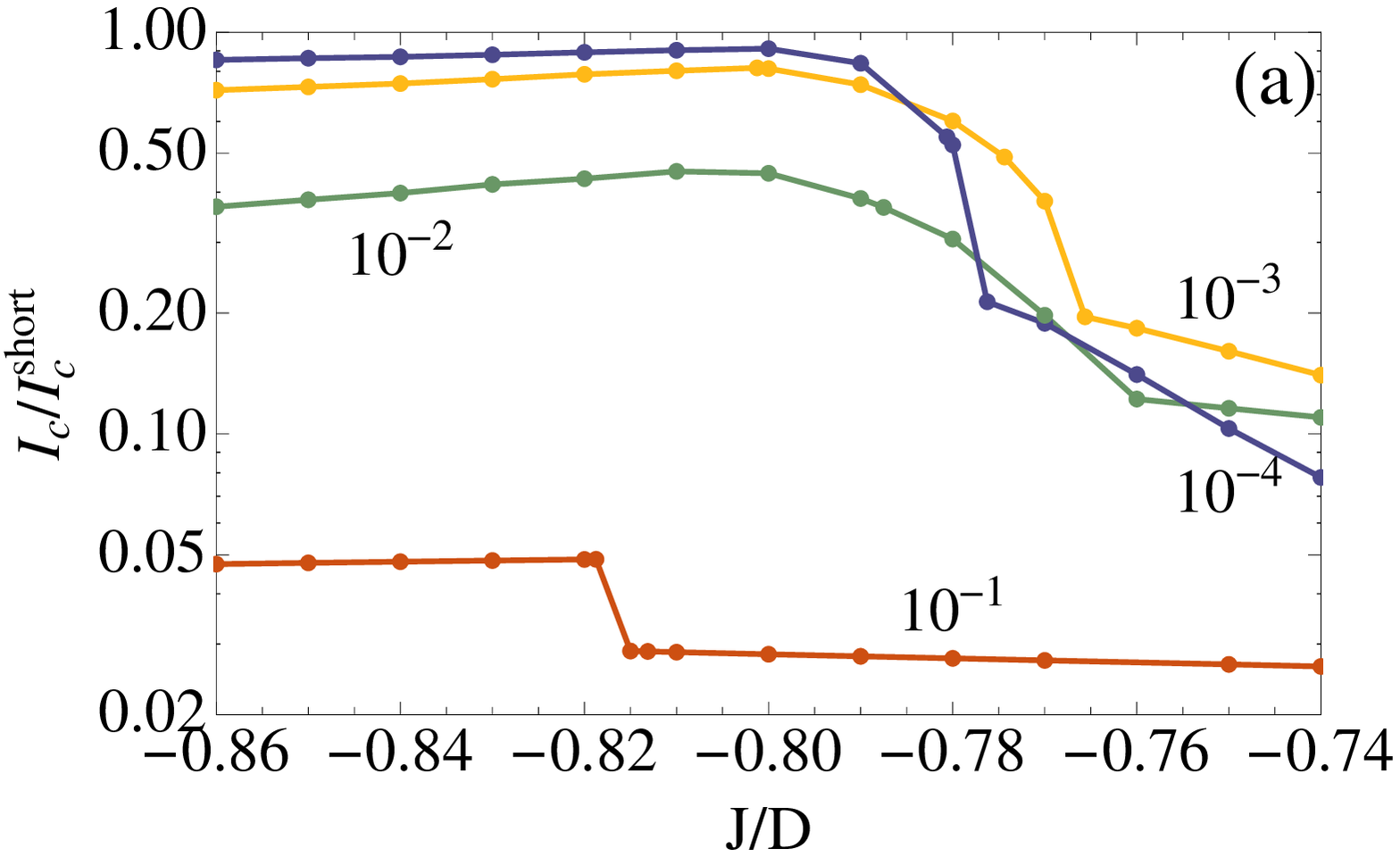}\\
  \includegraphics[width=7cm]{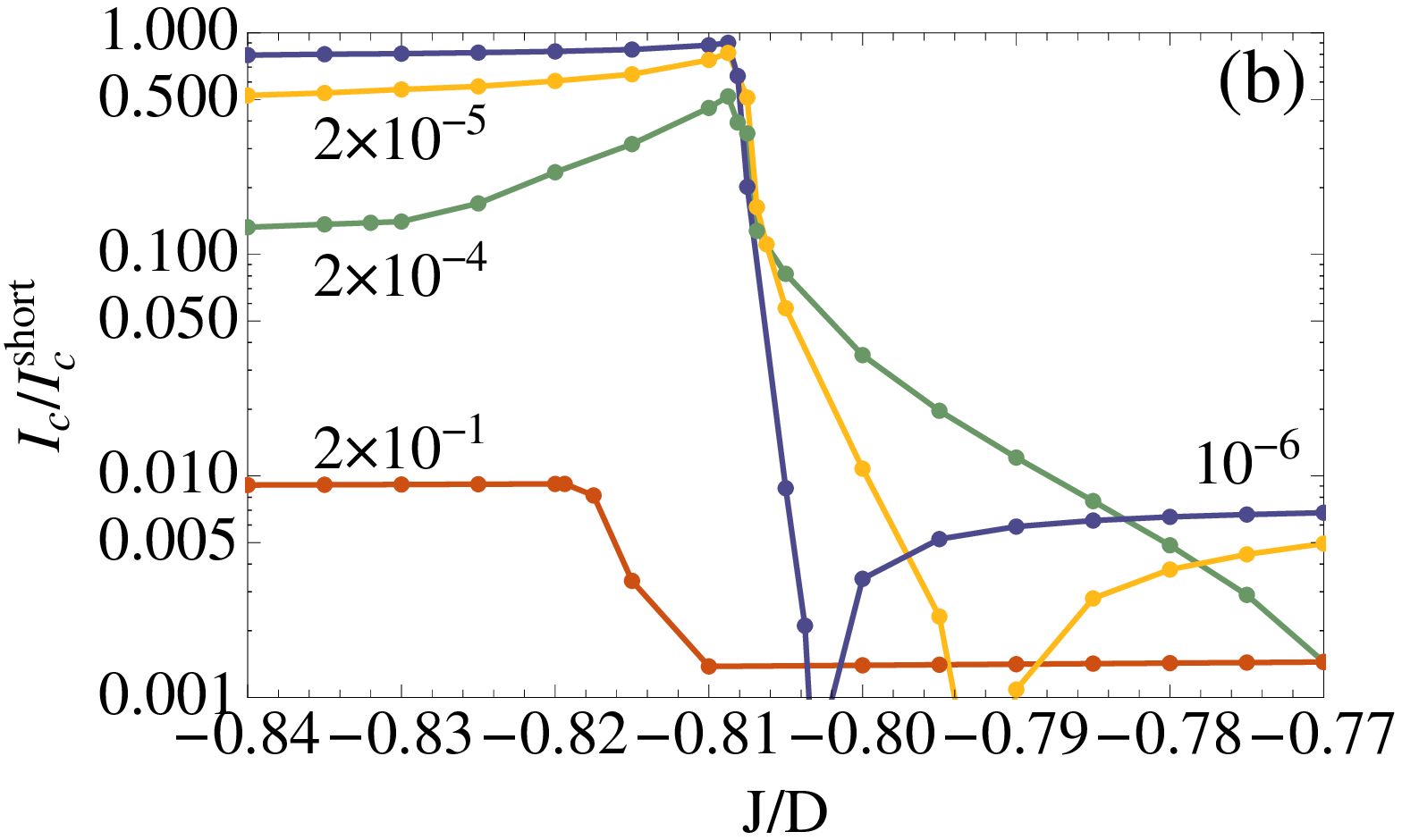}\\
  \includegraphics[width=7cm]{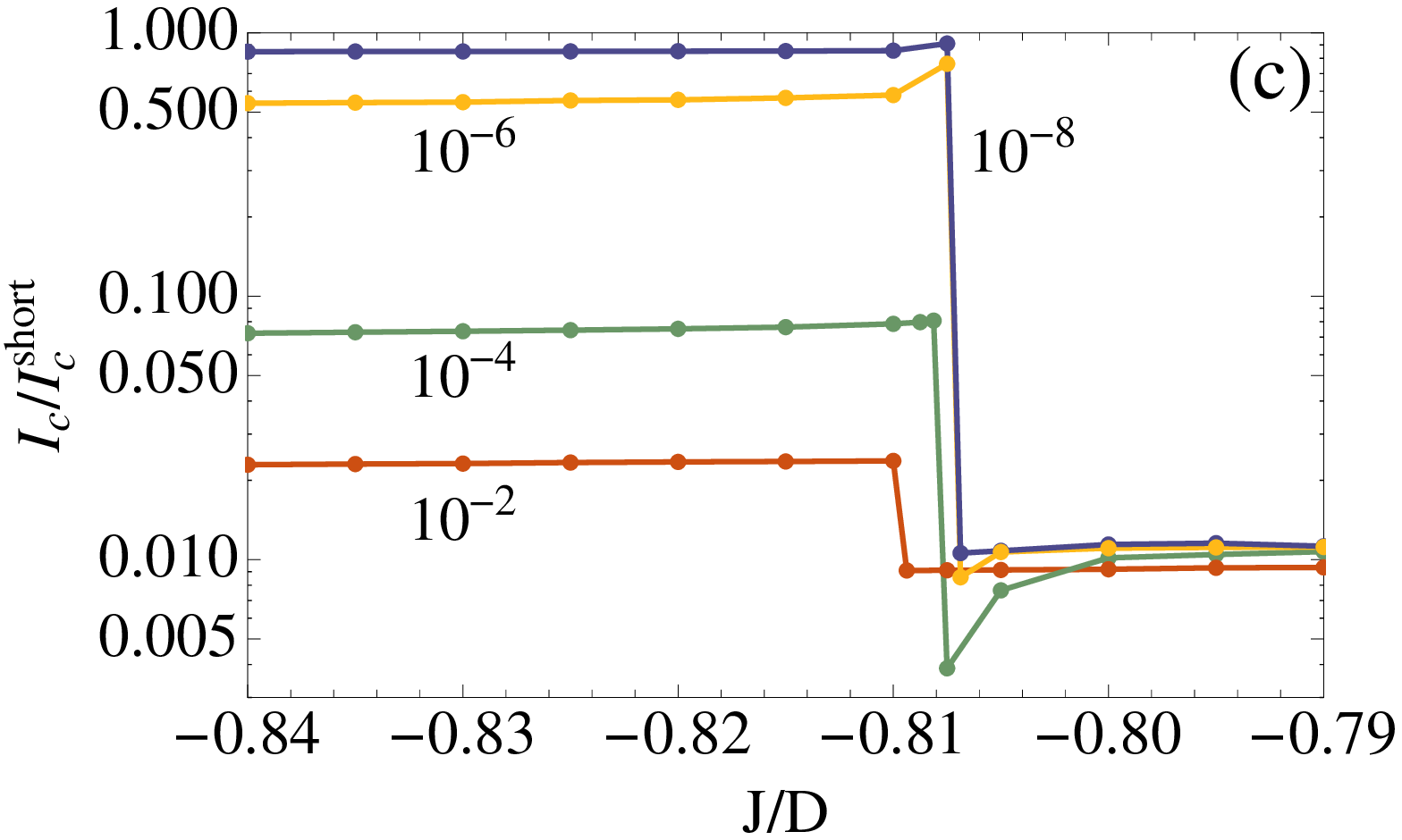}
  \caption{(color online) Critical currents as functions of $J/D$ in units of
    $I_c^{\rm short}$ for the case I with $\gamma = 1$ [(a)], 0.4 [(b)], and
    0.1 [(c)]. The numbers next to the lines represent the values of $\Delta/D$
    chosen for them. Here the solid lines are guide for eyes.}
  \label{fig:cci}
\end{figure}

\begin{figure}[!t]
  \centering
  \includegraphics[width=4.2cm]{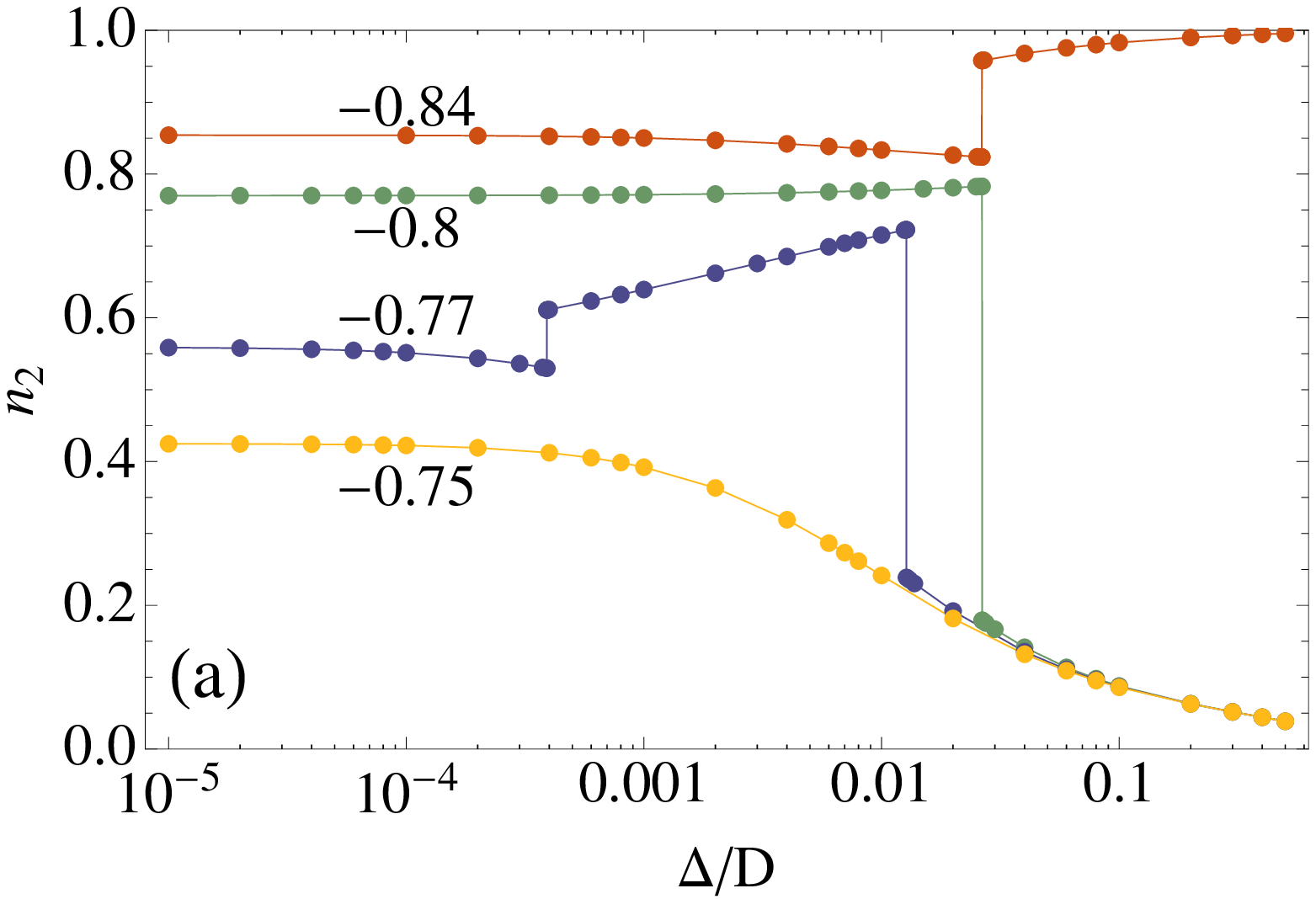}%
  \includegraphics[width=4.2cm]{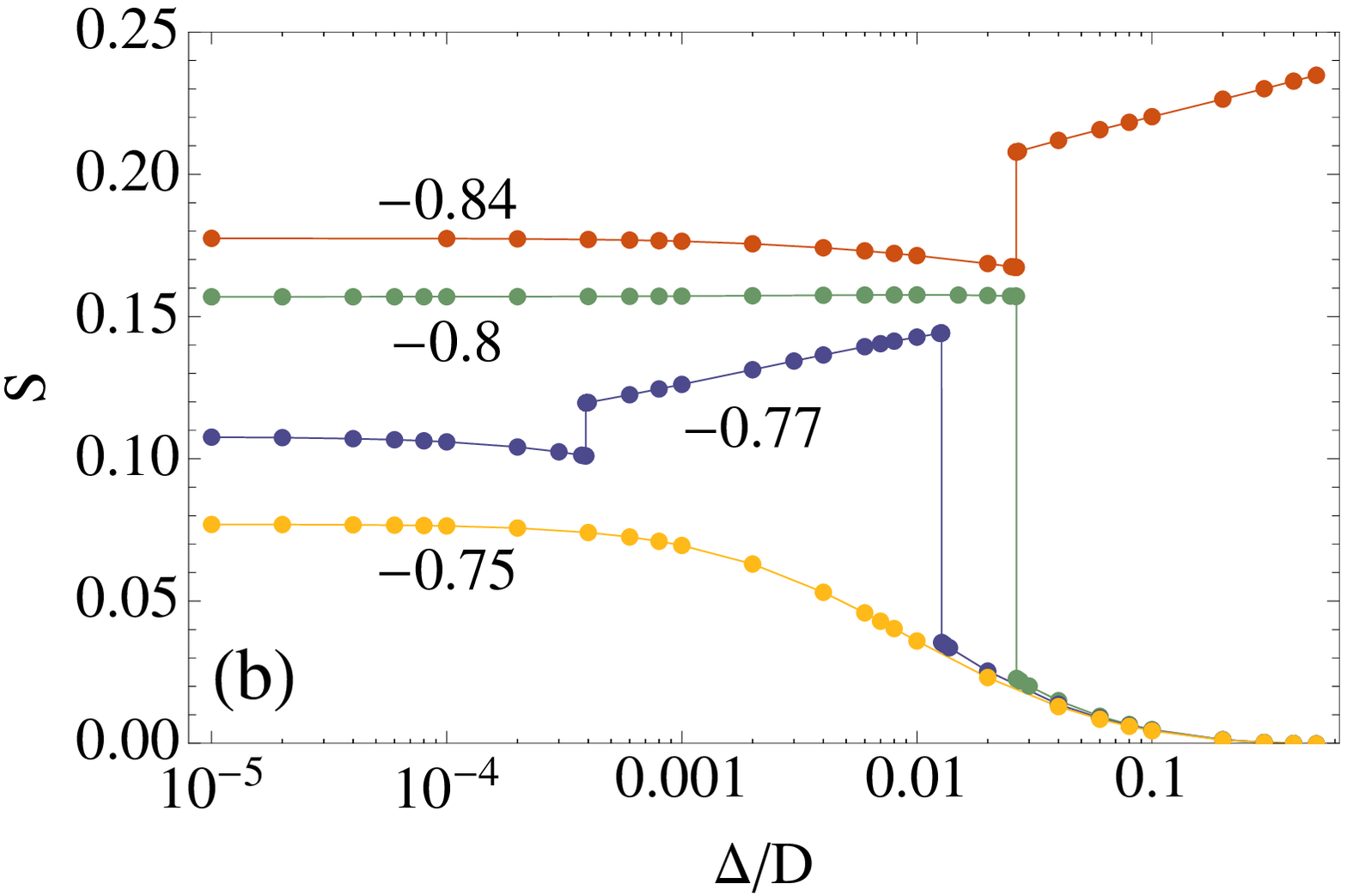}
  \caption{(color online) (a) Occupation $\avg{n_2}$ in the second level of the
    QD and (b) spin correlation $\avg{\bfS_1\cdot\bfS_2}$ as functions of
    $\Delta/D$ for the case I at $\gamma = 1$ and $\phi=0$. The numbers next to
    the lines represent the values of $J/D$ chosen for them. Due to the
    particle-hole symmetry, we have $\avg{n_1} = 2 - \avg{n_2}$. Here the solid
    lines are guide for eyes.}
  \label{fig:nsi}
\end{figure}

\textit{Critical Current}--- \Figref{fig:cci} shows the critical current as a
function of $J/D$ for given values of $\Delta$. For all values of $\gamma$, the
spin triplet state has a larger critical current than the spin singlet state,
which is consistent with the perturbation results: compare \figsref{fig:wccci}
and \ref{fig:cci}. The critical current is considerably boosted up in the
Kondo-dominant state and even approaches the ballistic short-junction value
$I_c^{\rm short}$ as it goes deep into the Kondo state. The critical current
has its maximum around the singlet-triplet transition point and decreases
rapidly in the antiferromagnetic side. A dip in the critical current is
observed in the antiferromagnetic side at moderate values of $\gamma$. This dip
happens at the boundary between the $\pi$ state and the second 0 state where
the current vanishes completely. Note that the Kondo-driven 0-$\pi$ transition
involves intermediate states so that the current does not vanish at the phase
boundaries.

\textit{Occupation and Spin Correlation}--- Finally, we'd like to mention about
the other ground-state properties such as the QD occupation and the spin
correlation between QD spins. For the isolated QD, the perfect spin singlet
state dictates $\avg{n_1} = 2$, $\avg{n_2} = 0$, and $\avg{\bfS_1\cdot\bfS_2} =
0$, while $\avg{n_1} = 1$, $\avg{n_2} = 1$, and $\avg{\bfS_1\cdot\bfS_2} = 1/4$
in the spin triplet state. This behavior is well reproduced for large values of
$\Delta$ [see \figref{fig:nsi}]. The strong superconductivity effectively
decouples the QD from the leads, and the correlations between QD spins are left
unpolluted. As $\Delta$ is decreased, on the other hand, the dot-lead
hybridization interferes the QD spin correlation and makes $\avg{n_i}$ and
$\avg{\bfS_1\cdot\bfS_2}$ deviate from their bare values. Furthermore, they
exhibit abrupt changes at phase transitions to the spin doublet state;
$\avg{n_2}$ and $\avg{\bfS_1\cdot\bfS_2}$ exhibit the same qualitative
dependence on $\Delta$ since the occupation in the orbital 2 is directly
related to the formation of the spin triplet state. Interestingly, the
emergence of the Kondo spin correlation which screens out one of the spins does
not completely suppress the local spin correlation between QD spins and even
help its recovery slightly as $\Delta$ is further decreased.

\subsubsection{Case II: Two Channels}

\begin{figure}[!t]
  \centering
  \includegraphics[width=7cm]{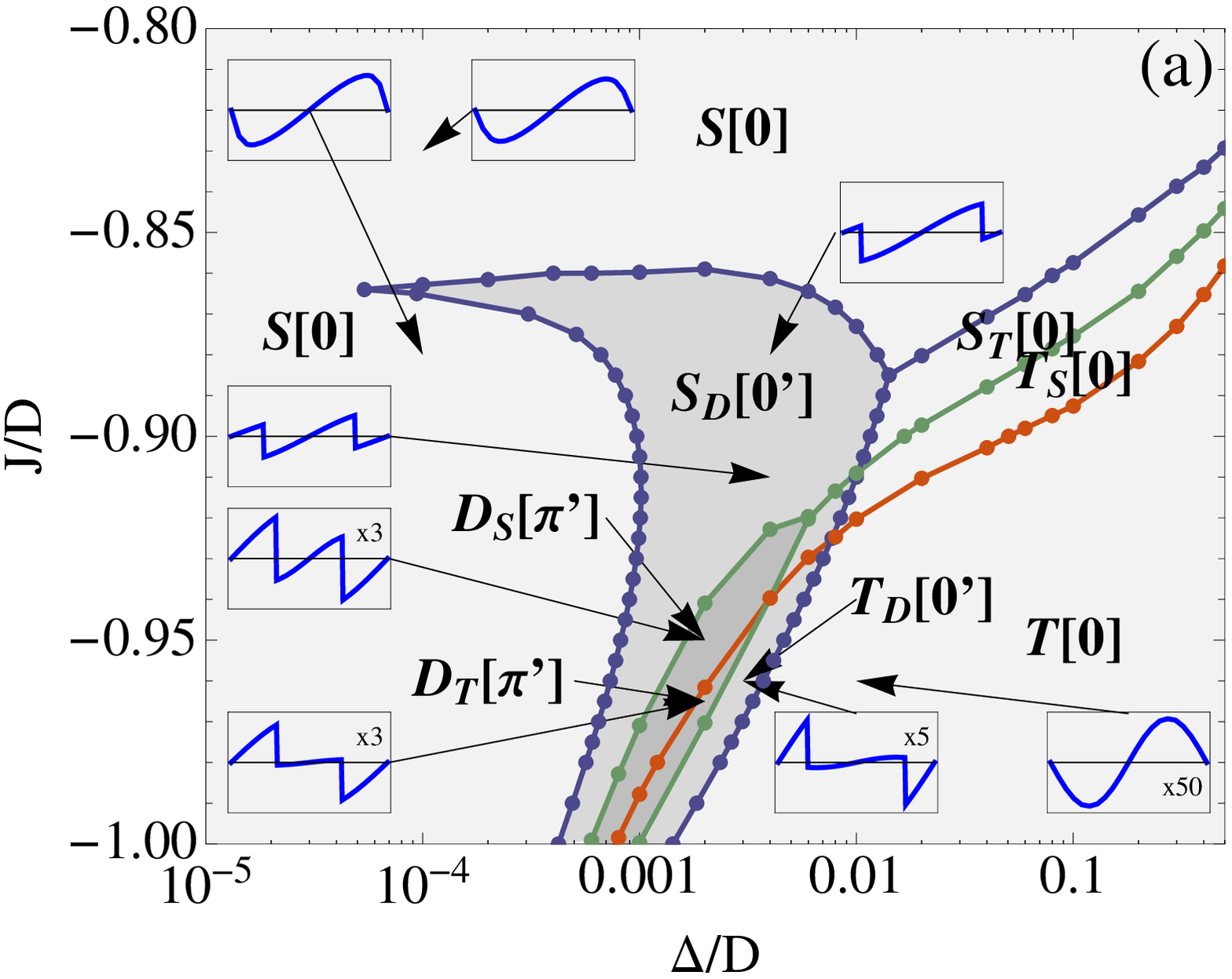}\\
  \includegraphics[width=7cm]{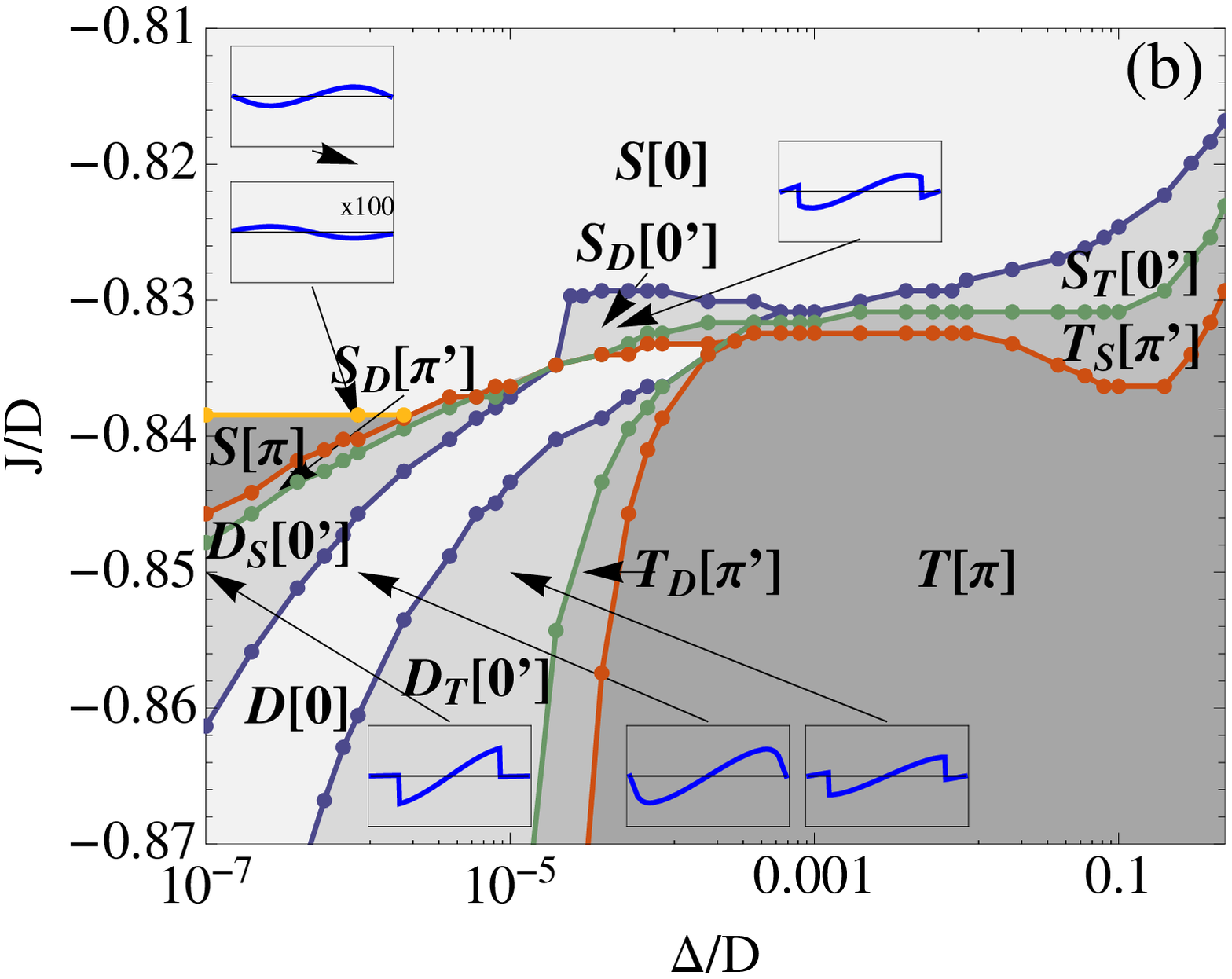}
  \caption{(color online) Phase diagrams in the $\Delta$-$J$ plane for the case
    II with $\gamma = 1$ [(a)] and 0.5 [(b)]. Refer to \figref{fig:pdi} for the
    details.}
  \label{fig:pdii}
\end{figure}

\textit{Weak Coupling Regime}--- Now we consider the two-channel case in the
presence of the superconductivity. \Figref{fig:pdii} displays the phase
diagrams in this case. In the large-$\Delta$ limit the phase boundaries and the
SPR characteristics coincide well with those obtained from the perturbative
theory: compare \figsref{fig:wcpdii} and \ref{fig:pdii}. At $\gamma=1$, the SPR
is of the 0-junction in both the spin singlet and triplet states, and that of
the state $T$ switches into the $\pi$-junction as $\gamma$ is decreased.

\textit{Strong Coupling Regime}--- The small-$\Delta$ part of the phase
diagram, on the other hand, features Kondo-oriented structures. First, consider
the asymmetric case ($\gamma \ne 1$) [see \figref{fig:pdii}~(b)] which exhibits
typical phase diagram in two-channel case. On the antiferromagnetic side no
transition is observed to take place. In this regime the antiferromagnetic
coupling is strong enough that the system is frozen in the spin singlet state
regardless of value of $\Delta$. On the ferromagnetic side, however, the system
experiences a double transition with decreasing $\Delta$: a transition from the
spin triplet to the spin doublet state is followed by a second transition to
the spin singlet state at smaller $\Delta$. It reflects the two-stage Kondo
effect discussed in \secref{sec:nlII}, where two Kondo scales $T_{K,1}$ and
$T_{K,2}$ are operating.  For $\Delta > T_{K,1/2}$, the conduction-band
electrons form the Cooper pairs by themselves without affecting the
ferromagnetic correlation formed in the QD. Once $\Delta$ is lowered below
the larger Kondo temperature $T_{K,1}$ so that $T_{K,2} < \Delta < T_{K,1}$,
one of the QD spins is screened out and the other spin that is left unscreened
defines the spin doublet state. For smaller $\Delta < T_{K,1/2}$, the remaining
QD spin is also screened out so that the entire system becomes of the spin
singlet. We have spotted such a double transition at any value of
$\phi$. \Figref{fig:pdii}~(b) shows that the phase difference affects the Kondo
temperatures in such a way that $T_{K,1}$ ($T_{K,2}$) decreases (increases)
with increasing $\phi$. The interval $T_{K,1}-T_{K,2}$ is the smallest at
$\phi=\pi$. As in the case I, the modulation of the Kondo temperatures can be
roughly understood from the dependence of the dot-lead coupling on $\phi$ [see
\eqnref{eq:dlc}]: The phase difference redistributes the amplitude of dot-lead
couplings so that the stronger one get weaker and vice versa as $\phi$ is
varied from 0 to $\pi$. As a result, with decreasing $\Delta$ the system
evolves from the spin triplet state $T$ to the singlet state $S$ via all the
possible intermediate states including the complete spin doublet state $D$.

\textit{Symmetric Case}--- In the symmetric case ($\gamma=1$), however, the
phase diagram exhibits quite different features [see \figref{fig:pdii}~(a)]. At
$\phi=0$ only one direct transition from the spin triplet to the spin singlet
is observed [see a red line]. It is surely due to the disappearance of the
two-stage Kondo effect at $\gamma = 1$: two QD spins are simultaneously
screened at a common Kondo temperature $T_K$ [refer to
\secref{sec:nlII}]. Hence the system passes from the spin triplet state to the
spin singlet state either by antiferromagnetic coupling or by Kondo
correlation. At finite $\phi$, however, the redistribution of the dot-lead
couplings deviate the Kondo couplings for two QD spins from their symmetric
point so that one becomes larger and the other smaller, recovering the
two-stage Kondo effect with two Kondo temperatures $T_{K,1} > T_{K,2}$
again. In this case $T_{K,1}$ ($T_{K,2}$) increases (decreases) with $\phi$ so
that the interval $T_{K,1} - T_{K,2}$ reaches its maximum at $\phi=\pi$. As a
result no complete $D$ state arises as the spin triplet state is changed into
the spin singlet state.

One more peculiarity in the symmetric case is a cusp in the phase boundary at
$\phi=\pi$ [see \figref{fig:pdii}~(a)]. Note that such a strange structure is
also observed in the asymmetric case [see \figref{fig:pdii}~(b)].  Analytical
theory to capture its physical origin is not available because the perturbative
scaling theory with finite $\Delta$ is hard to trace down. Instead, we draw a
tentative argument from its structural resemblance to that caused by the
single-channel two-stage Kondo effect. Our NRG calculation indicates that at
$\phi=\pi$ the lower Kondo temperature $T_{K,2}$ in the ferromagnetic side
exponentially decreases as $J$ approaches the antiferromagnetic region. As soon
as the effective singlet-triplet splitting $\Jst$ becomes antiferromagnetic,
the spin exchange coupling $\Jst \widetilde\bfS_a\cdot\widetilde\bfS_b$ can
then cause the second Kondo effect with the Kondo temperature $T_K^I$ between
the spin $\bfS_{\bar{q}}$ and the local Fermi liquid formed at the spin
$\bfS_q$ as explained in \secref{sec:nlI}.  Accordingly, we have again a
two-stage Kondo effect that explains the upward convex shape of the unusual
phase boundaries very well.  The condition that it can take place is that
$T_{K,2} < T_K^I$, that is, the second QD spin is screened by the continuous
degrees of freedom formed at the Kondo resonance level at the first QD spin
rather than by the conduction-band electrons in leads. In the absence of
superconductivity this does not arise because $T_{K,2}$ is always larger than
$T_K^I$. However, the NRG calculation suggests that this condition is satisfied
with finite $\Delta$ and more importantly with $\phi\approx\pi$.

\textit{SPR for the Asymmetric Case}--- The SPRs shown in the insets of
\figref{fig:pdii} clearly reflects the Kondo-driven phase transition. The
asymmetric case [see \figref{fig:pdii}~(b)] displays the similar evolution of
the SPR with respect to the alteration of the ground-state spin as observed in
the single-channel case. The formation of the Kondo-assisted resonant level
boosts up the tunneling of Cooper pairs and develops the ballistic
0-junction. Starting from the tunneling $\pi$-junction in the spin triplet
state, the SPR then has a shape of the three-segment structure with decreasing
$\Delta$. The central ballistic part enlarges further with lowering $\Delta$,
and the SPR becomes of the complete 0-junction as the system enters into the
state $D$. Passing through the second transition to the spin singlet state, the
SPR restores the three-segment form and eventually returns to the
$\pi$-junction. It is worth noting that while the ballistic 0-junction is
ascribed to the Kondo resonant tunneling, its suppression at smaller $\Delta$
is also due to the Kondo effect. The destructive interference between two
resonant tunnelings leads to a subduing of the tunneling current. The
interference is not complete (while it is the case in the single-channel case)
and becomes ineffective as $J$ becomes less negative, resulting in an increase
of the supercurrent.

The phase diagram in \figref{fig:pdii}~(b) shows that there
exist two spin singlet states $S$ with 0- and $\pi$-junctions, respectively. The
perturbative analysis manifests that the spin singlet state in the case II
pertains to the 0-junction behavior. Our NRG calculation shows that it is the
case even in the small $\Delta$ limit as long as the antiferromagnetic coupling
prevails. However, we found that in the Kondo-dominant singlet state the SPR
exhibits the tunneling $\pi$-junction behavior as shown in
\figref{fig:pdii}~(b), following the SPR of the spin triplet state in the
large-$\Delta$ regime.  The transition between two singlet state does not
involve any intermediate state so that the supercurrent completely vanishes at
the boundary (yellow line) between them.

\textit{SPR for the Symmetric Case}--- The symmetric case, on the other hand,
displays exotic features in the SPR as well as in the phase diagram. The most
intriguing observation is that the Kondo-assisted tunneling triggers the
ballistic $\pi$-junction behavior. As stated before, the Kondo-driven spin
doublet state does not form at $\phi=0$ but comes into being from $\phi=\pi$ as
$\Delta$ is decreased. The tunneling 0-junction SPR in the spin triplet state
then transforms into the three-segment one that, at this time, has the
Kondo-driven ballistic feature at its side segments, not in the central
part. In addition, the SPR in the side segments exhibits the $\pi$-junction
behavior: As observed in \figref{fig:ali}, the change of the ground-state spin
from 1 to 1/2 makes the Andreev levels cross at the Fermi level, which
accordingly induces sign change of the supercurrent across the crossing
point. This ballistic $\pi$-junction can develop only if the dot-lead couplings
are all comparable and their product is negative because only this condition
ensures that the spin triplet state has the 0-junction and the Kondo transition
starts at $\phi=\pi$ with $T_{K,1}(\phi=\pi) > T_{K,1}(\phi=0)$. The
Kondo-assisted $\pi$-junction does not reach its full strength, through, and
further lowering of $\Delta$ shrinks the side segments, finally restoring the
0-junction. Hence, in the small-$\Delta$ limit, the system is of the spin
singlet in the 0 state.

\begin{figure}[!t]
  \centering
  \includegraphics[width=7cm]{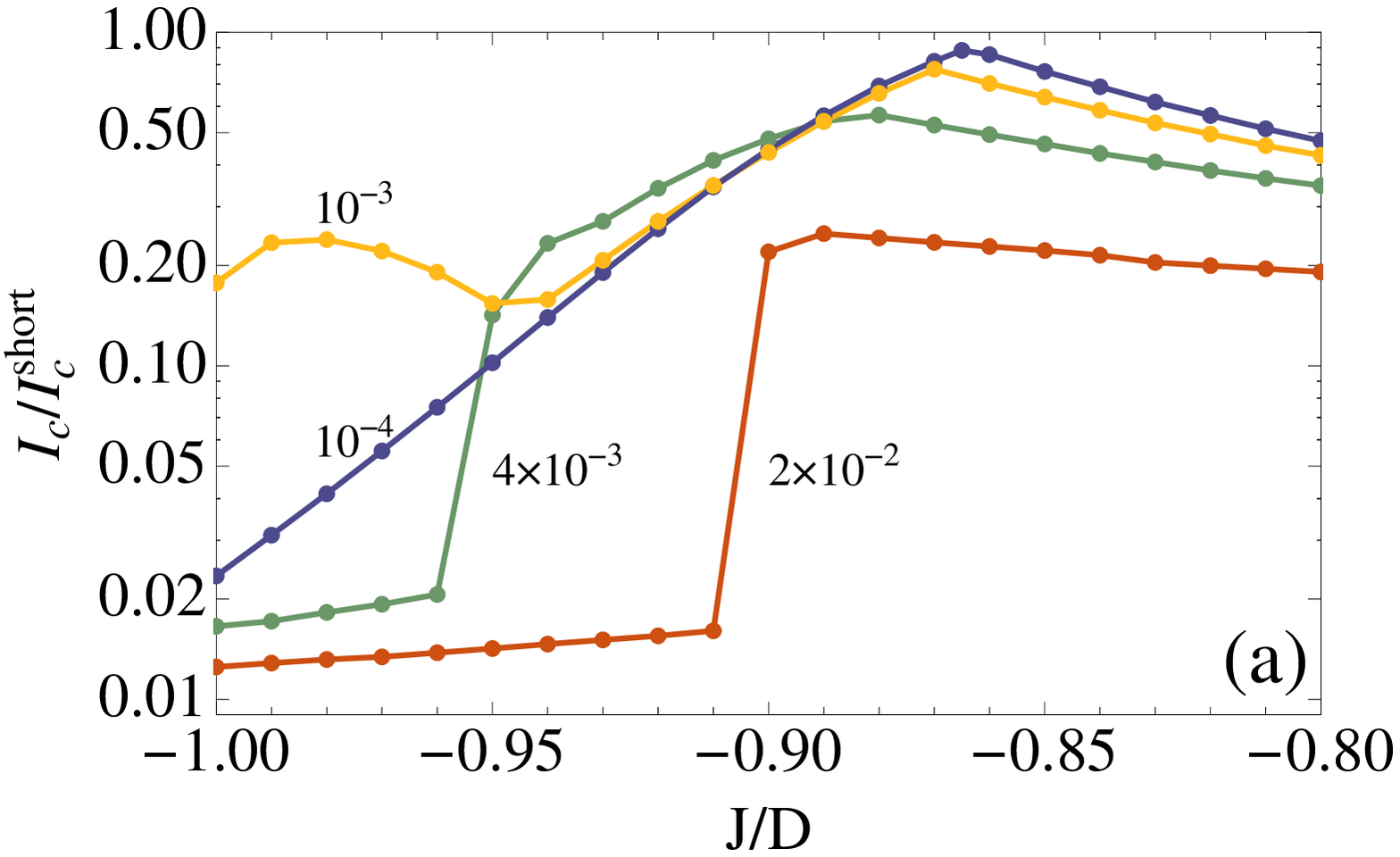}\\
  \includegraphics[width=7cm]{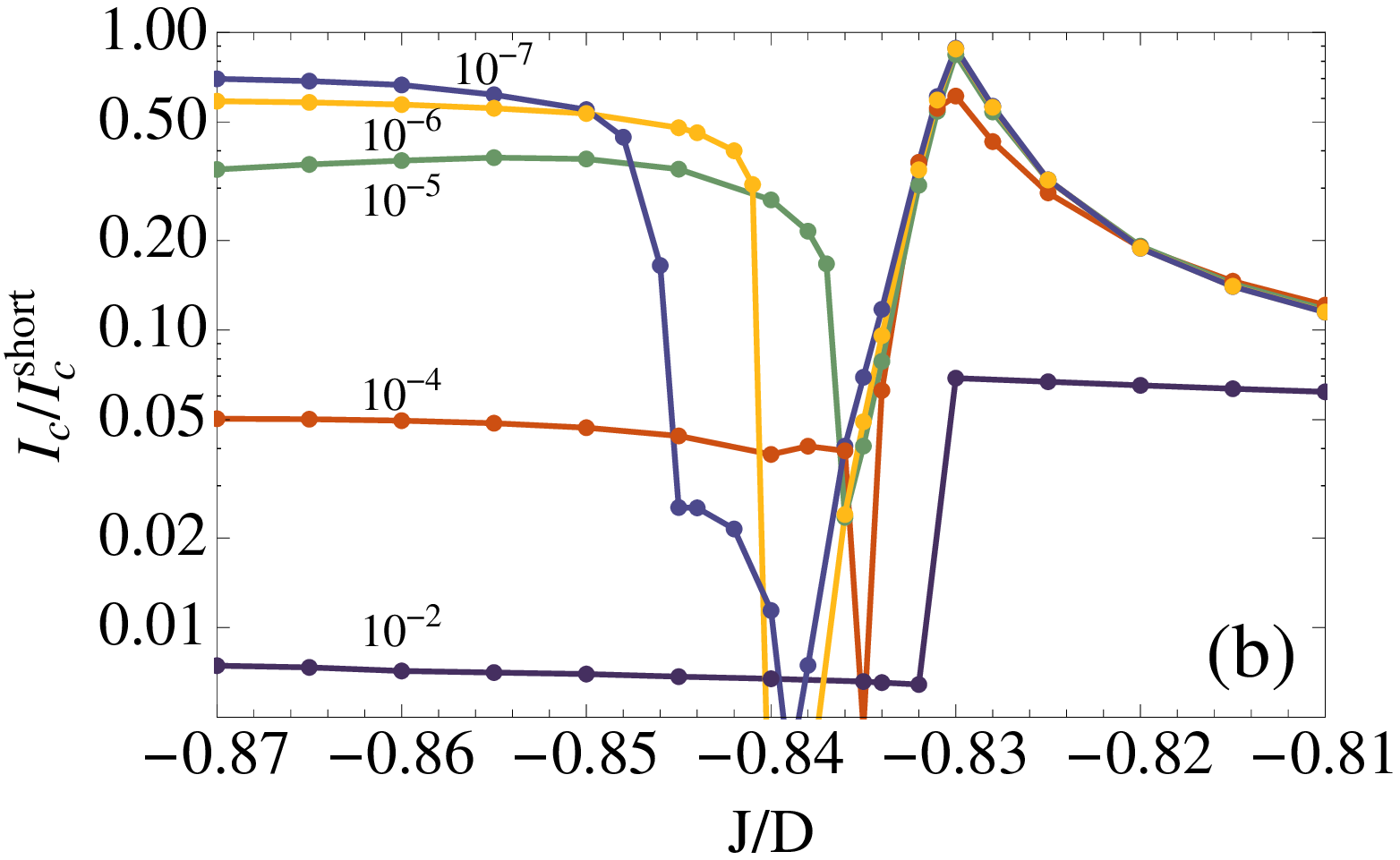}
  \caption{(color online) Critical currents as functions of $J/D$ in units of
    $I_c^{\rm short}$ for the case II with $\gamma = 1$ [(a)] and 0.5
    [(b)]. The numbers next to the lines represent the values of $\Delta/D$
    chosen for them. Here the solid lines are guide for eyes.}
  \label{fig:ccii}
\end{figure}

\textit{Critical Current}--- \Figref{fig:ccii} shows the critical current as a
function of $J/D$ for given values of $\Delta$. In the large-$\Delta$ limit,
the critical current is larger in the spin singlet state than in the spin
triplet state as predicted in the perturbation theory. On the other hand, the
small-$\Delta$ critical current is observed to follow the linear conductance
obtained in the normal-lead case: compare \figsref{fig:niic} and
\ref{fig:ccii}. The critical current exhibits a peak exactly where the linear
conductance reaches its maximum. The peak is not due to the Kondo boosting but
originates from the competition between the Kondo and the antiferromagnetic
correlations at the singlet-triplet transition as revealed in
\secref{sec:nlII}. However, the Kondo boosting enhances the critical current in
the spin doublet state. At a given $\Delta$, the system can be driven into the
spin doublet state as $J$ is decreased, where the unscreened Kondo correlation
opens a resonance tunneling. Hence, as can be seen in \figref{fig:ccii}~(b),
the small-$\Delta$ critical current features a peak and plateau as $J$ is
varied. A sharp dip is also identified between them. The Kondo-assisted plateau
is rather weak in the symmetric case [see \figref{fig:ccii}~(a)] and it
disappears in the small $\Delta$ limit since no spin doublet state exists in
this limit [see \figref{fig:pdii}~(a)]. Together with the Kondo-assisted
$\pi$-junction, this double-peak or peak-plateau structure of critical current
with respect to the spin exchange coupling $J$ contrast the two-channel case
with the single-channel one, so it provides a way to distinguish two cases in
experiments in which the amplitudes of dot-lead couplings are not known in
priori.

\section{Discussion and Conclusion}

We have investigated the physical properties and the electronic transport of
two-level quantum dot Josephson junctions by focusing on two representative
dot-lead coupling configurations, cases I and II [see \eqnref{eq:conf}]. The
fourth-order perturbation theory applied in the weak coupling limit has
revealed that the parities of dot orbital wavefunctions, that is, the sign of
the product of dot-lead tunneling amplitudes, can greatly affect the sign of
the supercurrent depending on the spin correlation present in the dot. The key
elements that determine the current characteristics are found to be the
existence of a localized moment in orbitals which reverses the order of
electrons in Cooper pairs and the competition between diagonal and offdiagonal
tunneling processes.

In the strong coupling limit the Kondo correlation competes with the
superconductivity and the spin exchange coupling and the system state is
determined by their relative strength. We have used the NRG method and the
scaling theory based on the Schrieffer-Wolff transformed Hamiltonian in order
to examine the Kondo effect in the normal-lead counterpart of our system. The
effective single-channel case (case I) exhibits three different states --
underscreened $S=1$ Kondo effect, two-stage Kondo effect, and spin singlet
state -- depending on the sign and strength of the spin exchange coupling,
while in the two-channel case (case II), the system displays two-stage Kondo
effect and spin singlet state. We have found that the numerical results from
the NRG method applied to superconducting case can be understood in terms of
comparison between the superconducting gap $\Delta$ and relevant Kondo
temperatures: The Kondo correlation found in the normal-lead case becomes
effective once $\Delta$ becomes smaller than the corresponding Kondo
temperature. In this way the superconducting gap acts like a coherent energy
probe for the excitations of the system otherwise. The competition between the
superconductivity and the many-body correlations present in the system gives
rise to phase transitions which accompany abrupt changes in physical properties
such as ground-state spin and SPR: Among the prominent changes that arise once
the Kondo effect prevails over superconductivity are the boost-up of the
supercurrent due to resonant tunneling and the appearance of strong 0-junction.
Knowing the magnitude of the superconducting gap that is under control,
therefore, it provides a way to measure the magnitude of the important
many-body correlations such as the Kondo temperature or vice versa.

Beside detecting the system excitations, the superconductivity directly alters
the system state. In the weak coupling limit it renormalizes the
singlet-triplet splitting and shifts the singlet-triplet transition point which
now depends on the phase difference as well. In the strong coupling limit the
finite phase difference influences the interference mechanism and suppress or
enhance the other many-body effects. Especially, at the maximally twisted
condition ($\phi=\pi$), the Kondo correlation is completely suppressed in the
case I and the Kondo-assisted $\pi$-junction is induced in the case II. Knowing
that the $\pi$-junction in most of cases is usually weak due to its
perturbative origin, the latter mechanism opens a way to have a strong
$\pi$-junction. In the light of quantum computational unit which is free of any
magnetic control, this $\pi$-junction is more promising because the state of
the junction can be controlled by electric manipulation: the spin exchange
coupling can be tuned by the gate voltage.

We expect that our theoretical prediction about phase diagrams and SPRs in
TLQD-JJs can be explored in experiment by using state-of-art fabrication
techniques. A recent experiment\cite{Roch08} measured electronic transport
through C$_{60}$-molecular junctions and detected the singlet-triplet
transition and the accompanying Kondo effects which are consistent with
existing theories. The same experiment group has extended their study to
superconducting case\cite{Winkelmann09} where the Al bars are attached on top
of Au leads coupled to C$_{60}$ and the superconductivity is induced in Au
leads by the proximity effect. The Josephson effect predicted in our paper can
be then investigated by forming a SQUID,\cite{Cleuziou06} one of which arms
contains the molecular junction. As implemented in Roch \textit{et
  al.},\cite{Roch08} the spin exchange coupling can be then controlled by an
externally applied gate voltage. Or, the superconducting gap can be tuned by
applying a magnetic field as long as it does not suppress the Kondo effect. In
this way the switching between 0 and $\pi$ states with respect to the tuning of
$J$ and/or $\Delta$ predicted in our calculations could be confirmed in
experiment.  Even without using the SQUID geometry, our theory could be tested
by measuring a critical current through the junction, which should exhibit
nontrivial dependence on the spin exchange coupling as discussed above.

\acknowledgements

The authors thank W. Wernsdorfer and F. Balestro for helpful discussions. This
work is supported by ANR-PNANO Contract MolSpintronics No. ANR-06-NANO-27 and
NRF-2009-0069554.

\appendix

\section{Energy Shift Coefficients in Perturbation Theory}

Here we present the detailed expressions of the coefficients $\beta_{ai}$ for
$a=S,T$ are defined in the energy shifts, \eqnref{eq:pert} in the fourth-order
perturbation theory. On behalf of readability, we introduce the following
integrals:
{\allowdisplaybreaks
\begin{align}
  A_i
  & \equiv \int \frac{dx/\pi}{f(x;\xi_{1i})}
  \\
  A'_i
  & \equiv \int \frac{dx/\pi}{f(x;\xi_{1i}) f(x)}
  \\
  A_{ij}
  & \equiv \int \frac{dx/\pi}{f(x;\xi_{1i}) f(x;\xi_{1j})}
  \\
  A_{ij}(\xi)
  & \equiv
  \int \int \frac{dxdy/\pi^2}{f(x,y;\xi) f(x;\xi_{1i}) f(x;\xi_{1j})}
  \\
  A'_{ij}(\xi)
  & \equiv
  \int \int
  \frac{dxdy/\pi^2}{f(x,y;\xi) f(x) f(y) f(x;\xi_{1i}) f(x;\xi_{1j})}
  \\
  B_{ij}(\xi)
  & \equiv
  \int \int \frac{dxdy/\pi^2}{f(x,y;\xi) f(x;\xi_{1i}) f(y;\xi_{1j})}
  \\
  B'_{ij}(\xi)
  & \equiv
  \int \int
  \frac{dxdy/\pi^2}{f(x,y;\xi) f(x) f(y) f(x;\xi_{1i}) f(y;\xi_{1j})},
\end{align}
}
where we have defined $f(x) \equiv \sqrt{1+x^2}$, $f(x;\xi) \equiv f(x) + \xi$,
and $f(x,y;\xi) \equiv f(x) + f(y) + \xi$ and all the integration should be
done over the region $[-D/\Delta,D/\Delta]$. All the charge excitation energies
that appear in the expressions are made dimensionless:
\begin{align}
  \xi_{11} & = (\epsilon_1 - E_a^{(0)})/\Delta,
  &
  \xi_{12} & = (\epsilon_2 - E_a^{(0)})/\Delta,
  \\
  \xi_{21} & = -E_a^{(0)}/\Delta,
  &
  \xi_{22} & = (E_{0;1} - E_a^{(0)})/\Delta,
  \\
  \xi_{23} & = (E_{0;2} - E_a^{(0)})/\Delta,
  &
  \xi_{24} & = (E_{0;3} - E_a^{(0)})/\Delta,
  \\
  \xi_{25} & = (E_1 - E_a^{(0)})/\Delta,
\end{align}
for $a={S,T}$, respectively. In terms of the integral macros, the coefficients
$\beta_{Si}$ for the singlet energy shift are then expressed as
\allowdisplaybreaks
\begin{align}
  \beta_{S0}
  & = -A_1
  \\
  \beta_{S1}
  & =
  A_1 A_{11}
  -
  \frac{[A'_1]^2}{\xi_{21}}
  \\
  \nonumber
  & \quad\mbox{}
  -
  \frac{A_{11}(\xi_{21}) + B_{11}(\xi_{21})}{2}
  -
  \frac{A_{11}(\xi_{22}) + B'_{11}(\xi_{22})}{2}
  \\
  \beta_{S2}
  & =
  2 A_1 A_{11}
  +
  A'_{11}(\xi_{22}) + B_{11}(\xi_{22})
  \\
  \nonumber
  & \quad\mbox{}
  -
  \frac{A_{11}(\xi_{24}) + B_{11}(\xi_{24})
    + A'_{11}(\xi_{24}) + B'_{11}(\xi_{24})}{2}
  \\
  \nonumber
  & \quad\mbox{}
  -
  \frac{5(A_{11}(\xi_{25}) + B_{11}(\xi_{25})
    - A'_{11}(\xi_{25}) - B'_{11}(\xi_{25}))}{2}
  \\
  \beta_{S3}
  & =
  2 A_1 A_{11}
  -
  \frac{A_{11}(\xi_{24}) + B_{11}(\xi_{24})}{2}
  \\
  \nonumber
  & \quad\mbox{}
  -
  \frac{3 (A_{11}(\xi_{25}) + B_{11}(\xi_{25}))}{2}
  \\
  \beta_{S4}
  & =
  2 A_1 A_{11} - A_{11}(\xi_{21}) - B_{11}(\xi_{21}) - A_{11}(\xi_{22})
  \\
  \beta_{S5}
  & = 2 B_{11}(\xi_{22})
  \\
  \beta'_{S4}
  & = B'_{11}(\xi_{22}) + 2 \frac{[A'_1]^2}{\xi_{21}}
  \\
  \beta'_{S5}
  & =
  - 2 A'_{11}(\xi_{22}) + A'_{11}(\xi_{24}) + B'_{11}(\xi_{24})
  \\
  \nonumber
  & \quad\mbox{}
  - 3 (A'_{11}(\xi_{25}) + B'_{11}(\xi_{25})),
\end{align}
where the scaled excitation energies are calculated with $E_a = E_S$. The
coefficients $\beta_{Ti}$ for the triplet energy shift are given by
\begin{align}
  \beta_{T0}
  & = - \frac{A_1 + A_2}{2}
  \\
  \beta_{T1}
  & =
  \frac{(A_1 + A_2) (A_{11} + A_{22})}{4}
  \\
  \nonumber
  & \quad\mbox{}
  - \frac{A_{11}(\xi_{24}) + A_{22}(\xi_{24}) + 2 B_{12}(\xi_{24})}{4}
  + \frac{A'_{12}(\xi_{24})}{2}
  \\
  \nonumber
  & \quad\mbox{}
  + \frac{B'_{11}(\xi_{24}) + B'_{22}(\xi_{24})}{4}
  \\
  \nonumber
  & \quad\mbox{}
  - \frac{A_{11}(\xi_{25}) + A_{22}(\xi_{25}) + B_{12}(\xi_{25})}{2}
  + A'_{12}(\xi_{25})
  \\
  \nonumber
  & \quad\mbox{}
  + \frac{B'_{11}(\xi_{25}) + B'_{22}(\xi_{25})}{4}
  \\
  \beta_{T2}
  & =
  \frac{(A_1 + A_2) (A_{11} + A_{22})}{2}
  \\
  \nonumber
  & \quad\mbox{}
  - A_{11}(\xi_{21}) + B_{11}(\xi_{21})
  - A_{22}(\xi_{21}) + B_{22}(\xi_{21})
  \\
  \nonumber
  & \quad\mbox{}
  + 2 (A_{12}(\xi_{21}) - B_{12}(\xi_{21}))
  \\
  \nonumber
  & \quad\mbox{}
  - A_{11}(\xi_{22}) - B_{11}(\xi_{22})
  + A'_{11}(\xi_{22}) + B'_{11}(\xi_{22})
  \\
  \nonumber
  & \quad\mbox{}
  - A_{22}(\xi_{23}) + B_{22}(\xi_{23})
  + A'_{22}(\xi_{23}) + B'_{22}(\xi_{23})
  \\
  \nonumber
  & \quad\mbox{}
  + A_{12}(\xi_{24})
  - B'_{12}(\xi_{24})
  \\
  \nonumber
  & \quad\mbox{}
  + \frac{B_{11}(\xi_{24}) + B_{22}(\xi_{24}) - A'_{11}(\xi_{24}) - A'_{22}(\xi_{24})}{2}
  \\
  \nonumber
  & \quad\mbox{}
  - 2 A_{12}(\xi_{25})
  + A'_{11}(\xi_{25})
  + A'_{22}(\xi_{25})
  + B'_{12}(\xi_{25})
  \\
  \nonumber
  & \quad\mbox{}
  - \frac{B_{11}(\xi_{25}) + B_{22}(\xi_{25})}{2}
  \\
  \beta_{T3}
  & =
  \frac{(A_1 + A_2) (A_{11} + A_{22})}{2}
  \\
  \nonumber
  & \quad\mbox{}
  - \frac{A_{11}(\xi_{21}) + A_{22}(\xi_{21})}{2} - B_{12}(\xi_{21})
  \\
  \nonumber
  & \quad\mbox{}
  - \frac{A_{11}(\xi_{22}) + B_{11}(\xi_{22}) + A_{22}(\xi_{23}) + B_{22}(\xi_{23})}{2}
  \\
  \beta_{T4}
  & =
  \frac{(A_1 + A_2) (A_{11} + A_{22})}{2}
  \\
  \nonumber
  & \quad\mbox{}
  - \frac{A_{11}(\xi_{24}) + A_{22}(\xi_{24}) + 2 B_{12}(\xi_{24})}{4}
  \\
  \nonumber
  & \quad\mbox{}
  - \frac{3 A_{11}(\xi_{25}) + 3 A_{22}(\xi_{25}) + 2 B_{12}(\xi_{25})}{4}
  \\
  \beta_{T5}
  & =
  2 A_{12}(\xi_{21}) + B_{11}(\xi_{21}) + B_{22}(\xi_{21})
  \\
  \nonumber
  & \quad\mbox{}
  + A_{12}(\xi_{24})
  + \frac{B_{11}(\xi_{24}) + B_{22}(\xi_{24})}{2}
  \\
  \nonumber
  & \quad\mbox{}
  - 3 A_{12}(\xi_{25})
  - \frac{B_{11}(\xi_{25}) + B_{22}(\xi_{25})}{2}
  \\
  \beta'_{T4}
  & =
  - \frac{A'_{12}(\xi_{24})}{2}
  - \frac{B'_{11}(\xi_{24}) + B'_{22}(\xi_{24})}{4}
  \\
  \nonumber
  & \quad\mbox{}
  - \frac{3A'_{12}(\xi_{25})}{2}
  - \frac{B'_{11}(\xi_{25}) + B'_{22}(\xi_{25})}{4}
  \\
  \beta'_{T5}
  & =
  - A'_{11}(\xi_{22}) - B'_{11}(\xi_{22})
  - A'_{22}(\xi_{23}) - B'_{22}(\xi_{23})
  \\
  \nonumber
  & \quad\mbox{}
  + \frac{A'_{11}(\xi_{24}) + A'_{22}(\xi_{24})}{2}
  + B'_{12}(\xi_{24})
  \\
  \nonumber
  & \quad\mbox{}
  - \frac{3 (A'_{11}(\xi_{25}) + A'_{22}(\xi_{25}))}{2}
  - B'_{12}(\xi_{25}),
\end{align}
where the scaled excitation energies are calculated with $E_a = E_T$. For better
estimation of the sign and the magnitude of the supercurrent, full expressions
for $\beta'_{S4/5}$ and $\beta'_{T4/5}$ are given below: \onecolumngrid
\begin{align}
  \label{eq:bs4}
  \beta'_{S4}
  & =
  \int \int
  \frac{dxdy/\pi^2}{f(x,y;\xi_{22}) f(x;\xi_{11}) f(y;\xi_{11}) f(x) f(y)}
  +
  \frac{2}{\xi_{21}}
  \left(\int \frac{dx/\pi}{f(x;\xi_{11})f(x)}\right)^2
  \\
  \label{eq:bs5}
  \beta'_{S5}
  & =
  \int \int \frac{dxdy/\pi^2}{f(x)f(y)}
  \Bigg[
  \left(\frac{1/2}{f(x,y;\xi_{24})} - \frac{3/2}{f(x,y,;\xi_{25})}\right)
  \left(\rcp{f(x;\xi_{11})} + \rcp{f(y;\xi_{11})}\right)^2
  -
  \frac{2}{f(x,y;\xi_{22}) [f(x;\xi_{11})]^2}
  \Bigg]
  \\
  \beta'_{T4}
  & =
  - \frac14 \int \int \frac{dxdy/\pi^2}{f(x) f(y)}
  \Bigg[
  \left(\rcp{f(x,y;\xi_{24})} + \rcp{f(x,y;\xi_{25})}\right)
  \left(
    \rcp{f(x;\xi_{11})f(y;\xi_{11})} + \rcp{f(x;\xi_{12})f(y;\xi_{12})}
  \right)
  \\
  \nonumber
  & \quad\qquad\qquad\qquad\qquad\qquad\mbox{}
  +
  \left(\rcp{f(x,y;\xi_{24})} + \frac{3}{f(x,y;\xi_{25})}\right)
  \frac{2}{f(x;\xi_{11})f(x;\xi_{12})}
  \Bigg]
  \\
  \beta'_{T5}
  & =
  - \frac12 \int \int \frac{dxdy/\pi^2}{f(x) f(y)}
  \Bigg[
  \rcp{f(x,y;\xi_{22})}
  \left(\rcp{f(x;\xi_{11})} + \rcp{f(y;\xi_{11})}\right)^2
  +
  \rcp{f(x,y;\xi_{23})}
  \left(\rcp{f(x;\xi_{12})} + \rcp{f(y;\xi_{12})}\right)^2
  \\
  \nonumber
  & \qquad\mbox{}
  -
  \rcp{f(x,y;\xi_{24})}
  \left(\rcp{f(x;\xi_{11})} + \rcp{f(y;\xi_{12})}\right)^2
  +
  \rcp{f(x,y;\xi_{25})}
  \left(
    \frac{3}{[f(x;\xi_{11})]^2}
    +
    \frac{2}{f(x;\xi_{11})f(y;\xi_{12})}
    +
    \frac{3}{[f(y;\xi_{12})]^2}
  \right)
  \Bigg].
\end{align}
\twocolumngrid


\begin{thebibliography}{99}

\bibitem{Josephson}
  B. D. Josephson, Phys. Lett. \textbf{1}, 251 (1962);
  Rev. Mod. Phys. \textbf{46}, 251 (1974).

\bibitem{Tinkham96}
  M. Tinkham, \textit{Introduction to Superconductivity} (McGraw-Hill, Singapore, 1996).

\bibitem{Bardeen57}
  J. Bardeen, L. N. Cooper, and J. R. Schrieffer, Phys. Rev. \textbf{106}, 162 (1957); \textbf{108}, 1175 (1957).


\bibitem{Shiba69}
  H. Shiba and T. Soda, Prog. Theor. Phys. \textbf{41}, 25 (1969).

\bibitem{Glazman89}
  L. I. Glazman and K. A. Matveev, Pis'ma Zh. Teor. Fiz. \textbf{49}, 570 (1989) [JETP Lett. \textbf{49}, 659 (1989)].

\bibitem{Spivak91}
  B. I. Spivak and S. A. Kivelson, Phys. Rev. B \textbf{43}, 3740 (1991).

\bibitem{Yeyati97}
  A. Levy Yeyati, J. C. Cuevas, A. Lopez-Davalos, and A. Mart\'in-Rodero, Phys. Rev. B \textbf{55}, 6137 (1997).

\bibitem{Shimizu98}
  Y. Shimizu, H. Horii, Y. Takane, and Y. Isawa, J. Phys. Soc. Jpn. \textbf{67}, 1525 (1998).

\bibitem{Rozhkov99}
  A. V. Rozhkov and D. P. Arovas, Phys. Rev. Lett. \textbf{82}, 2788 (1999);
  A. V. Rozhkov and D. P. Arovas, Phys. Rev. B \textbf{62}, 6687 (2000).

\bibitem{ChoiMS00}
  M.-S. Choi, C. Bruder, and D. Loss, Phys. Rev. B \textbf{62}, 13569 (2000).

\bibitem{Rozhkov01}
  A. V. Rozhkov, D. P. Arovas, and F. Guinea, Phys. Rev. B \textbf{64}, 233301 (2001).

\bibitem{Makhlin01}
  Y. Makhlin, G. Sch\"on, and A. Shnirman, Rev. Mod. Phys. \textbf{73}, 357 (2001).

\bibitem{Zaikin04}
  A. D. Zaikin, Low Temp. Phys. \textbf{30}, 568 (2004).

\bibitem{Choi04}
  M.-S. Choi, M. Lee, K. Kang, and W. Belzig, Phys. Rev. B \textbf{70}, 020502 (2004).

\bibitem{Siano04}
  F. Siano and R. Egger, Phys. Rev. Lett. \textbf{93}, 047002 (2004).

\bibitem{Karrasch08}
  C. Karrasch, A. Oguri, and V. Meden, Phys. Rev. B \textbf{77}, 024517 (2008).

\bibitem{Lee08}
  M. Lee, T. Jonckheere, and T. Martin, Phys. Rev. Lett. \textbf{101}, 146804 (2008).


\bibitem{Kasumov99}
  A. Yu. Kasumov, R. Deblock, M. Kociak, B. Reulet, H. Bouchiat, I. I. Khodos, Yu. B. Gorbatov, V. T. Volkov, C. Journet, and M. Burghard, Science \textbf{284}, 1508 (1999).

\bibitem{Kasumov01}
  A. Yu. Kasumov, M. Kociak, S. Gu\'eron, B. Reulet, V. T. Volkov, D. V. Klinov, and H. Bouchiat, Science \textbf{291}, 280 (2001).

\bibitem{Scheer01}
  E. Scheer, W. Belzig, Y. Naveh, M. H. Devoret, D. Esteve, and C. Urbina, Phys. Rev. Lett. \textbf{86}, 284 (2001).

\bibitem{Ryazanov01}
  V. V. Ryazanov, V. A. Oboznov, A. Yu. Rusanov, A. V. Veretennikov, A. A. Golubov, and J. Aarts, Phys. Rev. Lett. \textbf{86}, 2427 (2001).

\bibitem{Kontos02}
  T. Kontos, M. Aprili, J. Lesueur, F. Gen\^et, B. Stephanidis, and R. Boursier, Phys. Rev. Lett. \textbf{89}, 137007 (2002).

\bibitem{Buitelaar02}
  R. Buitelaar, T. Nussbaumer, and C. Sch\"onenberger, Phys. Rev. Lett. \textbf{89}, 256801 (2002);
  M. R. Buitelaar, W. Belzig, T. Nussbaumer, B. Babi\'{c}, C. Bruder, and C. Sch\"onenberger, Phys. Rev. Lett. \textbf{91}, 057005 (2003).

\bibitem{Kasumov05}
  A. Yu. Kasumov, K. Tsukagoshi, M. Kawamura, T. Kobayashi, Y. Aoyagi, K. Senba, T. Kodama, H. Nishikawa, I. Ikemoto, K. Kikuchi, V. T. Volkov, Yu. A. Kasumov, R. Deblock, S. Gu\'eron, and H. Bouchiat, Phys. Rev. B \textbf{72}, 033414 (2005).

\bibitem{vanDam06}
  J. A. van Dam, Y. V. Nazarov, E. P. A. M. Bakkers, S. De Franceschi, and L. P. Kouwenhoven, Nature \textbf{442}, 667 (2006).

\bibitem{Jarillo-Herrero06}
  P. Jarillo-Herrero, J. A. van Dam, and L. P. Kouwenhoven, Nature \textbf{439}, 953 (2006).

\bibitem{Jorgensen06}
  H. I. J{\o}rgensen, K. Grove-Rasmussen, T. Novotn, K. Flensberg, P. E. Lindelof, Phys. Rev. Lett. \textbf{96}, 207003 (2006).

\bibitem{Cleuziou06}
  J.-P. Cleuziou, W. Wernsdorfer, V. Bouchiat, T. Ondar\c{c}uhu, and M. Monthioux, Nature Nanotechnology \textbf{1}, 53 (2006).

\bibitem{Eichler07}
  A. Eichler, M. Weiss, S. Oberholzer, C. Sch\"onenberger, A. Levy Yeyati, J. C. Cuevas, and A. Mart\'in-Rodero, Phys. Rev. Lett. \textbf{99}, 126602 (2007).

\bibitem{Sand-Jespersen07}
  T. Sand-Jespersen, J. Paaske, B. M. Andersen, K. Grove-Rasmussen, H. I. Joergensen, M. Aagesen, C. B. Soerensen, P. E. Lindelof, K. Flensberg, and J. Nygard, Phys. Rev. Lett. \textbf{99}, 126603 (2007).

\bibitem{Eichler09}
  A. Eichler, R. Deblock, M. Weiss, C. Karrasch, V. Meden, C. Sch\"onenberger, and H. Bouchiat, Phys. Rev. B \textbf{79}, 161407(R) (2009).

\bibitem{Winkelmann09}
  C. B. Winkelmann, N. Roch, W. Wernsdorfer, V. Bouchiat, and F. Balestro, Nature Physics \textbf{5}, 876 (2009).


\bibitem{Golubov04}
  A. A. Golubov, M. Yu. Kupriyanov, and E. Il'ichev, Rev. Mod. Phys. \textbf{76}, 411 (2004).

\bibitem{Grabert92}
  H. Grabert and M. H. Devoret, \textit{Single Charge Tunneling: Coulomb Blockade Phenomena in Nanostructures} (Plenum, New York, 1992).

\bibitem{Goldhaber-Gordon98}
  D. Goldhaber-Gordon, H. Shtrikman, D. Mahalu, D. Abusch-Magder, U. Meirav, and M. A. Kastner, Nature \textbf{391}, 156 (1998).

\bibitem{Cronenwett98}
  S. M. Cronenwett, T. H. Oosterkamp, and L. P. Kouwenhoven, Science \textbf{281}, 540 (1998).

\bibitem{vanderWiel00}
  W. G. van der Wiel, S. De Franceschi, T. Fujisawa, J. M. Elzerman, S. Tarucha, and L. P. Kouwenhoven, Science \textbf{289}, 2105 (2000).

\bibitem{Haldane78}
  F. D. M. Haldane, Phys. Rev. Lett. \textbf{40}, 416 (1978).

\bibitem{Izumida01}
  W. Izumida, O. Sakai, and S. Tarucha, Phys. Rev. Lett. \textbf{87}, 216803 (2001).

\bibitem{Hofstetter02}
  W. Hofstetter and H. Schoeller, Phys. Rev. Lett. \textbf{88}, 016803 (2002).

\bibitem{Quay07}
  C. H. L. Quay, J. Cumings, S. J. Gamble, R. de Picciotto, H. Kataura, and D. Goldhaber-Gordon, Phys. Rev. B \textbf{76}, 245311 (2007).

\bibitem{Roch08}
  N. Roch, S. Florens, V. Bouchiat, W. Wernsdorfer, and F. Balestro, Nature \textbf{453}, 633 (2008).

\bibitem{Elste05}
  F. Elste and C. Timm, Phys. Rev. B \textbf{71}, 155403 (2005);
  C. Timm and F. Elste, \textit{ibid.} \textbf{73}, 235304 (2006);
  \textbf{73}, 235305 (2006).

\bibitem{Kogan03}
  A. Kogan, G. Granger, M. A. Kastner, D. Goldhaber-Gordon, and Hadas
  Shtrikman, Phys. Rev. B \textbf{67}, 113309 (2003).

\bibitem{Holm08}
  J. V. Holm, H. I. J{\o}rgensen, K. Grove-Rasmussen, J. Paaske, K. Flensberg, and P. E. Lindelof, Phys. Rev. B \textbf{77}, 161406(R) (2008).

\bibitem{Wilson75}
  K. G. Wilson, Rev. Mod. Phys. \textbf{47}, 773 (1975).

\bibitem{Krishnamurthy80}
  H. R. Krishnamurthy, J. W. Wilkins, and K. G. Wilson, Phys. Rev. B \textbf{21}, 1003 (1980);
  H. R. Krishnamurthy, J. W. Wilkins, and K. G. Wilson, Phys. Rev. B \textbf{21}, 1044 (1980).

\bibitem{Yoshioka00}
  T. Yoshioka and Y. Ohashi, J. Phys. Soc. Jpn. \textbf{69}, 1812 (2000).

\bibitem{Campo05}
  V. L. Campo and L. N. Oliverira, Phys. Rev. B \textbf{72}, 104432 (2005).

\bibitem{Schrieffer66}
  J. R. Schrieffer and P. A. Wolff, Phys. Rev. \textbf{149}, 491 (1966).

\bibitem{Vojta02}
  M. Vojta, R. Bulla, and W. Hofstetter, Phys. Rev. B \textbf{65}, 140405 (2002).

\bibitem{Cragg79}
  D. M. Cragg and P. Lloyd, J. Phys. C \textbf{12}, L215 (1979).

\bibitem{SCQD}
  P. S. Cornaglia and D. R. Grempel, Phys. Rev. B \textbf{71}, 75305 (2005);
  R. Zitko and J. Bonca, Phys. Rev. B \textbf{73}, 35332 (2006).

\bibitem{vanderWiel02}
  W. G. van der Wiel, S. De Franceschi, J. M. Elzerman, S. Tarucha, L. P. Kouwenhoven, J. Motohisa, F. Nakajima, and T. Fukui, Phys. Rev. Lett. \textbf{88}, 126803 (2002).

\bibitem{Pustilnik01}
  M. Pustilnik and L. I. Glazman, Phys. Rev. Lett. \textbf{87}, 216601 (2001).

\bibitem{Hofstetter04}
  W. Hofstetter and G. Zarand, Phys. Rev. B \textbf{69}, 235301 (2004).

\end{thebibliography}
\end{document}